\documentclass[usenatbib]{mnras}

\usepackage[T1]{fontenc}
\usepackage{ae,aecompl}

\usepackage{graphicx,subfigure}	
\usepackage{amsmath}	
\usepackage{amssymb}
\usepackage{booktabs}
\usepackage{pdflscape} 
\usepackage{longtable}
\usepackage{graphicx}
\usepackage{multirow}
\usepackage{xspace}
\usepackage{hyperref}
\usepackage[table]{xcolor}
\usepackage[normalem]{ulem}
\usepackage{siunitx}
\usepackage{bm}

\usepackage{newtxtext,newtxmath} 

\makeatletter
\newcommand{\groupedRowColors}[5][0]{
    \global\rownum=\z@
    \global\@rowcolorstrue
    \@ifxempty{#4}%
        {\def\@oddrowcolor{\@norowcolor}}%
        {\def\@oddrowcolor{\gdef\CT@row@color{\CT@color{#4}}}}%
    \@ifxempty{#5}%
        {\def\@evenrowcolor{\@norowcolor}}%
        {\def\@evenrowcolor{\gdef\CT@row@color{\CT@color{#5}}}}%
    \def\@rowcolors{%
        \if@rowcolors
            \noalign{%
                \relax
                \ifnum\rownum<#3
                    \@norowcolor
                \else \ifodd \numexpr (\rownum-#1)/#2\relax
                    \@oddrowcolor
                \else
                    \@evenrowcolor
                \fi \fi
            }%
        \fi
    }%
    \CT@everycr{\@rowc@lors\the\everycr}%
    \ignorespaces
}
\makeatother

\graphicspath{{./figs/}}
\newcommand{\beq}{\begin{equation}}
\newcommand{\eeq}{\end{equation}}
\def\eq#1{{Eq.~(\ref{#1})}}

\def\sec#1{{Sec.~\ref{#1}}}
\def\tab#1{{Table~\ref{#1}}}
\def\fig#1{{Fig.~\ref{#1}}}
\def\app#1{{Appendix~\ref{#1}}}

\newcommand{\PySM}{{\tt PySM}\xspace}
\newcommand{\Websky}{{\tt WebSky}\xspace}
\newcommand{\CLASS}{{\tt CLASS}\xspace}

\newcommand{\Planck}{{\it Planck}\xspace}
\newcommand{\pixie}{{\it PIXIE}\xspace}

\newcommand{\LiteBIRD}{{\it Litebird}\xspace}

\newcommand{\fnl}{{$f_{\rm NL}$}\xspace}
\newcommand{\fNL}{f_{\rm NL}}
\newcommand{\fsky}{f_{\rm sky}}
\newcommand{\diff}{\mathop{}\!\mathrm{d}}
\newcommand\Diff[1]{\mathop{}\!\mathrm{d^#1}}
\newcommand{\SW}{Sachs-Wolfe\xspace}
\renewcommand{\vec}[1]{{\bm#1}}

\newcommand{\hCl}{\hat{C}_\ell}

\graphicspath{{"./figures/"}}

\title[$f_{NL}$ constraints from \Planck]{Non-Gaussianity constraints with anisotropic $\mu$ distortion measurements from \Planck}

\author[Rotti, Ravenni and Chluba]{
Aditya Rotti$^1$\thanks{aditya.rotti@manchester.ac.uk},
Andrea Ravenni$^{2,3}$\thanks{andrea.ravenni@unipd.it} and
Jens Chluba$^1$\thanks{jens.chluba@manchester.ac.uk}
\\
$^1$ Jodrell Bank Centre for Astrophysics, Alan Turing Building, University of Manchester, Manchester M13 9PL
\\
$^2$ Dipartimento di Fisica e Astronomia ``G. Galilei'', Universit{\`a} degli Studi di Padova, via~Marzolo~8, I-35131, Padova, Italy.\\
$^3$ INFN, Sezione di Padova, via~Marzolo~8, I-35131, Padova, Italy.
}

\date{\vspace{-5mm}{Accepted 2022 --. Received 2022 --}}

\pubyear{2019}

\begin{document}

\maketitle

\begin{abstract}
Primordial non-Gaussianity can source $\mu$-distortion anisotropies that are correlated with the large-scale temperature and polarization signals of the cosmic microwave background (CMB). A measurement of $\mu T$ and $\mu E$ correlations can therefore be used to constrain it 
on wavelengths of perturbations not directly probed by the standard CMB anisotropies. 
In this work, we carry out a first rigorous search for $\mu$-type spectral distortion anisotropies with \Planck data, applying the well-tested constrained ILC component-separation method combined with the needlet framework. We reconstruct a $\mu$ map from \Planck data, which we then correlate with the CMB anisotropies to derive constraints on the amplitude $\fNL$ of the local form bispectrum, specifically on the highly squeezed configurations with effective wavenumbers $k_s \simeq \SI{740}{Mpc^{-1}}$ and $k_L \simeq \SI{0.05}{Mpc^{-1}}$. 
We improve previously estimated constraints by more than an order of magnitude. This enhancement is owing to the fact that for the first time we are able to use the full multipole information by carefully controlling biases and systematic effects in the final analysis. We also for the first time incorporate constraints from measurements of $\mu E$ correlations, which further tighten the limits. A combination of the derived \Planck $\mu T$ and $\mu E$ power spectra yields $|\fNL| \lesssim 6800$ (95\% c.l.) on this highly squeezed bispectrum.
This is only $\simeq 3$ times weaker than the anticipated constraint from \LiteBIRD alone. Our analysis highlights the importance of low and high-frequency channels. We show that a combination of \LiteBIRD with \Planck will improve the expected future constraint by $\simeq 20\%$ over \LiteBIRD alone. 
These limits can be used to constrain multi-field inflation models and primordial black hole formation scenarios, thus providing a promising novel avenue forward in CMB cosmology.
\end{abstract}


\vspace{0mm}

\section{Introduction}
\label{sec:intro}
In the past decades, data from the \Planck surveyor and other cosmic microwave background (CMB) experiments has been thoroughly analyzed to extract the vast majority of the available cosmological information. In particular, the analyses of the CMB temperature and polarization anisotropies have yielded an unprecedented understanding of the $\Lambda$CDM model \citep{Planck_2018_results_IX}. However, at the same time it pointed us towards possible inconsistencies \citep[e.g.,][]{Planck2014anomalies, Planck2016SZ, Valentino2021Hubble}, that might become a gateway to the next major discoveries in cosmology.

Even if a bright future in CMB cosmology lies ahead with \LiteBIRD \citep{Hazumi2019,LiteBIRD:2022cnt}, The Simons Observatory \citep{so_forecast_2019} and CMB-S4 \citep{Abazajian2016S4SB}, to make progress we ultimately need to access new observables beyond the CMB temperature and polarization signals. One of these new observables is from spectral distortions of the CMB \citep{Chluba:2019nxa}. The information encoded in the pixel-by-pixel deviations of the sky-signal from a blackbody has already delivered us an exquisite model of the local universe, by mapping, for example, extragalactic objects via the Sunyaev-Zeldovich effect \citep{Zeldovich1969}, and our galaxy's dust. However, with $\mu$-type distortions we can glean information from the primordial Universe, probing processes that occurred just a few months after the big bang through measurements of the average CMB spectrum \citep[see][for modern reviews of CMB spectral distortion physics]{Chluba2011therm, Sunyaev2013, Lucca2020}.

While \Planck and other CMB imagers are not able to provide measurements of the average spectral distortion --- a spectrometer akin to \pixie \citep{Kogut2011PIXIE, Kogut2016_pixie} is needed for that --- it can be used to measure their spatial variation \citep{Pajer:2012vz,Ganc:2012ae}.
In $\Lambda$CDM and also in most extensions being discussed in the literature, the power spectrum of primordial spectral distortions is currently too faint to be directly observed \citep{Pajer:2012vz, Chluba:2016aln}.
However, if the distortion signals were correlated with CMB temperature or polarization, both much more intense signals which would increase the signal-to-noise ratio, we could possibly have better chances of detection.

One source of spectral distortions is the dissipation of primordial acoustic modes on small scales, $\SI{50}{Mpc^{-1}} < k < \SI{2e4}{Mpc^{-1}}$ for $\mu$-type distortions \citep{Sunyaev1970diss, Daly1991, Hu1994, Chluba:2012gq}.
Thus, models in which the small-scale power is modulated by the long-wavelength perturbations are the perfect candidate to be studied with spectral distortion cross correlations.
The first model to be considered for this kind of analysis is local non-Gaussianity \citep{Pajer:2012vz, Emami:2015xqa, Ota:2016mqd, Chluba:2016aln, Ravenni:2017lgw, Cabass:2018jgj}.
Even though many other interesting models have also been put forward \citep[e.g.,][]{Ganc:2012ae, Ozsoy:2021qrg, Zegeye2021, Orlando:2021nkv, Ozsoy:2021pws}, here we shall focus on the former.
Local model non-Gaussianity has been studied, among other templates, with measurements of the CMB temperature and polarization bispectrum, the Fourier transform of the 3-point correlation function.
The tightest constraint to date on its amplitude (more formally defined in the next section) is $\fNL=-0.9 \pm 5.1$, which have been set by \cite{Planck_2018_results_IX}.
This limit is valid on CMB anisotropy scales which have typical wavenumbers of $k_0\simeq 0.05\,{\rm Mpc}^{-1}$.
In fact, $\fNL$ does not need to be constant in general; various models predict some scale dependence \citep{Dimastrogiovanni:2016aul, Byrnes:2010ft, 2011JCAP...03..017S, Chen:2005fe}.
Insight on much smaller scales can be gained by looking at the aforementioned  cross correlations of spectral distortions anisotropies with the temperature and polarization anisotropies, which we pursue here using \Planck data.

In this paper we provide the first constraints on the \fnl parameter that make use of the whole $\mu T$ and $\mu E$ cross-correlation information. 
This improves the limit quoted in previous attempts of carrying out a similar task \citep{Khatri2015mu} by more than one order of magnitude (see Sect.~\ref{sec:planck_muT_compare} for a  detailed comparison).
As easily expected, and as we will show here, pursuing a faint signal such as the $\mu$ distortion anisotropies in data from a survey that was not optimized for this goal is challenging, and requires us to take into account many usually neglected details to avoid biasing the results. To this end we validate and tune our whole pipeline on state-of-the-art simulations, that we use to motivate all our assumptions and analysis decisions. 
As shown in \cite{Remazeilles2018}, the leading source of bias is leakage of temperature anisotropies in the $\mu$ maps, which we address with constrained needlet ILC (cNILC), as proposed there.
We prove that leakage of other foregrounds, while present, do not impact the analysis in a significant way and shall be neglected for \Planck.
Other sources of bias can be introduced by mis-modelling of the instrument beam, and in much smaller part by channel mis-calibration.
As argued in \cite{Ravenni:2017lgw, Remazeilles2021mu}, the cross-correlation with CMB polarization represents a much cleaner signal to work with; the agreement in the results we obtain with the two methods provides an important sanity check for the analysis of temperature maps.

Our work is structured as follows:
the theoretical cross correlations sourced in the presence of local non-Gaussianity are reviewed in \sec{sec:signals}.
The next two sections are devoted to the statistical tools used in our pipeline and how they were optimized.
In \sec{sec:fnl_lkl} we define and discuss the likelihood and the estimator we use to constrain \fnl.
We validate our findings with two preliminary analysis, a Fisher matrix forecast described in \sec{sec:fisher_forecast}, and by running our whole pipeline on simulated \Planck maps in \sec{sec:planck_forecast}.
In \sec{sec:comp_sep} we highlight a few details about the component separations and their expected interplay with the signals we are trying to extract.
The actual analysis on \Planck data is presented in \sec{sec:planck_analysis}. In \sec{sec:NextGenerationForecast} we provide an outlook on future surveys, just before concluding in \sec{sec:conclusions}.


\vspace{-3mm}
\section{The $\mu T$ and $\mu E$ spectra}
\label{sec:signals}
If the primordial perturbation field were Gaussian it would be completely described by its power spectrum
\begin{equation}
    \langle
        \zeta_\vec{k_1} \zeta_\vec{k_2}
    \rangle
    =
    (2\pi)^3
    \delta^{(3)}(\vec{k_1} + \vec{k_2}) 
    P_\zeta(k)\, .
\end{equation}
Deviations from Gaussianity are encoded in higher order correlation functions. In the weakly coupled regime, the primordial potential perturbation can be parametrized in real space as a Gaussian term plus a quadratic correction whose amplitude is the non-linearity parameter $\fNL$ \citep{Salopek:1990jq}.
It is then easy to show that $\fNL$ controls the amplitude of the primordial bispectrum, which is defined as 
\begin{equation}
    \langle
        \zeta_\vec{k_1} \zeta_\vec{k_2}  \zeta_\vec{k_3}
    \rangle
    =
    (2\pi)^3
    \delta^{(3)}(\vec{k_1} + \vec{k_2} + \vec{k_3}) 
    B_\zeta(k_1, k_2, k_3) \, ,
\end{equation}
which reads \citep[e.g.,][]{Gangui:1993tt, Verde:1999ij, Komatsu:2001rj}
\begin{equation}
    B_\zeta^\text{loc}
    (k_1, k_2, k_3)
    = 
    \frac{6}{5} \fNL
    \left[
        P_\zeta(k_1) \, P_\zeta(k_2)
        + 2 \text{ perms.}
    \right] .
\end{equation}
This bispectrum shape peaks in the squeezed limit ($k_1 \ll k_2 \approx k_3$), which CMB-spectral distortion cross correlations are most sensitive to, and is especially relevant to distinguishing between single-field and multi-field inflationary models \citep[see][and references therein]{Planck_2018_results_IX} which predict $\fNL \lesssim 1$ and $\fNL \gtrsim 1$, respectively. 

We should bear in mind that very high values of $\fNL$ (i.e., $\simeq 10^4-10^5$) might invalidate the perturbative expansion in primordial fluctuation. Finding unexpectedly high power in the $\mu T$ and $\mu E$ measurements might require non-perturbative modelling of these non-Gaussian fluctuations in the early universe.
  Here we focus on building a reliable pipeline to analyse CMB data, and not on furthering the theoretical modelling of poly-spectra in the highly non-Gaussian regime. Given the need to test our framework against simulation with high signal to noise to guarantee a recovery, we assume that the results which are valid for relatively low values of $\fNL$ scale linearly to  higher values. We take for granted that a detection of high $\fNL$ would require further theoretical work to be interpreted consistently.

In such a model, the cross power spectra of $X = T,E$ and $\mu$-distortion anisotropies can be cast in the form
\begin{align}
\label{eq:xmu}
C_\ell^{\,\mu X}
&\approx
    12 \fNL^{\mu}
    \int \diff k \,
    \frac{2}{\pi} \,
    \frac{1}{5} \,
     k^2 \, 
    \mathcal{T}_\ell^{X/\zeta}(k) \,
    j_\ell(k \, r_\text{ls})\,
    P_\zeta(k)
\\ \nonumber
    &
    \qquad \times
    \int
    \frac{\Diff{3} \vec{k'}}{(2\pi)^{3}}
    f^{\mu}(k, k, k')
    P_\zeta(k')
    \,.
\end{align}
Here $\mathcal{T}_\ell^{X/\zeta}(k)$ is the transfer function of temperature or polarization, calculated using \CLASS \citep{Blas:2011JCAP...07..034B}.
The spherical Bessel functions $j_\ell(k \, r_\text{ls})$ account for the angular projection of SD inhomogeneities on the last scattering surface, at a comoving distance $r_\text{ls}$.
$f^\mu(k_1, k_2, k_3)$ is the $\mu$ window function that we take here to be \citep{Pajer:2012vz, Ganc:2012ae, Emami:2015xqa, Chluba:2016aln}
\begin{align}
    f^\mu(k_1, k_2, k_3)
    &\approx 
    2.27 
    \left[
        {\rm e}^{-(k_1^2 + k_2^2)/k_{\rm D}(z)}
    \right]_{z_{\mu y}}^{z_\mu}
    \Pi\left(\frac{k_3}{k_{\rm D}(z_{\mu y})}\right)
\\
\nonumber
    \Pi(x)
    &=
    \frac{3 j_1 (x)}
    {x},
\end{align}
where $k_{\rm D}(z_{\mu})\approx\SI{12000}{Mpc^{-1}}$ and $k_{\rm D}(z_{\mu y})\approx\SI{46}{Mpc^{-1}}$ are the diffusion damping scales at the beginning and end of the $\mu$-distortion era.
Improved expressions can be found in \cite{Chluba:2016aln}, however, for the purpose of the present analysis, these refinements will not affect the results significantly.

Notice that the first integral in \eq{eq:xmu} would match the temperature power spectrum in the \SW limit if one were to take $X=T$ and approximate the transfer function $\mathcal{T}_\ell^{X/\zeta}(k) \approx -j_\ell(k \, r_\text{ls})/5$.
Moreover, the integral in the second line is approximately equivalent to the sky averaged $\mu$ distortion sourced by dissipation of acoustic modes.
Thus one obtains the approximated relation \citep{Chluba:2016aln}
\begin{equation}
    C_\ell^{\,\mu X}
    \approx
    - 12 f_{\rm NL}^\mu\, \langle \mu \rangle\, \frac{2\pi}{25}\frac{A_\text{s}}{\ell(\ell+1)}\,.
\end{equation}
However, notice that we will always employ the more refined version presented before.

The $\mu$-distortion power spectrum has two contributions, a generally negligible Gaussian term \citep{Pajer:2012vz, Ganc:2012ae}, and a non-Gaussian one which, within the set of assumptions discussed above, reads \citep{Chluba:2016aln}
\begin{align}
C_\ell^{\mu \mu} & \approx 144\, \left(f_{\rm NL}^\mu\right)^2\, \langle \mu \rangle^2\, \frac{2\pi}{25}\frac{A_\text{s}}{\ell(\ell+1)}\,.
\end{align}
We note that the above description neglects any effect of perturbations on a possible average distortions of the CMB \citep{Chluba:2012gq}. This will source correlated $\mu$-distortion fluctuations that will become noticeable at the level $\fNL\simeq 1$ \citep{Kite2022}.


\vspace{-3mm}
\section{Likelihood analysis}
\label{sec:fnl_lkl}
We will employ a simple Gaussian likelihood to infer the measurement statistics on the \fnl parameter, 
\begin{align}
-2 \ln \mathcal{L}=\sum_{\ell}\left(\hat{C}_{\ell}^{X} - f_{\rm NL} C_{\ell}^{X,{\rm Thry.}} \right)\mathcal{C}_X^{-1}\left( \hat{C}_{\ell}^{X} - f_{\rm NL} C_{\ell}^{X,{\rm Thry.}} \right) \,,
\label{eq:fNL_likelihood}
\end{align}
where $\hat{C}_{\ell}$ denotes the power spectrum derived from the pseudo $C_{\ell}$ estimator.
In principle, the Gaussian assumption becomes inaccurate at low multipoles \citep{Hamimeche2008}. 
However, in our case it will still be valid by virtue of the fact that we work with spectra that are binned using wide multipole bins ($ \Delta \ell =32$), though we do not try to quantify this detail further.

We report the constraints on the \fnl  parameter derived from measurements of $\mu T$, $\mu E$ power spectra as well as the ultimate constraints resulting from appropriately combining the two measurements, duly accounting for the covariance between them. Applying Wick's theorem for simplifying the following covariance between the spectra: $\langle (\hat{C}_{\ell}^{X} - \langle \hat{C}_{\ell}^{X} \rangle ) (\hat{C}_{\ell}^{X'} - \langle \hat{C}_{\ell}^{X'} \rangle ) \rangle$ it can be shown that the covariance takes the general form,
\begin{align}
\mathcal{C} = \begin{bmatrix}
\frac{\hat{C}^{TT}_{\ell_{\rm bin}}\hat{C}^{\mu\mu}_{\ell_{\rm bin}} + (\hat{C}^{\mu T}_{\ell_{\rm bin}})^2}{(2 \ell_{\rm bin}+1) \Delta \ell f_{\rm sky}} & \frac{\hat{C}^{TE}_{\ell_{\rm bin}}\hat{C}^{\mu\mu}_{\ell_{\rm bin}} + \hat{C}^{\mu T}_{\ell_{\rm bin}}\hat{C}^{\mu E}_{\ell_{\rm bin}}}{(2 \ell_{\rm bin}+1) \Delta \ell f_{\rm sky}} \\ 
\frac{\hat{C}^{TE}_{\ell_{\rm bin}}\hat{C}^{\mu\mu}_{\ell_{\rm bin}} + \hat{C}^{\mu T}_{\ell_{\rm bin}}\hat{C}^{\mu E}_{\ell_{\rm bin}}}{(2 \ell_{\rm bin}+1) \Delta \ell f_{\rm sky}} & \frac{\hat{C}^{EE}_{\ell_{\rm bin}}\hat{C}^{\mu\mu}_{\ell_{\rm bin}} + (\hat{C}^{\mu E}_{\ell_{\rm bin}})^2}{(2 \ell_{\rm bin}+1) \Delta \ell f_{\rm sky}}
\end{bmatrix} \,.
\end{align}

Note that all the spectra appearing in the above equation correspond to those evaluated from data and therefore already encode the measurement noise. While this form of the covariance is relevant when estimating the \fnl constraints from combining $\mu T$ \& $\mu E$ spectra, we work with the appropriate diagonal terms when estimating constraints individually with $\mu T$ and $\mu E$ spectra. We evaluate the covariances using the respective power spectra estimated from the component separation maps.
Note that while the analysis of $\mu T$ spectrum measurements rely on cross correlating half mission data products, the $\mu E$ spectrum measurements rely on using full mission data products. Consequently, there are more details to the covariance estimates which we elucidate in  \app{app:FM_HM_comparison}.

\vspace{-3mm}
\section{Fisher forecasts}
\label{sec:fisher_forecast}
\begin{figure*}
\subfigure[\label{fig:ideal_ilc_noise}]{\includegraphics[width=\columnwidth]{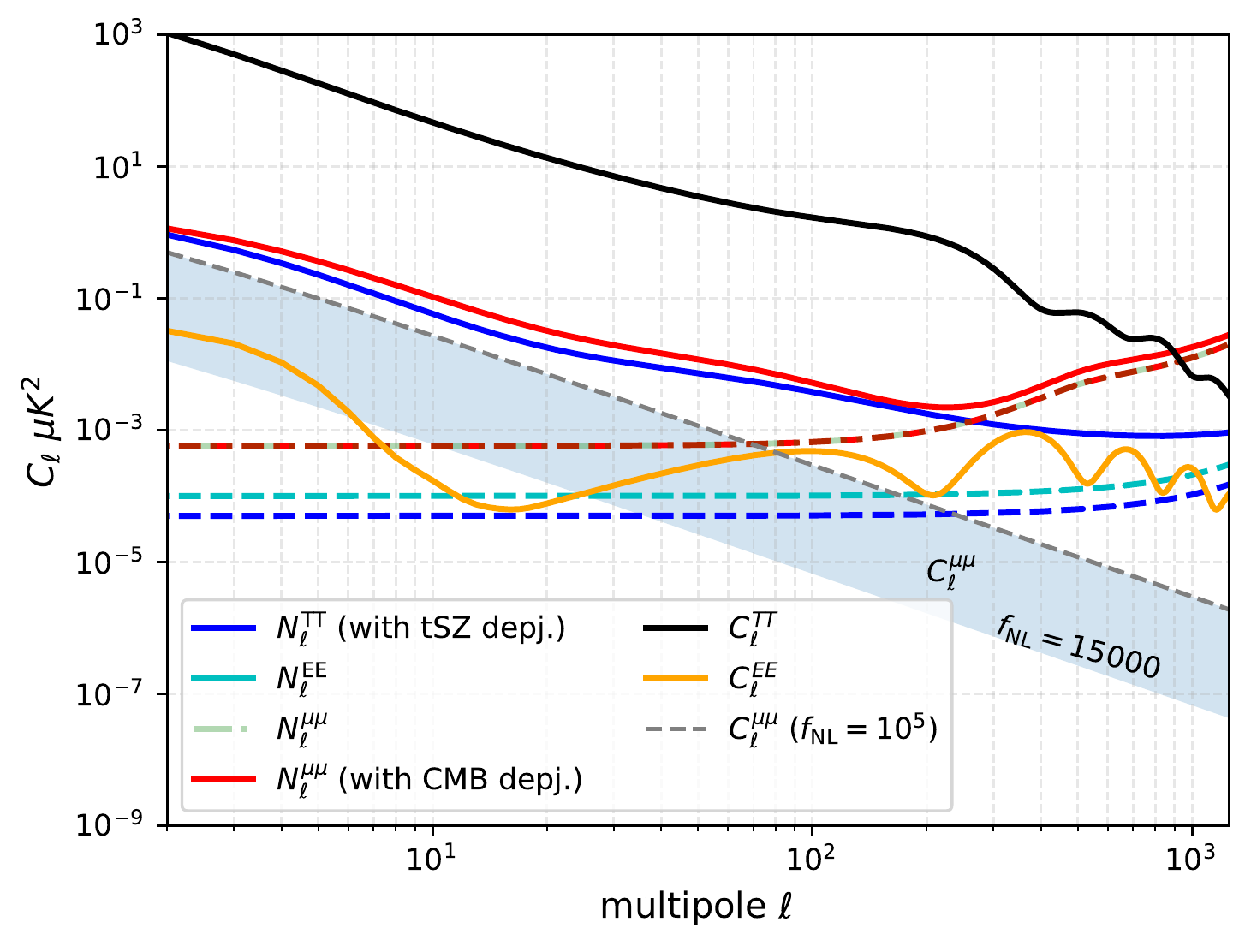}}
\subfigure[\label{fig:ideal_muT_muE_det}]{\includegraphics[width=\columnwidth]{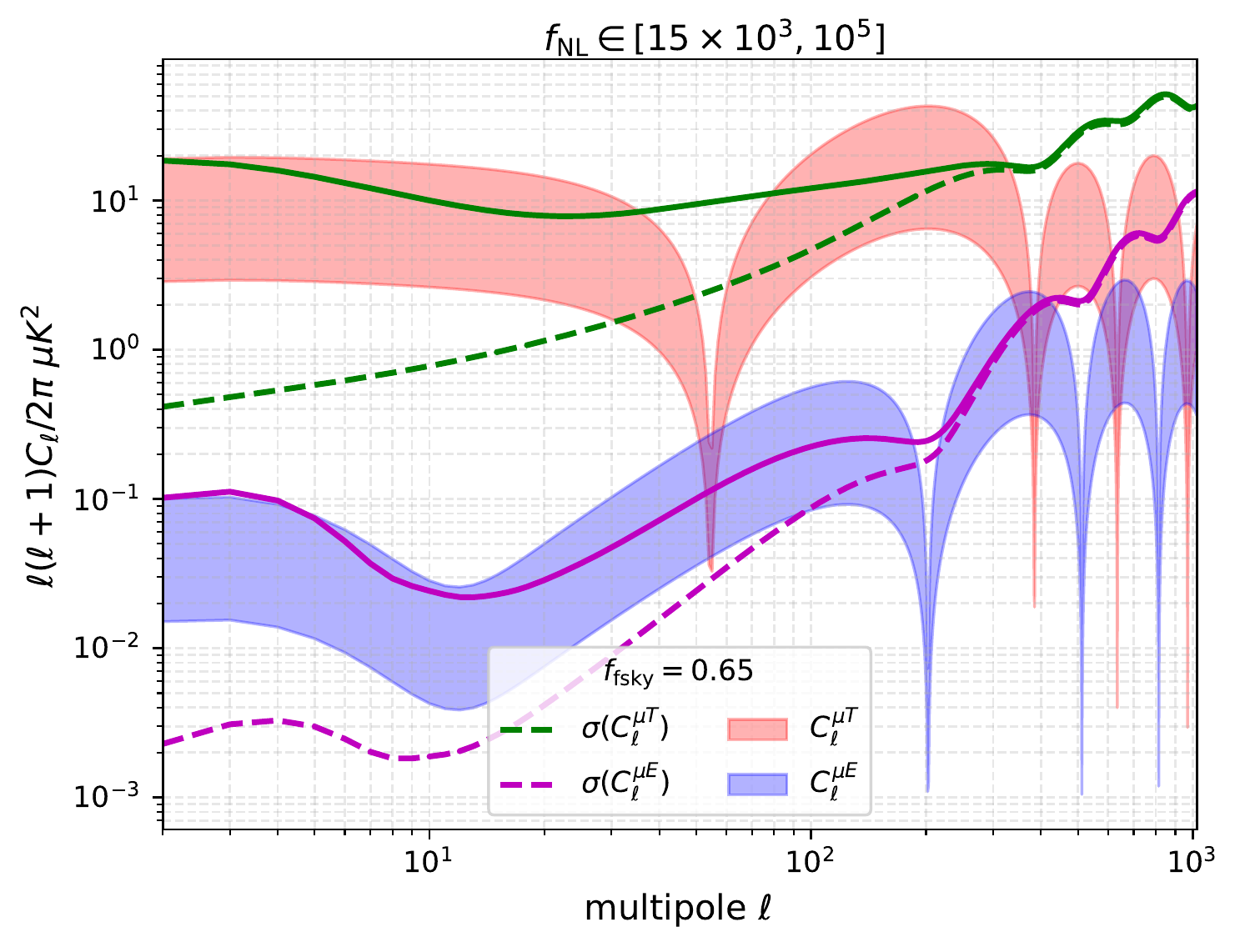}}
\caption{Left : This figure depicts the ideal ILC noise in the reconstructed T, $E$ and $\mu$ maps, expected for the \Planck instrument. While the dashed lines depict the ideal noise power spectra, the solid lines includes the estimated degradation due to realistic foregrounds. The temperature, E-mode and anisotropic $\mu$ map power spectra for a range of \fnl values are also shown for context. Right : This figure depicts the expected $C_{\ell}^{\mu T}$ and $C_{\ell}^{\mu E}$ spectra, the top and lower edged of the bands corresponding to $f_{\rm NL}=10^5$ and $f_{\rm NL}=15\times10^3$ respectively. Also shown are the expected errors on the measured power spectra using the ILC noise estimates, which assume a $f_{\rm sky}=0.65$. While the dashed lines show the ideally expected noise power spectrum, the solid lines incorporate the foreground degradation factor.} 
\end{figure*}
Before venturing into presenting results from actual component separated maps, we carry out error estimates on $f_{\rm NL}$ through a Fisher matrix forecast based on simple simulations of noise in the component separated maps. 
The Fisher information --- for the single parameter \fnl in our case --- is defined as 
\begin{equation}
    F_{\fNL}
    =
    \left \langle \left (
        \frac{\partial \ln \mathcal{L}}{\partial \fNL}
    \right )^{\! 2} \right \rangle
\label{eq:Fisher_matrix_definition}
\end{equation}
which is equivalent to the expectation value of the Hessian of $\mathcal{L}$.
The Cramer-Rao bound:  $\sigma_{\fNL}^{2} \geq F_{\fNL}^{-1}$ sets a lower limit on the variance of a parameter's unbiased estimator. Despite its limitation of being an inequality, this bound is still a valuable way to very inexpensively forecast errors. Using Eq. \eqref{eq:fNL_likelihood} in Eq. \eqref{eq:Fisher_matrix_definition}, one arrives at an estimate of the error on \fnl parameter
\begin{align}
    \sigma_{\fNL}= \left[ \sum_\ell C_\ell^{X,\text{Thry.}} \mathcal{C}_X^{-1} C_\ell^{X,\text{Thry.}} \right]^{-1/2}\, .
\end{align}
Evaluating the above equation requires the theoretical $\mu T$ and $\mu E$ spectra which were discussed in \sec{sec:signals} as well as an estimate of the covariance matrix $\mathcal{C}_X$.

The noise in any component separated maps at a multipole $\ell$ is given by \citep{Tegmark2003_ilc}
\begin{align}
\label{eq:error_ILC}
\sigma_\text{ILC} = \left [ s_{\nu} \mathcal{C}^{-1}_{\ell, \nu\nu'} s_{\nu'} \right] ^{-1} \, ,    
\end{align}
where $\mathcal{C}$ is the data covariance matrix, and $s_\nu$ is the spectral energy density vector of the target component, calculated at the observed channels. We derive the forecasts in two cases. In the first, idealized case we only consider CMB, tSZ, $\mu$, and instrumental noise as components of the data.
In this case the covariance matrix expressed in thermodynamic temperature units takes the simple form: $\mathcal{C}_{\ell, \nu \nu'} =N_{\ell}^{\nu} \delta_{\nu \nu'} + C^{\nu \nu', \rm CMB}_{\ell} + C^{\nu \nu', \rm tSZ}_{\ell}$, where the $\delta$-function encodes the fact that the measurement noise in different channels is uncorrelated. 
In the second, more realistic case, we include an exhaustive list of cosmological and galactic foregrounds: dust, synchrotron, free-free, radio and infrared sources \citep{Abitbol:2017vwa, Hill:2013baa, Tegmark:1999ke, Dunkley:2013vu}. Specifically this is done by including a multi-frequency power spectral model of each of the foreground components into the estimate of the data covariance matrix as prescribed in  \citep{Cooray2001, Hill:2013baa}. We employ the implementation of \cite{Ravenni:2020rzd}, which was developed to assess the role of higher order statistics in future SZ analyses, and refer the interested reader to for a more detailed discussion.

As we will see, deprojecting CMB when reconstructing the $\mu$ map is essential and the noise in the reconstructed $\mu$ maps in principle differs from the standard ILC noise estimate discussed above.  In the case of constrained ILC, the noise in the reconstructed $\mu$ map is given by the following expression,
\begin{equation}
\label{eq:error_cILC}
\sigma_\text{cILC} = 
\left [ \left(s^{\mu}_{\nu},s^{\Delta T}_{\nu}\right)^{T} \mathcal{C}^{-1}_{\ell, \nu\nu'}  \left(s^{\mu}_{\nu'},s^{\Delta T}_{\nu'}\right) \right]^{-1}_{0,0} \, .    
\end{equation}
One expects these spectral deprojection procedures to result in a noise penalty as one is simultaneously solving for additional parameters. However, we find that at the most relevant multipoles (i.e., $\ell \lesssim 1000$), spectral deprojection does not result in any additional noise penalty as seen in \fig{fig:ideal_ilc_noise}. We do find that for multipoles $\ell \gtrsim$1000, the noise in the CMB deprojected $\mu$ map is mildly enhanced compared to the conventional ILC noise, but these multipoles will not play a significant role in the analysis presented here.

An equivalent procedure is also used to recover the noise spectrum of temperature maps where tSZ has been deprojected, as this is again potentially important to avoid biases. However, since the dominant contribution to the noise is from the CMB cosmic variance, any subtle changes in the CMB measurement noise levels is expected to have an insignificant effect.

We begin by comparing the estimated noise power spectra with the $\mu$ power spectra that can be expected from a range of values for the \fnl parameter in \fig{fig:ideal_ilc_noise}. We note the $\mu$ auto power spectrum corresponding to an $f_{\rm NL}=10^5$ could already be detected by \Planck at high significance, even conservatively assuming the connected component of the primordial trispectrum to be zero. 
This simple assessment already indicates that the analysis of \cite{Khatri2015mu} may have been too conservative.
Note, however, that this detection would only be possible on large angular scales, corresponding to $\ell \lesssim 100$. On incorporating the effects of foregrounds, we observe that there the noise is significantly enhanced, greatly diminishing the ability to extract this possible signal.

Our main goal in this work is to estimate the $f_{\rm NL}$ parameter via measurements of the $\mu T$ and $\mu E$ cross power spectra. These are stronger signals, owing to the large $T$ \& $E$ contributions and as such can be expected to be measured with greater ease compared to the $\mu$ auto-correlation signal. This is quantified in \fig{fig:ideal_muT_muE_det}, where we see that $C_{\ell}^{\mu T}$ and $C_{\ell}^{\mu E}$ spectra corresponding to $f_{\rm NL}=15 \times 10^3 - 10^5$ are clearly above or comparable to the expected level of noise, on a much larger set of multipoles. Note that here the error estimates already account for the $f_{\rm sky}$ degradation factor. This continues to be true even on taking into account the degradation due to foregrounds, as can be inferred from comparing the signal power spectra with the respective solid lines in \fig{fig:ideal_muT_muE_det}.
%
%
%
\begin{table}
\centering
\groupedRowColors{3}{2}{gray!10}{white!10}
\begin{tabular}{lllll}
\toprule
$f_{\rm NL}$       &      Data              &  $\sigma_{f_{\rm NL}}$ &  $\sigma_{f_{\rm NL}}$ (frg. dgr.) \\
\midrule
25000 & $\mu T$ &  1542 &              2980 \\
      & $\mu E$ &  1236 &              2847 \\
      & $\mu T$ \& $\mu E$ &  1096 &              2334 \\
12500 & $\mu T$ &  1301 &              2975 \\
      & $\mu E$ &  1113 &              2841 \\
      & $\mu T$ \& $\mu E$ &   969 &              2329 \\
0     & $\mu T$ &   937 &              2973 \\
      & $\mu E$ &   915 &              2839 \\
      & $\mu T$ \& $\mu E$ &   761 &              2328 \\
\bottomrule
\end{tabular}
\caption{Fisher error estimates for different values of $f_{\rm NL}$ using the ILC noise estimate for \Planck, assuming $\ell \in [2,1024]$ and $f_{\rm sky}=0.65$.}
\label{tab:ilc_fnl_forecast}
\end{table}

We further quantify the expected errors on a range of values for the $f_{\rm NL}$ parameters. These are estimated by passing the fiducial spectra together with the ILC noise estimate in appropriate combinations to our likelihood module. We derive the \fnl error estimates in the ideal case as well as in the case including the effect of foregrounds, the results are summarized in \tab{tab:ilc_fnl_forecast}. Since \Planck data yields cosmic variance limited temperature anisotropy measurements to very high multipoles $\ell \simeq 1000$, we find that the degradation in $N_{\ell}^{\rm TT}$ due to foregrounds does not play an important role in the \fnl noise estimates. The E-mode maps being noise dominated at all multipole, we assume the simple ideal ILC noise estimates to be valid in both the cases we study here. We find the dominant factor for the increase of \fnl errors is the foreground degradation of $N_{\ell}^{\mu \mu}$. Note that the errors change significantly as a function of $f_{\rm NL}$ in the ideal noise case and this is a reflection of the change in the signal amplitude which contributes to the covariance. On the contrary on inclusion of foreground degradation in the noise estimate, the changes to the covariance due to changing \fnl are significantly suppressed due to enhanced noise levels caused by foregrounds [see \fig{fig:ideal_muT_muE_det}].

\begin{figure}
\includegraphics[width=\columnwidth]{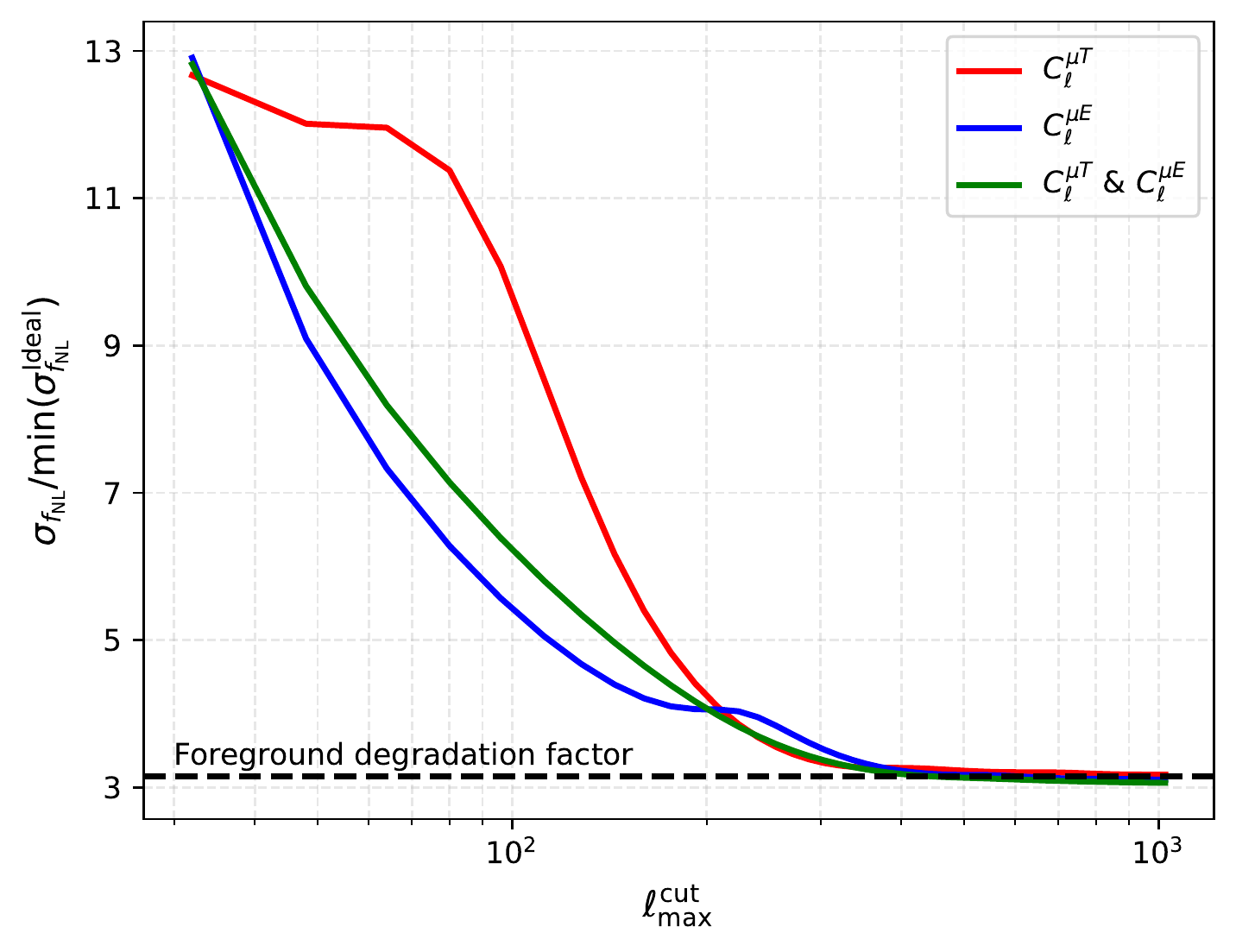}
\caption{This figure depicts the percentage change in error on \fnl with respect to the ideal errors (estimated assuming only noise limited measurements, using $\ell_{\rm max}^{\rm cut}=1024$), as a function of the maximum multipole used in the likelihood analysis.}
\label{fig:ideal_ilc_err_evol}
\end{figure}

Finally we use this setup to assess the multipoles that can be expected to provide the most constraining power from $C_{\ell}^{\mu T}$, $C_{\ell}^{\mu E}$ and a combination of the two measurements. We estimate ``ideal'' errors on \fnl assuming only measurement noise as well as the ``realistic'' error where we additionally account for the foregrounds. \fig{fig:ideal_ilc_err_evol} depicts how the realistic errors evolve as a function of the maximum multipole used in the analysis expressed in units of the ideal \fnl errors estimated assuming $\ell_{\rm max}=1024$. We observe that with $C_{\ell}^{\mu T}$ spectra the $f_{\rm NL}$ errors are a factor of $\simeq$4 worse than the best possible errors and evolve only mildly up to to inclusion of $\ell \simeq 80$, there after which we see a rapid improvement on including higher multipoles in the analysis. For errors derived from $C_{\ell}^{\mu E}$ we observe an almost monotonic decline on progressively increasing the maximum multipole used in the likelihood analysis. The combined errors are seen to follow the same trend as those derived from $C_{\ell}^{\mu E}$. Finally we note that the error reduction is saturated on inclusion of multipoles up to $\ell \simeq 400$. Overall we see the \fnl errors are degraded by a factor of $\sim 3$  due to presence of foregrounds compared to the ideal errors set by measurement noise alone as seen in \fig{fig:ideal_ilc_err_evol}. Having set some benchmarks from these simple exercises, we now shift our attention to discussing more realistic analyses.

\vspace{-3mm}
\section{Component separation}
\label{sec:comp_sep}
We employ the Needlet Internal Linear Combination (NILC) component separation pipeline in this work. The implementation details follow closely those discussed in the original papers \citep{Delabrouille2009, Basak2012}. However, we note that the details provided in these papers are not sufficient for one to be able to develop a setup with similar performance. A needlet ILC implementation essentially requires making choices on the shape and size of needlets and the sizes of pixel domains over which the direction dependent data covariance is estimated. These choices in general can vary for the different components that need to be extracted \citep[e.g., see for][]{planck_tsz_2015}.  We emphasize that there is significant amount of tuning needed to arrive at the optimal set-up. In particular, no formal procedure for obtaining these optimal choices is available, indicating that more theoretical work is required on this front.

For temperature map reconstruction we adjust NILC parameters and repeatedly apply the pipeline to simulated data to yield the injected power spectrum at approximately percent level precision at most multipoles. Temperature anisotropies being a relatively strong signal, suffer from issues of ILC bias, and these are mitigated by choosing relatively large pixel domains over which the covariance is estimated. For $\mu$ anisotropy reconstruction, we find that the signal is tiny and we need very good control on foreground residuals. To achieve this we find that estimating the covariance on relatively (w.r.t T) smaller pixel domains helps with better minimizing foregrounds. Since \Planck E-mode measurements are relatively noisy, and there are fewer polarized foregrounds, we find tuning NILC for E-mode reconstruction to be relatively simple and yields comparable performances for a range of pixel domain sizes. 
Eventually the robustness of the setup is ensured by checking that one can recovers the injected (i.e.x simulated) signal.

xAnother important detail of the NILC setup is that we mask a copy of the multi-frequency maps, excising regions with some of the brightest foregrounds. These masked multi-frequency maps are then individually processed, where in we add back some of the large angle power in  masked region, using a diffusive in-painting algorithm\footnote{This basically involves assigning to each masked pixel a value which is the mean of pixels within some radius. We iterate over this procedures a few times until convergence.} \citep{Planck2015_comp_sep}. We also reintroduce small angle power by injecting realizations of the noise simulated duly incorporating the instrument characteristics for each frequency. We perform this post processing in order to mitigate ringing close to the sharp edges of the mask, which reduces potential biases even in the covariance estimates for pixels that are far from these bright foreground regions, but are likely to be used in the final power spectrum analysis. This approach is expected to yield better foreground minimization, particularly in regions which will be used in the final analysis. Clearly the performance of NILC  in minimizing foregrounds is severely compromised in regions which were masked, but those regions are in any case not used in the final analysis. We stress that most of the tuning of our NILC algorithm is performed when analyzing the semi-realistic \Planck like simulations, thereafter we leave the pipeline almost untouched.

A more quantitative demonstration and discussion of the ideas presented above and their benefits is beyond the scope of this work, as it strongly deviates from the primary science objective. We will however demonstrate some important aspects of our NILC setup by showcasing analysis results on simulated \Planck data and a direct comparison to the injected component maps. Furthermore we also employ our NILC implementation on \Planck data to recover the temperature and E-mode maps from \Planck multi-frequency maps. Finding consistency between our component separated maps and those published by the \Planck collaboration will serve as an additional check on our component separation pipeline. Demonstrations validating our implementation of the NILC algorithm are presented in \app{app:sim_spec_compare} and \app{app:smica_spec_compare}.

\begin{figure}
\hspace*{-0.18cm} 
\includegraphics[width=\columnwidth]{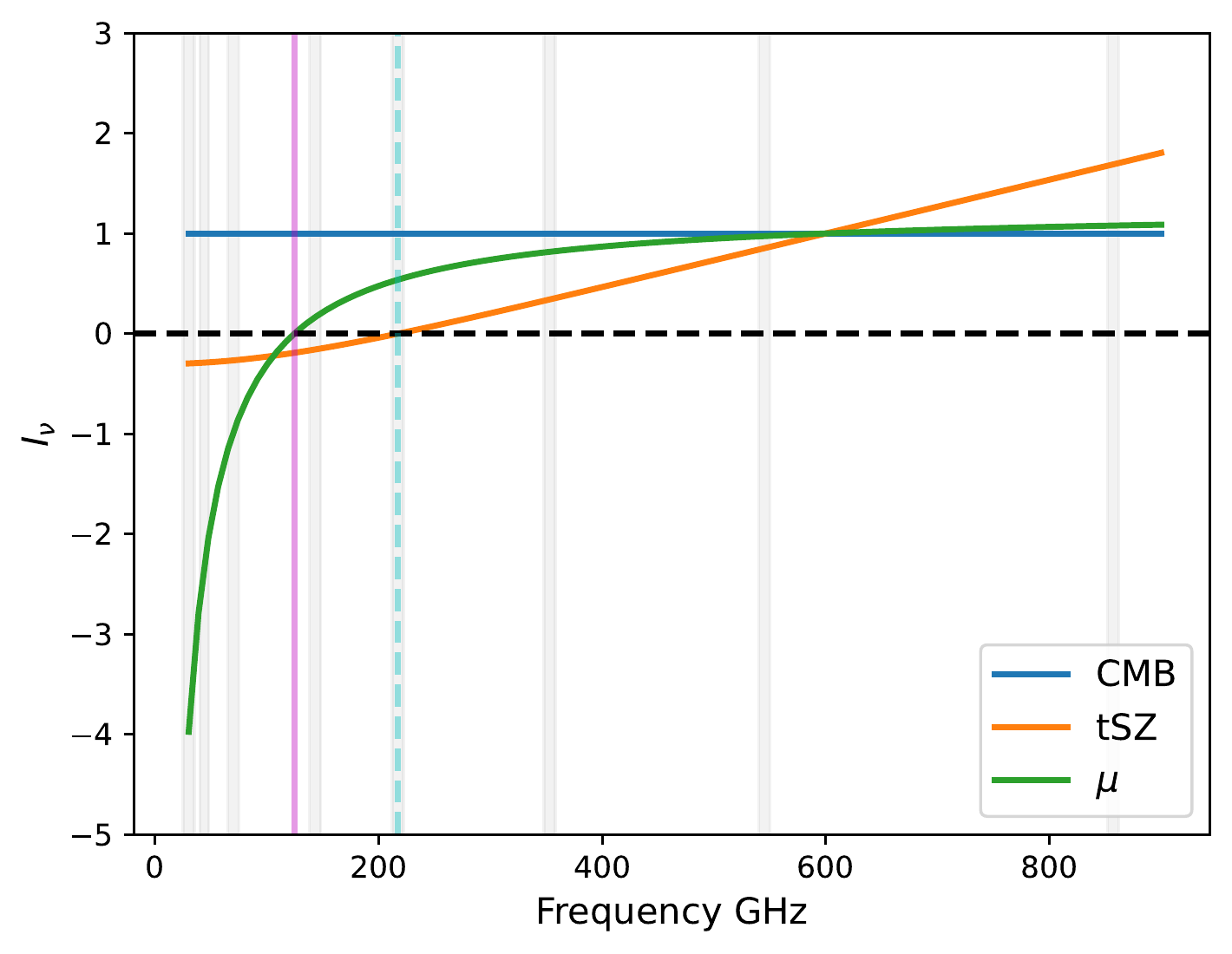}
\caption{This figure depicts the \textrm{CMB}, \textrm{tSZ} and $\mu$ spectra in thermodynamic units normalized to unity at 600 GHz. The gray bands mark the \textit{Planck} frequency bands and magenta and cyan vertical lines mark the nulls of the $\mu$ and \textrm{tSZ} spectra respectively.}
\label{fig:spectra}
\end{figure}
\begin{figure*}
\hspace*{-0.18cm} 
\subfigure[$\mu T$]{\includegraphics[width=0.95\columnwidth]{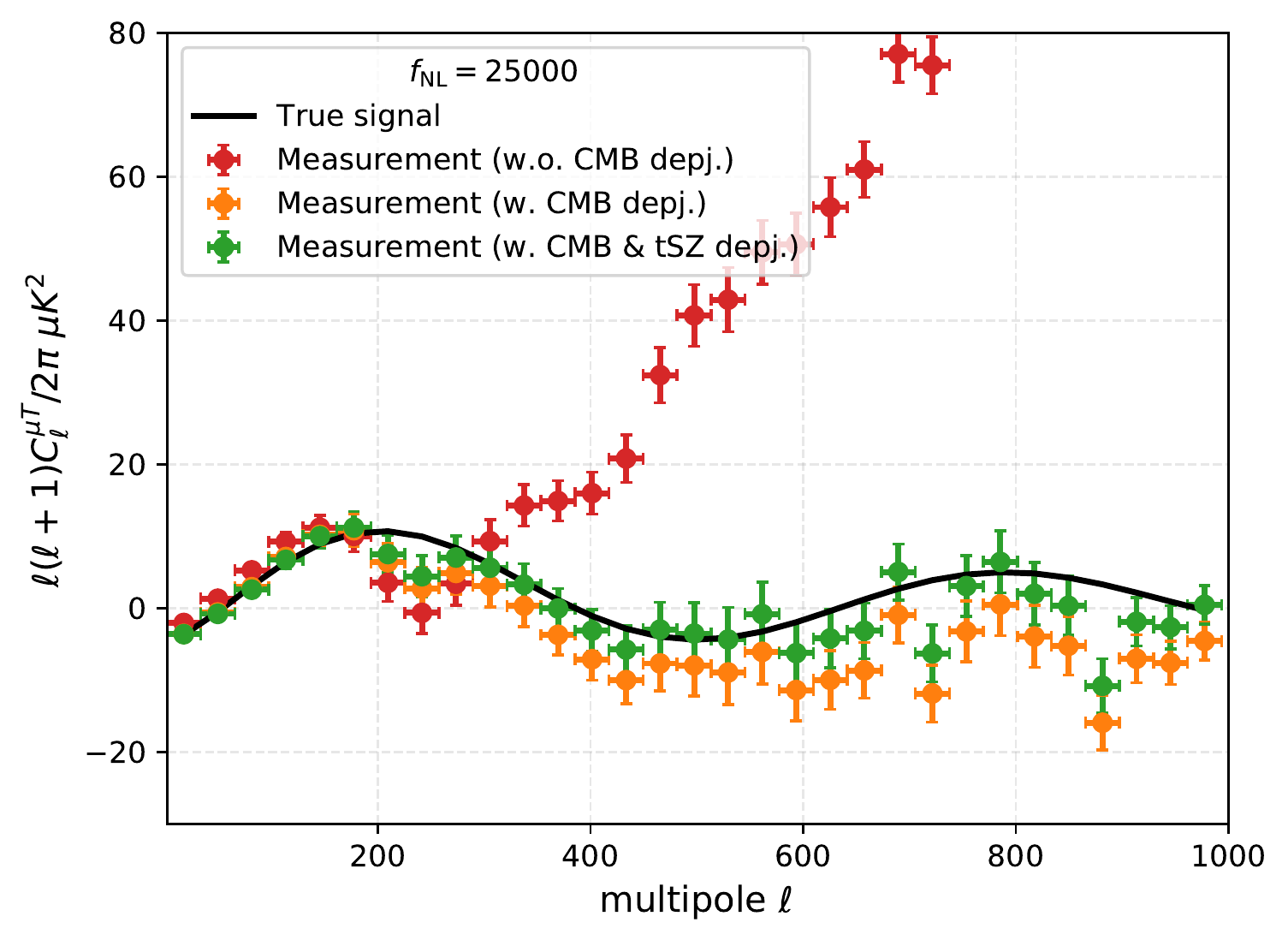}}
\hspace{4mm}
\subfigure[$\mu E$]{\includegraphics[width=0.95\columnwidth]{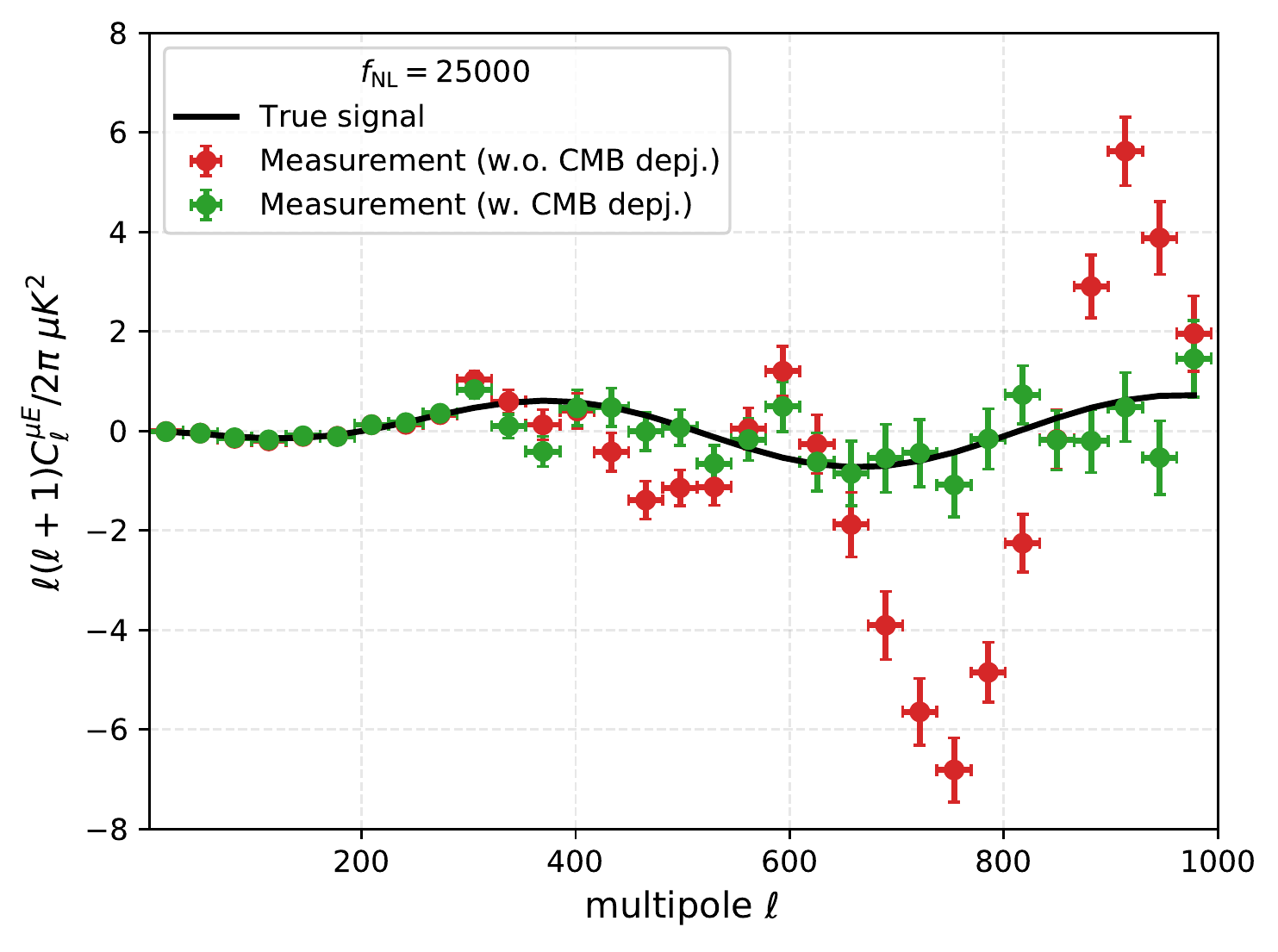}}
\caption{The top and bottom panels depict the measurements of the $C_{\ell}^{\mu T}$ \& $C_{\ell}^{\mu E}$ signals  respectively from ideal \Planck simulations. Note that the measurements are biased, particularly at high multipoles when CMB deprojection is not performed when reconstructing the $\mu$ map. Also note that tSZ deprojection in the reconstructed CMB maps is important to ensure unbiased measurement of the $C_{\ell}^{\mu T}$ spectrum.}
\label{fig:ideal_ilc_mt_me}
\end{figure*}

\subsection{Biases sourced by cosmological foregrounds}
\label{sec:bias_sourced_by_frg}
Before diving into the details of the analysis, we make some remarks regarding the various cosmological sources of biases that can be expected when recovering the $\mu$ map. We begin by drawing attention to  \fig{fig:spectra}, where we note that at high frequencies the $\mu$ spectrum resembles the CMB spectral energy density making the low-frequency channels particularly important for $\mu$ reconstruction \citep[see also][]{Abitbol2017, rotti2020milc}. Owing to these spectral properties the reconstructed $\mu$ anisotropy maps can be significantly contaminated by these temperature anisotropies. This subtle but important detail has been noted and the importance of CMB deprojection has been demonstrated in previous works \citep{Remazeilles2018mu,Remazeilles2021mu}. To avoid this potential contamination we employ the constrained ILC method, in which we de-project the CMB temperature anisotropies when reconstructing the $\mu$ map, thereby ensuring that the reconstructed $\mu$ maps are not contaminated by CMB temperature anisotropies.  This treatment is completely analogous to the technique of reconstructing the SZ deprojected CMB maps, now a routinely used technique \citep{Remazeilles2011cILC}. 

In addition to this, we also expect leakage of the tSZ signal, occurring in both the reconstructed CMB temperature map as well as the $\mu$ map. These as we will see appear as negative deficits in the temperature maps and as positive peaks in the reconstructed $\mu$ maps, prominently visible at the location of the brightest galaxy clusters. The net effect of these tSZ associated leakage is to introduce a negative bias in the $\mu T$ correlation signal, biasing the inferences. These biases are understandably associated only with the measurements of the $\mu T$ correlation signal while leaving the $\mu E$ measurements unaffected. We will demonstrate that this bias can again be mitigated by reconstructing the CMB temperature anisotropy maps while deprojecting the tSZ spectrum \citep{Remazeilles2011cILC}.We will demonstrate that not doing these spectral deprojections can introduce statistically significant biases in the measurement of $\mu T$ as well as the $\mu E$ cross power spectra. We will also show that introducing these additional spectral constraints to mitigate biases leads to an insignificant noise penalty.

\vspace{-3mm}
\section{Forecasts on simulated \Planck data}
\label{sec:planck_forecast}
We begin our more rigorous treatment by analyzing simulated \Planck data. With control on injected simulations, this exercise allows us to benchmark our component separation pipeline as well as the \fnl inference pipelines in realistic settings. 

We analyse sets of simulations with varying amplitudes of \fnl. The considered \fnl amplitudes cover values which can be expected to be seen with very high and moderately high significance with \Planck, informed by the Fisher forecasts, as well as the case of null detection i.e. $\fNL=0$. We also work specifically with half mission simulations to avoid the noise bias in $\mu T$ power spectrum estimates, as this will be our strategy when analysing \Planck data. The $\mu E$ spectrum, however, is not prone to this noise bias, since the noise in temperature and polarization measurements is independent and therefore we work with the full mission component maps in this case. Iterating and tuning our pipelines to yield optimal results on these suite of simulations allows us to construct a robust setup. Overall, thorough testing on simulations lends us confidence in the obtained analyses pipelines.

\vspace{-2mm}
\subsection{Simulating \Planck data}
\label{sec:planck_sims}
We simulate multi-frequency observations of the microwave sky as seen by \Planck using the Planck Sky model (\PySM) \citep{PySky}. This software incorporates most important foreground contaminants such as synchrotron, free-free and dust. We also generate simulations such that cases with $\mu$ anisotropies with sufficiently high values of \fnl which would be detectable by \Planck are included. For this we use in-built Healpix functionalities to generated the appropriately correlated $\mu$, $T$ and $E$ field, which are then passed to the \PySM software for appropriate inclusion in the multi-frequency sky simulations. The multi-frequency $\mu$ anisotropy signals are appended to the simulations delivered by \PySM. Similarly we also extend the foreground complexity of \PySM delivered skies by adding to them tSZ and CIB components for which we use \Websky simulations \citep{Stein_2020_websky}. Here we note that these CIB and tSZ simulations duly incorporate the correlations between these components.

\vspace{-2mm}
\subsection{Analysis on ideal \Planck simulations}
\label{sec:ideal_planck_sim}
We begin by analyzing idealized \Planck simulations that include only CMB, $\mu$, and \textrm{tSZ} anisotropies along with \Planck measurement noise, as this is the simplest test we can run on our pipelines. From running the component separation analysis on these simulations we expect to find error estimates that are consistent with the Fisher estimates presented in \sec{sec:fisher_forecast}. Additionally this also allows us to study and quantify the various biases arising due the presence of CMB and tSZ components (see  \sec{sec:bias_sourced_by_frg} for some discussion).

The importance of CMB deprojection when reconstructing the $\mu$ anisotropy map has been noted in previous works \citep{Remazeilles2018mu,Remazeilles2021mu}, however for completeness here we make demonstrations using our independent analysis pipeline. When reconstructing the $\mu$ map without performing CMB deprojection the $C_{\ell}^{\mu T}$ and $C_{\ell}^{\mu E}$ measurements are highly biased, as shown in \fig{fig:ideal_ilc_mt_me}. In particular note that while the measurements at the lowest multipoles ($\ell \lesssim 180$ for $\mu T$ and $\ell \lesssim 300$ for $\mu E$) seem reasonable, at high multipoles however the measurements are strongly biased. On passing these spectra through the likelihood module, we find significantly biased measurement of \fnl (see \tab{tab:ideal_analysis_fnl_stat}). Particularly for the $\mu T$ measurements the bias is highly statistically significant, while the biases in inferred \fnl from $\mu E$ measurements are less biased (see \tab{tab:ideal_analysis_fnl_stat}). The most stringent constraints derived from combining $\mu T$ \& $\mu E$ measurements are also biased strongly, but this is primarily owing to the $\mu T$ measurement. On reconstructing the $\mu$ map while simultaneously deprojecting the CMB spectrum, we see that the cross correlation measurements are less biased (\fig{fig:ideal_ilc_mt_me}), and this is immediately reflected in the measured values of \fnl (see \tab{tab:ideal_analysis_fnl_stat}).
While the bias in the $C_{\ell}^{\mu E}$ measurement has completely disappeared, there still appears to be a small negative deficit in the measurement of the  $C_{\ell}^{\mu T}$ spectrum as apparent from the left panel of \fig{fig:ideal_ilc_mt_me}. This is sourced by leakages of the tSZ signal in the reconstructed CMB and $\mu$ fields. While it is possible to remove this bias by additionally deprojecting the tSZ spectrum when reconstructing the $\mu$ field, this leads to a non-negligible noise penalty which affects both the $C_{\ell}^{\mu T}$ and $C_{\ell}^{\mu E}$ measurements. This is easily avoided by simply constructing the CMB temperature anisotropies while simultaneously deprojecting the tSZ spectrum. Not only does this remove the bias (see  \fig{fig:ideal_ilc_mt_me}) at nearly no additional noise cost, but it also leaves the $C_{\ell}^{\mu E}$ measurement unaffected.

We draw attention to the fact that the error estimates derived from the component separation analysis are fully consistent with the Fisher forecasts. For analysis on these idealized setting this is to be expected, as the Fisher setup in this case is strictly the ensemble averaged estimate of the respective noise estimates. We also note that the spectral deprojections do expectedly cause minor ($\simeq 5\%$) increases in the errors, but these are completely tolerable given the control this lends on bias.
\begin{table}
\centering
\groupedRowColors{3}{2}{gray!10}{white!10}
\begin{tabular}{lllllll}
\toprule
Type                        &      Data              &   $f_{\rm NL}$ & $\sigma_{f_{\rm NL}}$ &   SNR &  Bias \\
\midrule
NILC & $\mu T$ &  33487 &  1460 &  22.9 &   5.8 \\
                        & $\mu E$ &  25303 &  1268 &  20.0 &   0.2 \\
                        & $\mu T$ \& $\mu E$ &  28706 &  107 &  26.7 &   3.4 \\
CMB depj. & $\mu T$ &  21873 &  1534 &  14.3 &  -2.0 \\
                        & $\mu E$ &  23205 &  1293 &  17.9 &  -1.4 \\
                        & $\mu T$ \& $\mu E$ &  22475 &  1115 &  20.2 &  -2.3 \\
CMB \& tSZ depj. & $\mu T$ &  23119 &  1536 &  15.1 &  -1.2 \\
                        & $\mu T$ \& $\mu E$ &  22985 &  1116 &  20.6 &  -1.8 \\
\bottomrule
\end{tabular}


\caption{Derived \fnl statistics derived from analysis on ideal \Planck simulations generated assuming $\fNL = 25000$. All analyses use multipoles $\ell \in [2,1024]$ and $f_{\rm sky}=0.67$.}
\label{tab:ideal_analysis_fnl_stat}
\end{table}
As we have seen in \fig{fig:ideal_ilc_err_evol} and surrounding discussion, \Planck is expected to gain from multipoles up to $\ell \simeq 400$ and therefore to fully exploit the information encoded in the data it is important to mitigate these bias which can be achieved by carrying out the spectral deprojections emphasized above. Since this bias is a consequence of the interplay between $\mu$, CMB and tSZ spectra, one may expect it to equally relevant when we work with realistic sky simulations that include foregrounds, which we study next.

\begin{figure}
\hspace*{-0.18cm} 
\includegraphics[width=\columnwidth]{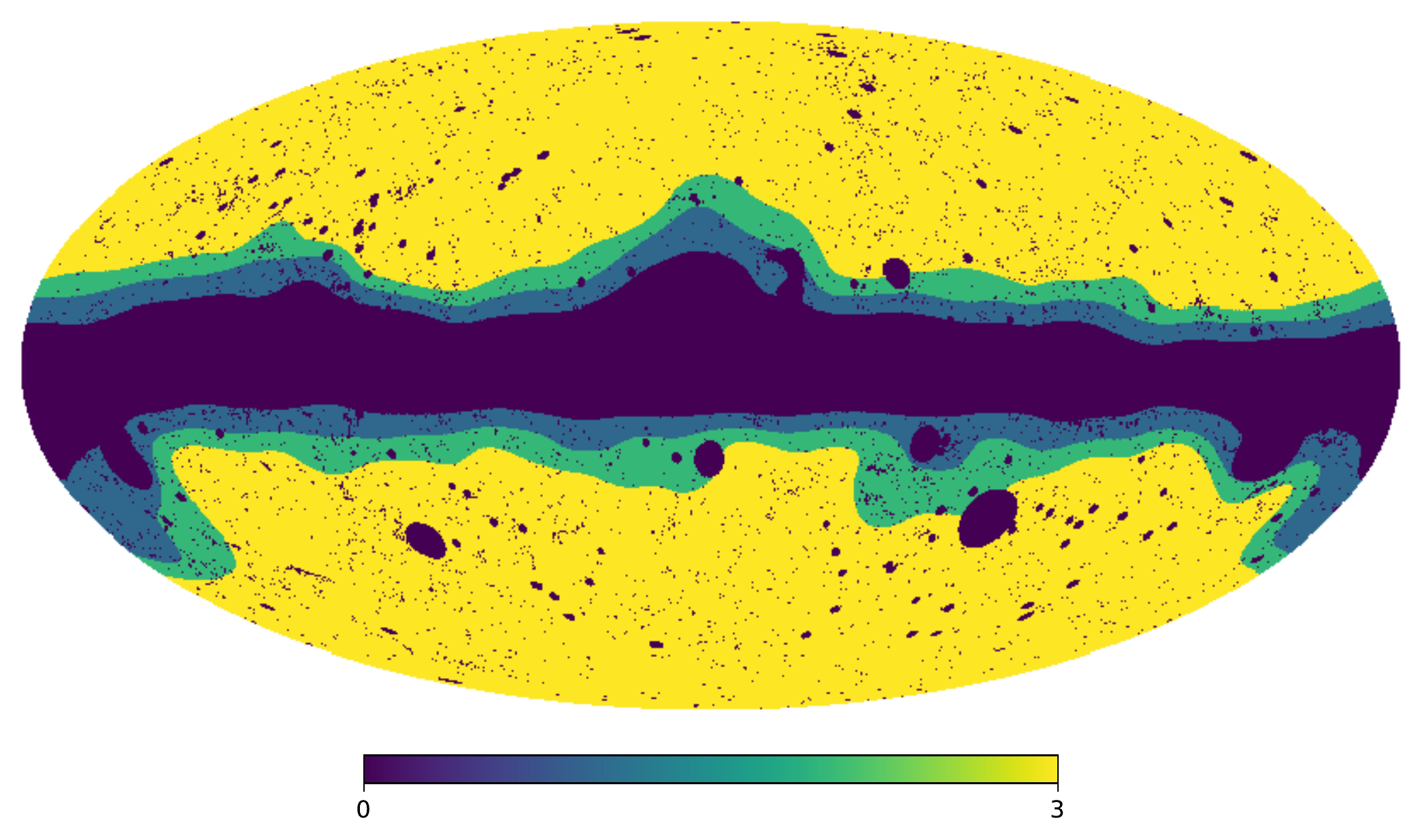}
\caption{This figure depicts the three different masks with $\fsky \in [0.58,0.67,0.75]$ used in the analyses.}
\label{fig:masks}
\end{figure}
\begin{figure*}
\hspace*{-0.18cm} 
\subfigure[\label{fig:sim_auto_mumu_fnl}]{\includegraphics[width=\columnwidth]{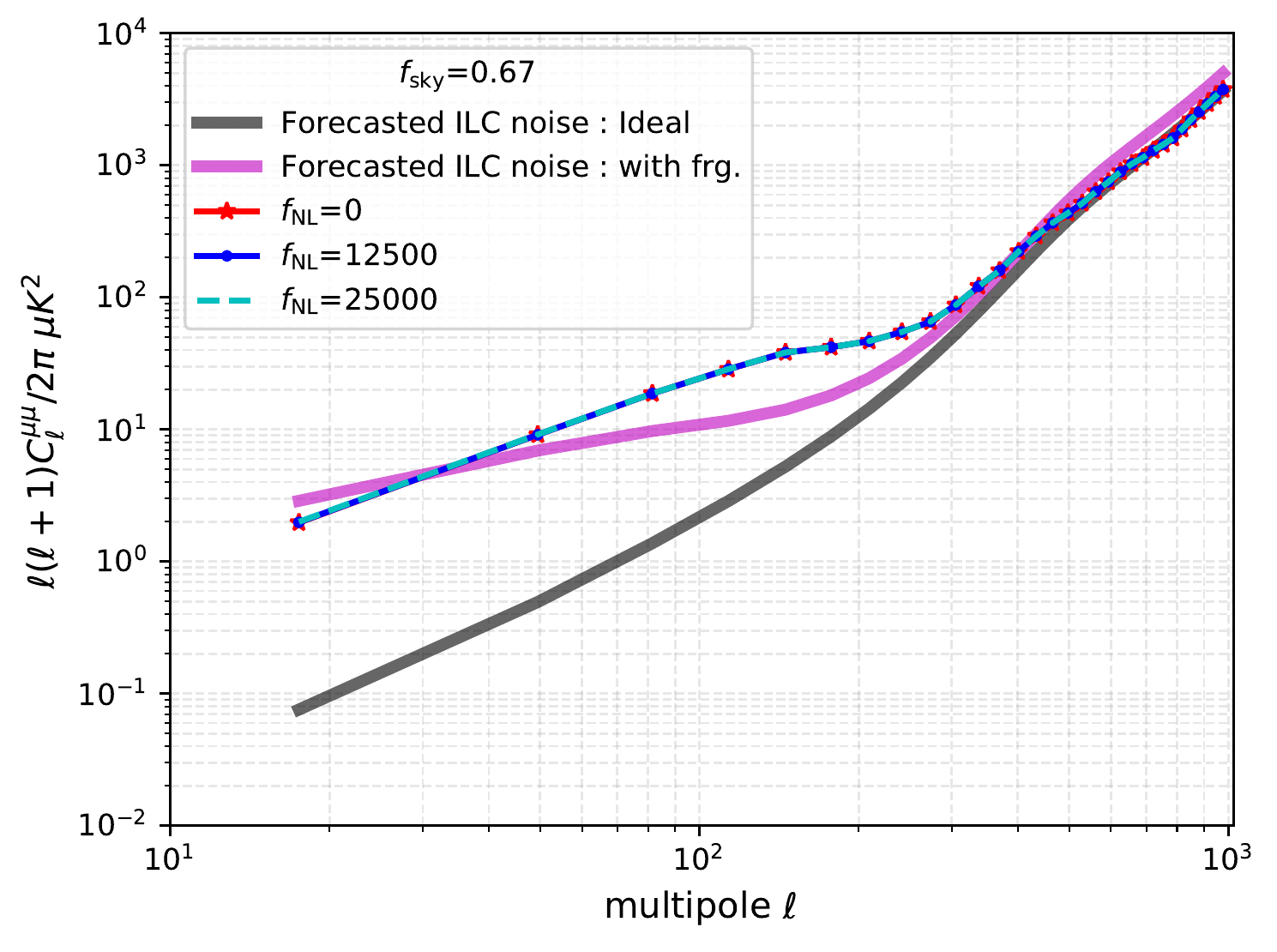}}
\subfigure[\label{fig:sim_auto_mumu_fsky}]{\includegraphics[width=\columnwidth]{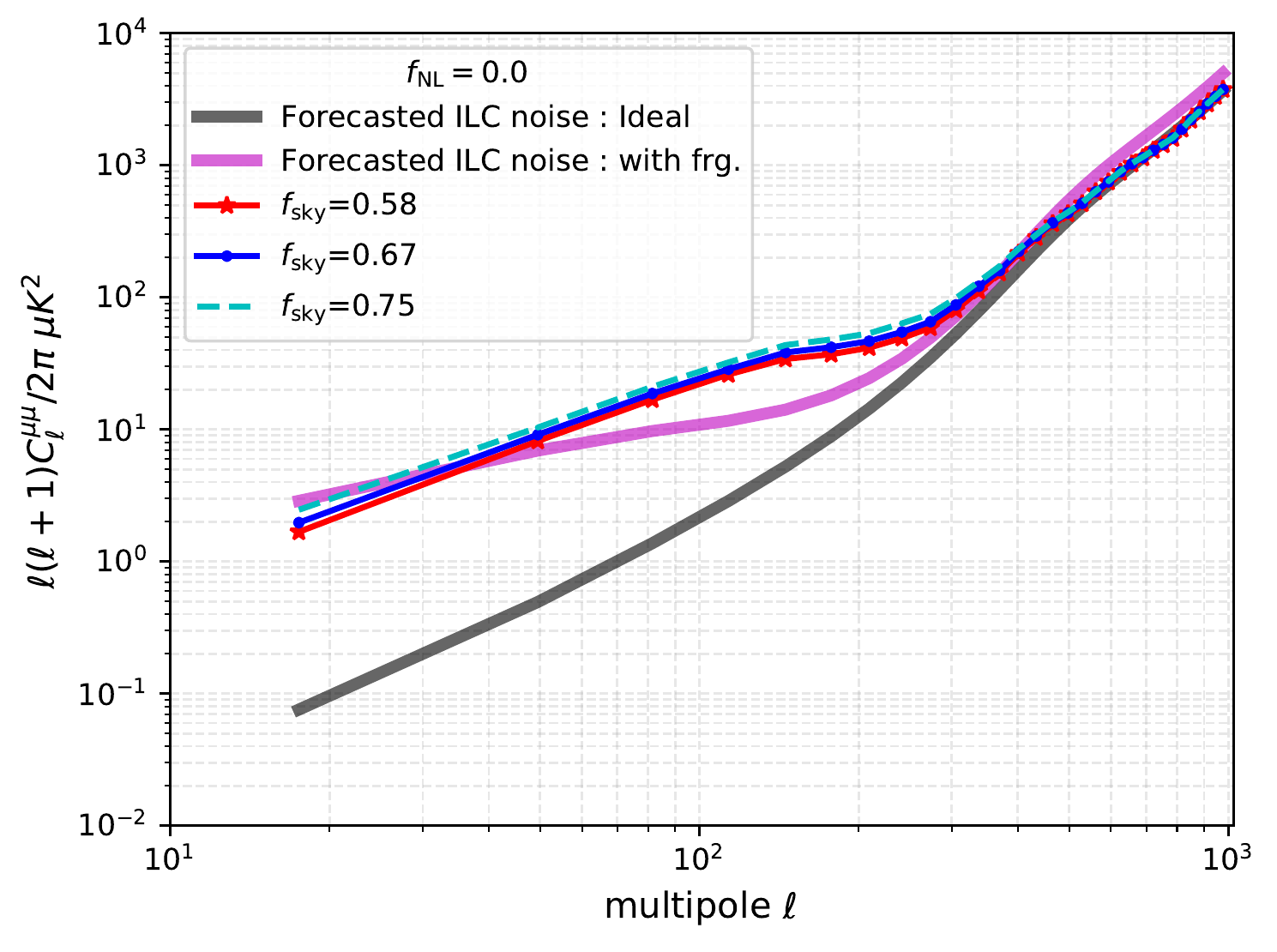}}
\caption{The left panel depicts the reconstructed $\mu$-map power spectrum estimated from analyses carried out on the different set of simulations with varying values of the \fnl parameter. The right panel depicts how it varies on using different fraction of the sky used in the power spectrum analyses.}
\label{fig:sim_auto_mumu}
\end{figure*}
%

\subsection{Analysis on realistic \Planck simulations}
\label{sec:realistic_planck_sim}
Based on our Fisher estimates of expected error on \fnl~ parameter, including the foreground degradation factor (see \sec{sec:fisher_forecast}), we choose to work with three separate sets of simulation which incorporate $\mu$ anisotropies corresponding to \fnl = 0, 12500 \& 25000. The details of the generation of these simulations were discussed in \sec{sec:planck_sims}. Including $\mu T$ and $\mu E$, we expect the analysis on these simulations to yield a detection of the \fnl parameter with roughly 0, 5 and 10 $\sigma$, respectively (see Table~\ref{tab:ilc_fnl_forecast}).

\subsubsection{Tuning of NILC pipeline}
As noted previously, we tune our NILC pipeline on these simulations. This procedure is carried out by demanding that our pipelines yield measurements of the \fnl parameter that are statistically consistent with the injected values as well as consistent measurements of the standard CMB spectra.  In fact we demand that the {\it same} optimization yields statistically consistent results for the variety of \fnl values injected in the different simulation sets. When performing these NILC optimization and tests we also varied the seed of the random number generator, when simulating the set CMB and $\mu$ skies for different set of simulations corresponding to different value of \fnl parameter, so as to avoid the potential caveat of tuning to a particular random realization. We do however keep the foregrounds fixed. 

The NILC pipeline requires most tuning when reconstructing $\mu$ maps. Since we use different NILC optimization when reconstructing $T$ and $E$ maps, these do not directly interfere with the NILC optimization of the $\mu$-map. In addition to \fnl measurements we also assess the performance of our component separation pipeline by ensuring that the power spectra of our component separated maps, specifically $T$ and E, are consistent with those injected into simulations as summarized in \app{app:sim_spec_compare}.  Finally we also check that our results are stable against variations in the sky fraction retained in the final analysis. The different masks used here are shown in \fig{fig:masks}.

\subsubsection{Analysis strategy}
We apply an identical pipeline to all three sets of simulations corresponding to \fnl = 0, 12500, 25000. When reconstructing the $\mu$-map, we deproject CMB to avoid contamination by CMB anisotropies as discussed and demonstrated in the \sec{sec:ideal_planck_sim}. Not doing this yields a highly biased $\mu T$ and $\mu E$ spectrum in all our simulations, as expected. Given the analysis on idealized simulations discussed in \sec{sec:ideal_planck_sim}, we also anticipate SZ deprojection to be important element of the analysis. However, when performing the same analysis on the realistic simulations, we find contrary to expectations, that performing a SZ deprojection yields a biased measurement of the $\mu T$ spectrum. In fact we find unbiased $\mu T$ spectral measurements when not performing a SZ deprojection. This intriguing observation warrants a dedicated discussion that we present in \sec{sec:sz_cib_role}. 

\begin{table}
\centering
\groupedRowColors{3}{2}{gray!10}{white!10}
\begin{tabular}{lllllll}
\toprule
 $f_{\rm NL}$      &           Data         &   $f_{\rm NL}$ & $\sigma_{f_{\rm NL}}$ &  SNR &  Bias \\
\midrule
25000 & $\mu T$ &  24001 &  3853 &  6.2 &  -0.3 \\
      & $\mu E$ &  24749 &  3529 &  7.0 &  -0.1 \\
      & $\mu T$ \& $\mu E$ &  24574 &  2762 &  8.9 &  -0.2 \\
12500 & $\mu T$ &   9859 &  3826 &  2.6 &  -0.7 \\
      & $\mu E$ &  11567 &  3521 &  3.3 &  -0.3 \\
      & $\mu T$ \& $\mu E$ &  10267 &  2742 &  3.7 &  -0.8 \\
0     & $\mu T$ &  -2699 &  3857 & -0.7 &  -0.7 \\
      & $\mu E$ &  -2406 &  3529 & -0.7 &  -0.7 \\
      & $\mu T$ \& $\mu E$ &  -2495 &  2762 & -0.9 &  -0.9 \\
\bottomrule
\end{tabular}

\caption{This table summarizes the statistics of the measured \fnl parameters from analyses on the different set of simulations. These likelihood analyses used multipoles $\ell \in [2,1024]$ and $f_{\rm sky}=0.67$.}
\label{tab:real_analysis_vary_fnl}
\end{table}
\begin{table}
\centering
\groupedRowColors{3}{2}{gray!10}{white!10}
\begin{tabular}{lllllll}
\toprule
 $f_{\rm sky}$     &       Data             &   $f_{\rm NL}$ & $\sigma_{f_{\rm NL}}$ &  SNR &  Bias \\
\midrule
0.58 & $\mu T$ &   9986 &  3907 &  2.6 &  -0.6 \\
     & $\mu E$ &  11399 &  3837 &  3.0 &  -0.3 \\
     & $\mu T$ \& $\mu E$ &  10537 &  3057 &  3.4 &  -0.6 \\
0.67 & $\mu T$ &   9859 &  3826 &  2.6 &  -0.7 \\
     & $\mu E$ &  10774 &  3745 &  2.9 &  -0.5 \\
     & $\mu T$ \& $\mu E$ &  10193 &  2983 &  3.4 &  -0.8 \\
0.75 & $\mu T$ &  10501 &  3855 &  2.7 &  -0.5 \\
     & $\mu E$ &   9225 &  3762 &  2.5 &  -0.9 \\
     & $\mu T$ \& $\mu E$ &  10064 &  2998 &  3.4 &  -0.8 \\
\bottomrule
\end{tabular}

\caption{This table presents a quantitiative summary of the stability of the statistics of the inferred \fnl parameter from analysis on simulations with \fnl$=12500$, on using different fractions of the sky in the likelihood analysis.}
\label{tab:real_analysis_vary_fsky}
\end{table}
\begin{figure*}
\hspace*{-0.18cm} 
\subfigure[\label{fig:sim_muT_fnl25k}]{\includegraphics[width=\columnwidth]{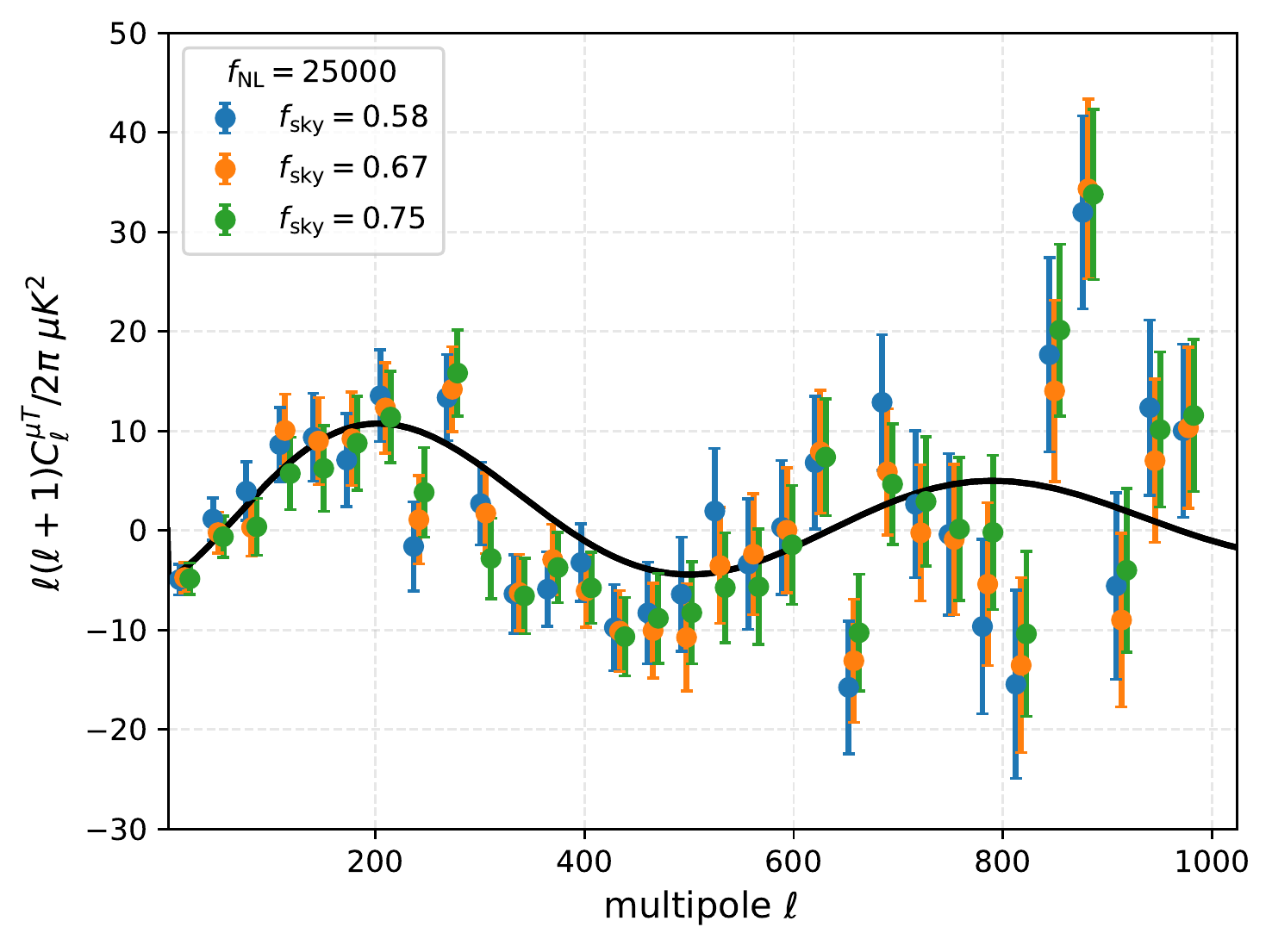}}
\subfigure[\label{fig:sim_muE_fnl25k}]{\includegraphics[width=\columnwidth]{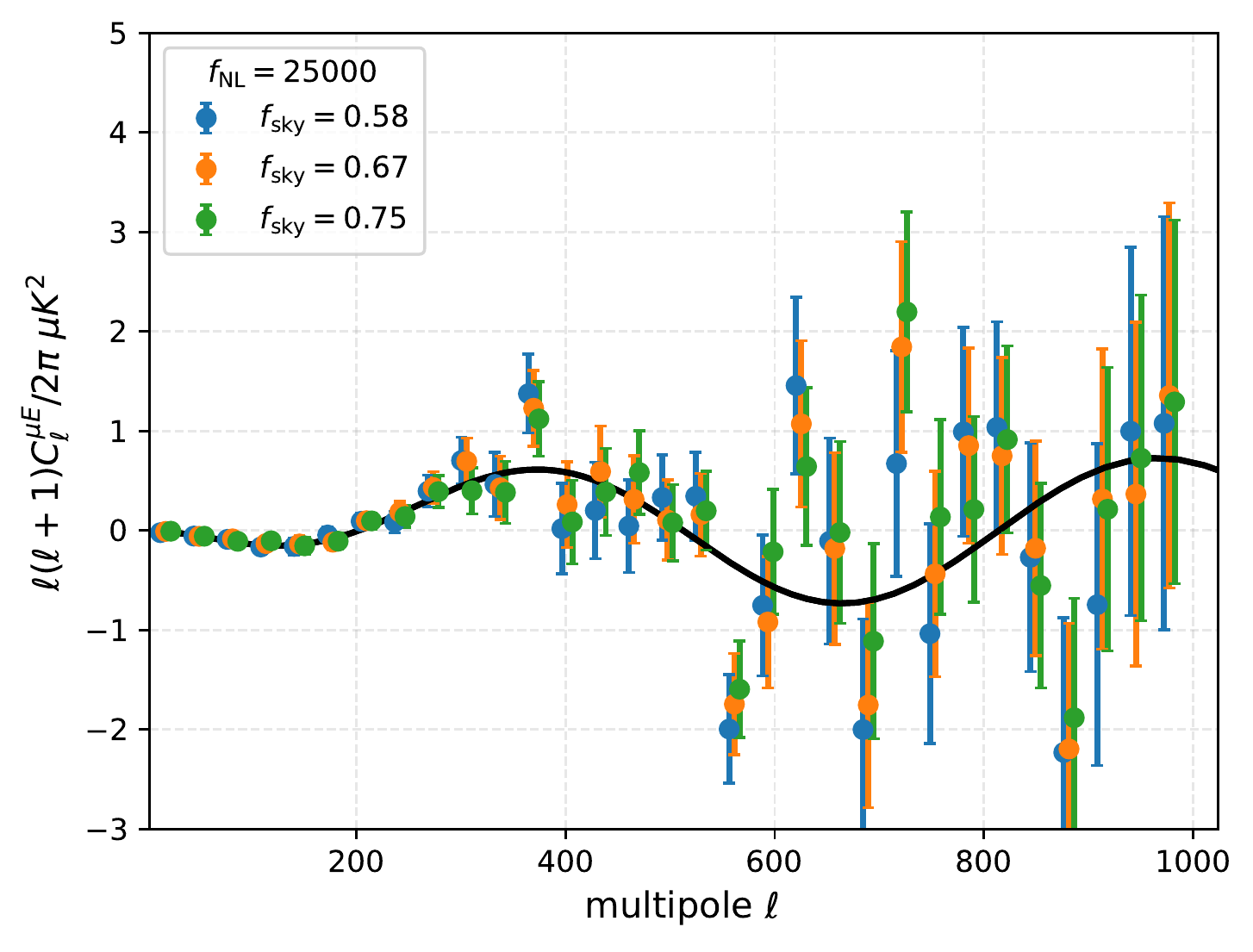}}
\subfigure[\label{fig:sim_muT_fnl12p5k}]{\includegraphics[width=\columnwidth]{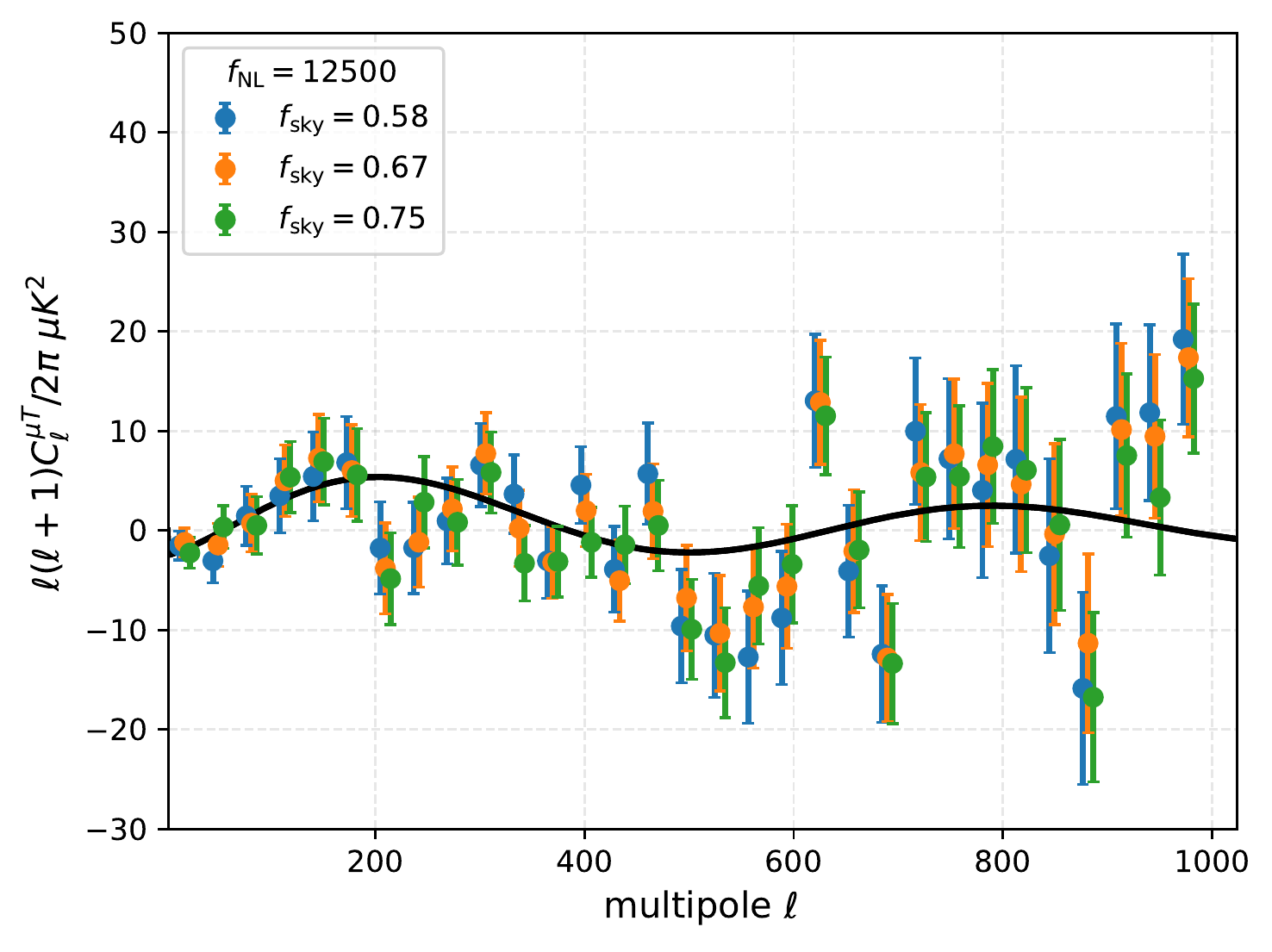}}
\subfigure[\label{fig:sim_muE_fnl12p5k}]{\includegraphics[width=\columnwidth]{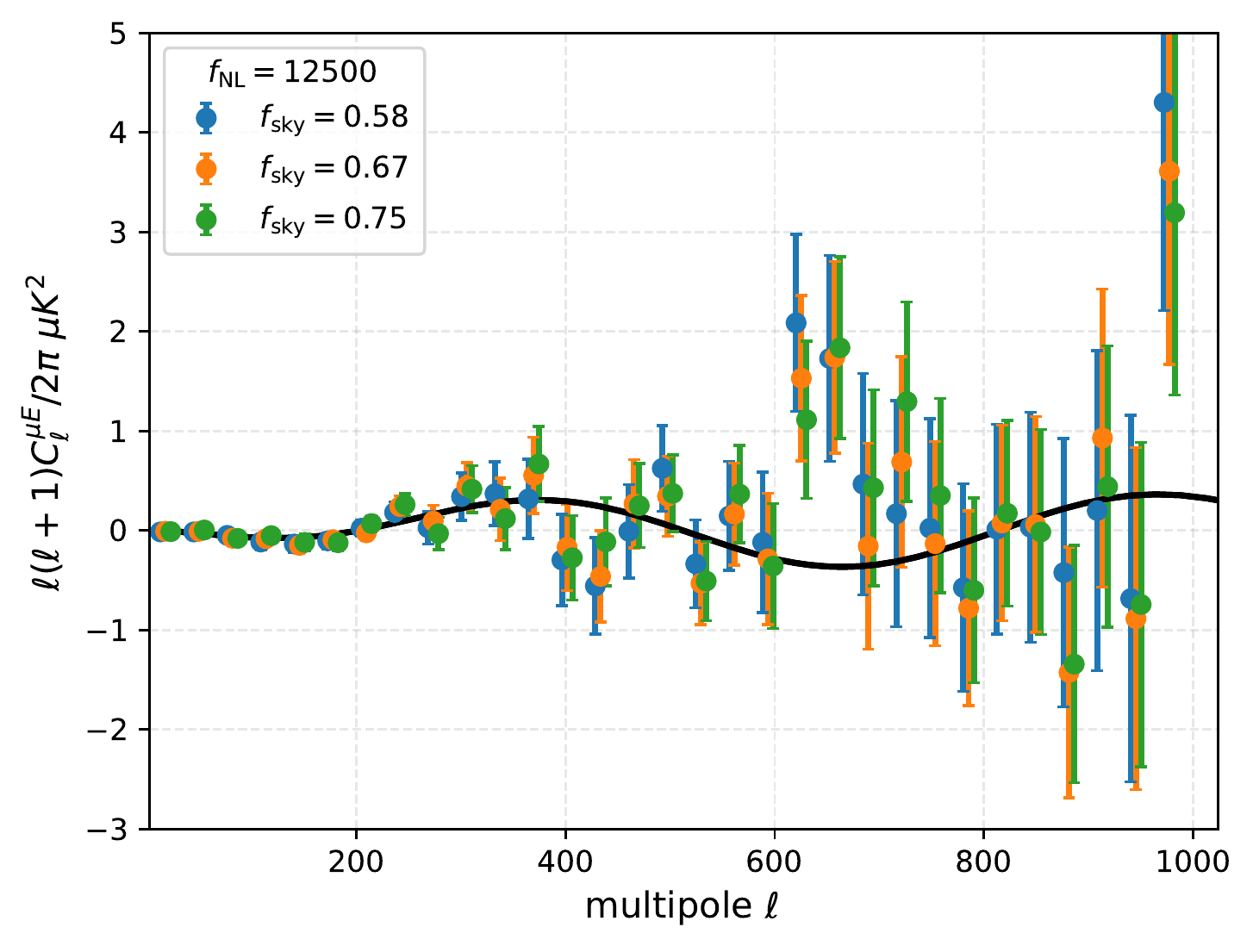}}
\subfigure[\label{fig:sim_muT_fnl0}]{\includegraphics[width=\columnwidth]{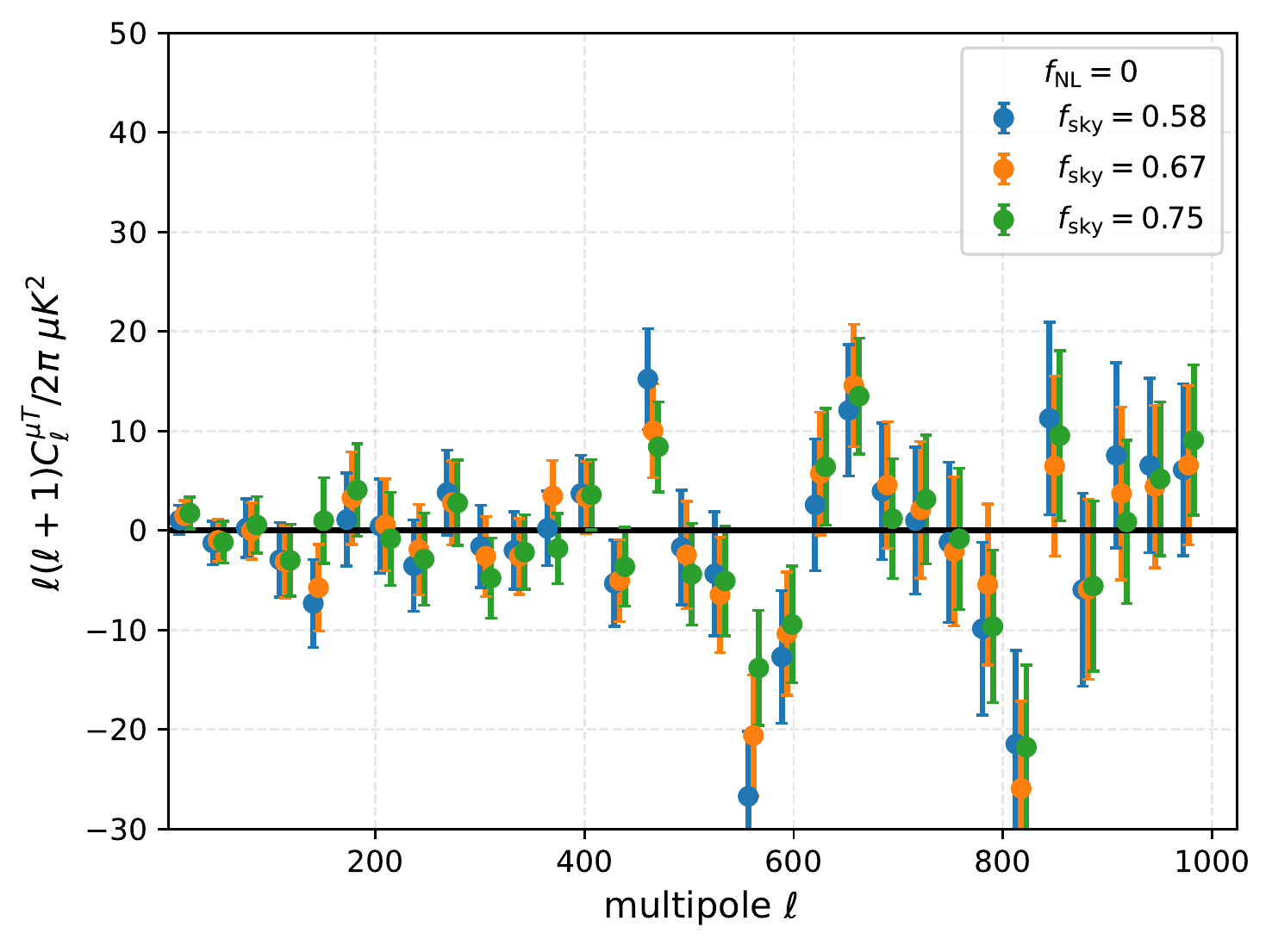}}
\subfigure[\label{fig:sim_muE_fnl0}]{\includegraphics[width=\columnwidth]{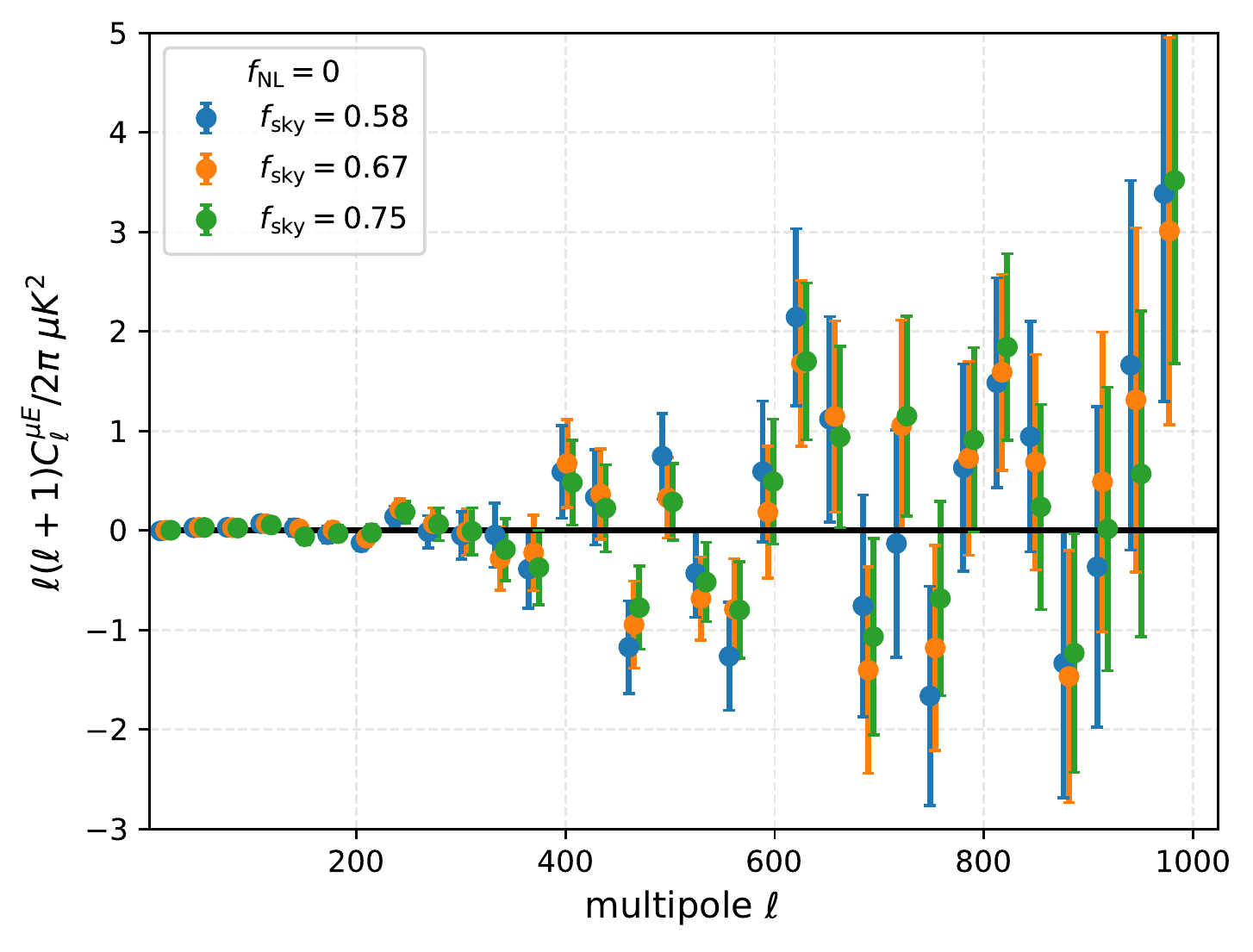}}
\caption{The figures depict the measurement of the $C_{\ell}^{\mu T}$ \& $C_{\ell}^{\mu E}$ spectra derived from analysis on different suite of simulations with \fnl$\in [0,12500,25000]$. Each panel also shows the stability of the measurements on varying the fraction of the sky used in the analysis.}
\label{fig:sim_muX_spec}
\end{figure*}

\subsubsection{Results from simulated analysis}
While $C_{\ell}^{\mu \mu}$ from \Planck are not expected to reveal interesting cosmological insights, it plays a crucial role in determining the noise in the $C_{\ell}^{\mu T}$ \& $C_{\ell}^{\mu E}$ measurements. We begin by noting that the angular power spectra of the recovered $\mu$ maps are independent of the value of the injected \fnl parameter as seen in \fig{fig:sim_auto_mumu_fnl}. The constancy of the power spectrum at low multipoles and it being strongly deviant from the expected ideal ILC noise is clear indication that the power is dominated by foreground residuals in the reconstructed $\mu$ map. This is further supported by the observations that on varying the fraction of the sky, the power spectrum amplitude at low multipoles is seen to systematically increase with increasing $f_{\rm sky}$ as seen in \fig{fig:sim_auto_mumu_fsky}. These observations clearly suggests that for the values of \fnl considered in this exercise, the measurement of the true $C_{\ell}^{\mu \mu}$ is highly unlikely as we had anticipated in \sec{sec:fisher_forecast}. Finally we also note that the spectrum matches the ideal ILC noise estimates at multipoles above $\ell \simeq 400$. This is reassuring, as the sophisticated NILC method yields results that are consistent with very simple noise estimates which only require the instrument resolution and noise properties as a function of frequency as inputs. 

\begin{figure}
\hspace*{-0.18cm} 
\includegraphics[width=\columnwidth]{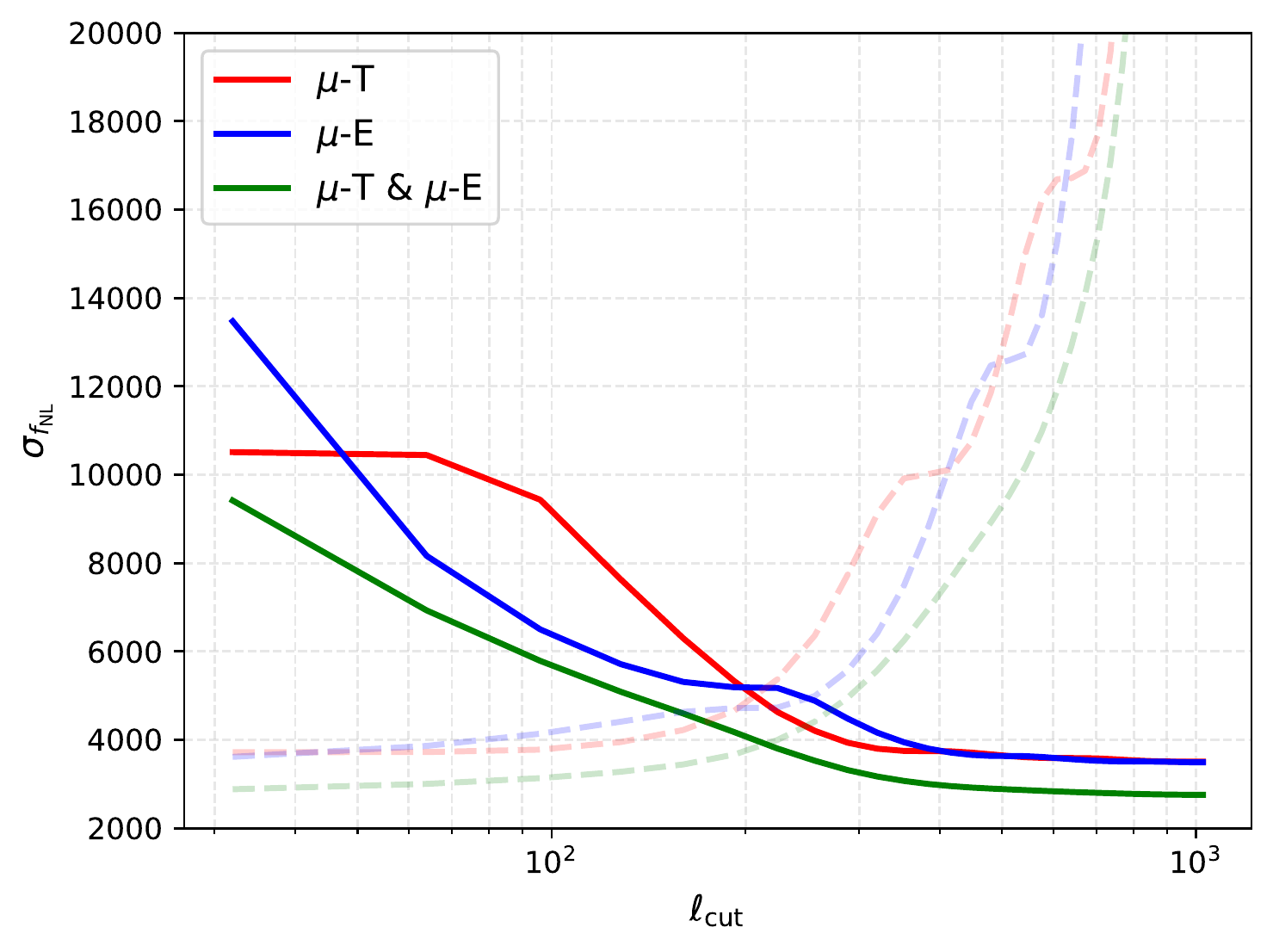}
\caption{This figure depicts the evolution of \fnl~ error as a function of multipoles. Solid lines $\ell_{\rm cut}$ corresponds to the maximum multipole used with $\ell_{\rm min}=2$ and for dashed lines $\ell_{\rm cut}$ corresponds to the minimum multipole used in the analysis with $\ell_{\rm max}=1024$.}
\label{fig:fnl_err_evol_ell}
\end{figure}
%

The measured $C_{\ell}^{\mu T}$ and $C_{\ell}^{\mu E}$ spectra from the set of simulated observations are shown in \fig{fig:sim_muX_spec}. 
Observing the different panels of  \fig{fig:sim_muX_spec} we also note that these measured spectra are nearly invariant on changing the fraction of the sky used in the analysis, which indicates that the measured correlations are dominantly sourced by signals of cosmological origins.
We quantify these observations further by passing the corresponding spectra through the likelihood pipeline. We find the measurements on the \fnl parameter to be statistically consistent with injected values in analysis of all different simulations. The inferred statistics on the \fnl parameter for $f_{\rm sky}=0.67$ are summarized in \tab{tab:real_analysis_vary_fnl}. We also find excellent consistency of measurements across varying fraction of the sky used in the analysis; the inferred \fnl measurement statistics presented in \tab{tab:real_analysis_vary_fsky}. From these tables it is clear that the measurements on the \fnl parameter are statistically consistent with input values as inferred from noting that the bias in the measured \fnl values are well with in statistical uncertainties.

The errors estimated on the \fnl parameter are expected to be independent of the \fnl parameter, as we had anticipated from analysis presented in \sec{sec:fisher_forecast}. In this simulated analysis, this is supported by noting that the $C_{\ell}^{\mu \mu}$, estimated from analysis with different injected \fnl~ parameters is invariant under changes to \fnl as seen in \fig{fig:sim_auto_mumu}. This is reflected in the the errors inferred from analysis on different set of simulations being nearly constant as seen in \tab{tab:real_analysis_vary_fnl}. One may naively expect the errors on \fnl to improve as $\sqrt{f_{\rm sky}}$. On the other hand $C_{\ell}^{\mu \mu}$ at low multipoles increase monotonically on increasing $f_{\rm sky}$. This systematic increase in power can be attributed to enhanced contribution from residual foreground contamination at low galactic latitude regions which are retained in the analysis on increasing $f_{\rm sky}$ as seen in \fig{fig:masks}. Therefore the net error behaviour will be determined by these two competing factors. Error reduction from increased sky coverage may be partially compensated or even over compensated by the increase in foreground power resulting in poorer error estimates and the exact trend cannot be easily anticipated. As seen in \tab{tab:real_analysis_vary_fsky} we find that the error estimates marginally improve on increasing the $f_{\rm sky}$ from 0.58 to 0.75. However these changes are not in line with $\sqrt{f_{\rm sky}}$ improvements in error estimates for reasons discussed above. 

Here it is also interesting to note that the estimated errors on the \fnl parameter are in the ball park set by the Fisher forecasts but are $\simeq$30\% worse. Since the Fisher approach only employs an approximate power spectral model for foregrounds, which does not capture details such as the foreground non-Gaussianity, their highly anisotropic nature, the spatio-spectral correlations across different foreground component etc, one may expect it to underestimate error. These missing factors may be attributed to the excess power seen in $C_{\ell}^{\mu \mu}$ around $\ell \simeq 150$ measured from simulations as compared to that estimated from the ILC noise estimates as seen in \fig{fig:sim_auto_mumu}.

Finally, we quantify the multipole ranges which provide the most constraining power on the \fnl parameter. For this we estimate \fnl~ error by progressively increasing the maximum (minimum) multipole used in the analysis (see \fig{fig:fnl_err_evol_ell}). Since the errors are independent of the injected \fnl~ and vary only mildly with changes in $f_{\rm sky}$, we carry out this study only for a single case. We confirm with our simulated analysis that the most constraining power is provided by multipoles below $\ell \lesssim 400$ as we had projected using the Fisher analysis (see \sec{sec:fisher_forecast}). While the lowest multipoles are important, the multipoles in the range $\ell \simeq 100-300$ play the most critical role as the \fnl errors start seeing a steep rise on excluding these multipoles from the analysis as seen in the dashed lines in \fig{fig:fnl_err_evol_ell}.
%


\vspace{-3mm}
\section{Analysis of \Planck data}
\label{sec:planck_analysis}
Having extensively tested our analysis pipelines on simulations, we now shift our focus to discussing details of the analysis carried out on \Planck data. 
We work with both the full mission as well as the half mission \Planck data. For the high frequency instrument (HFI) the half mission data sets are provided by the \Planck collaboration. For the low frequency instrument (LFI), no specific half mission data set is provided and hence we specifically choose to work with maps produced by combining $1^{\rm st}$ year \& $3^{\rm rd}$ year data as part of half mission 1 data set and maps produced by combining $2^{\rm nd}$ year \& $4^{\rm th}$ year data as part of the half mission 2 data set. Since a part of the analysis involves reconstruction of a $\mu$ map from \Planck data, the LFI covering 30 GHz, 44 GHz and 70 GHz plays a very important role as was discussed in \sec{sec:bias_sourced_by_frg} (see \fig{fig:spectra}).  

Since we are cross correlating the reconstructed $\mu$ map with temperature \& E-mode of polarization while seeking signal of  order $10 \mu K^2$ in $\mu T$ and $ 0.5 \mu K^2$ in the $\mu E$ cross correlation, a mere $\simeq 0.5\%$ leakage of $T$ to $\mu$ is enough to strongly bias the inference on these signals of interest. This requires very good precision, which not only emphasizes the need of deprojecting temperature when reconstructing the $\mu$ map but also demands that care is taken when undoing some of the instrumental effects. In this regard, we find using the RIMO (Reduced Instrument Model) beams to deconvolve the maps to a common resolution of 5 arcminutes to be critical to the analysis. We specifically note that using the effective Gaussian \textrm{FWHM} models for characterizing the instrument beams is not sufficiently accurate for this analysis. We also find correcting for the pixel window function to be crucial to our analysis. In fact ignoring these detailed corrections leads to a measurement of $\mu T$ and $\mu E$ correlation that resemble the $C_{\ell}^{\rm TT}$ and $C_{\ell}^{\rm TE}$ in shape, a sign of $T$ leakage to $\mu$. While these corrections are small ($\lesssim 1 \%$) at the multipoles of interest ($\ell \lesssim 400$), these act like multipole dependent calibration effects facilitating $T$ to $\mu$ leakages. We discuss aspects of these corrections and show the $\mu T$ \& $\mu E$ measurements when using effective Gaussian beam in \app{app:instr_model}. 

The $\mu T$ measurement is expected to be more challenging, since both $T$ \& $\mu$ are subject to same foregrounds and possible systematics in the data. On the contrary the polarization E-mode measurements are independent both in terms of noise as well as systematics and hence expected to be more robust. Since these two measurements are independent, they serve as a crucial litmus test of a potentially-detected signal having primordial origin. Since \Planck data is expected to deliver equally competent constraints on \fnl from both $\mu T$ and $\mu E$ measurements, a detection in one and not in another could point to some systematic in the data or a potential issue in the analysis. Since the $\mu T$ and $\mu E$ spectra we seek have the same primordial origin, when appropriately combined they are expected to present a stronger joint constraints on the \fnl~ parameter as demonstrated in the simulated analysis (see \sec{sec:realistic_planck_sim}). However this coaddition is only sensible if the two measurements are compatible with each other. We therefore discuss the measurement of $\mu T$ and $\mu E$ spectra independently and assess the results before combining them.

\begin{figure}
\hspace*{-0.18cm} 
\includegraphics[width=\columnwidth]{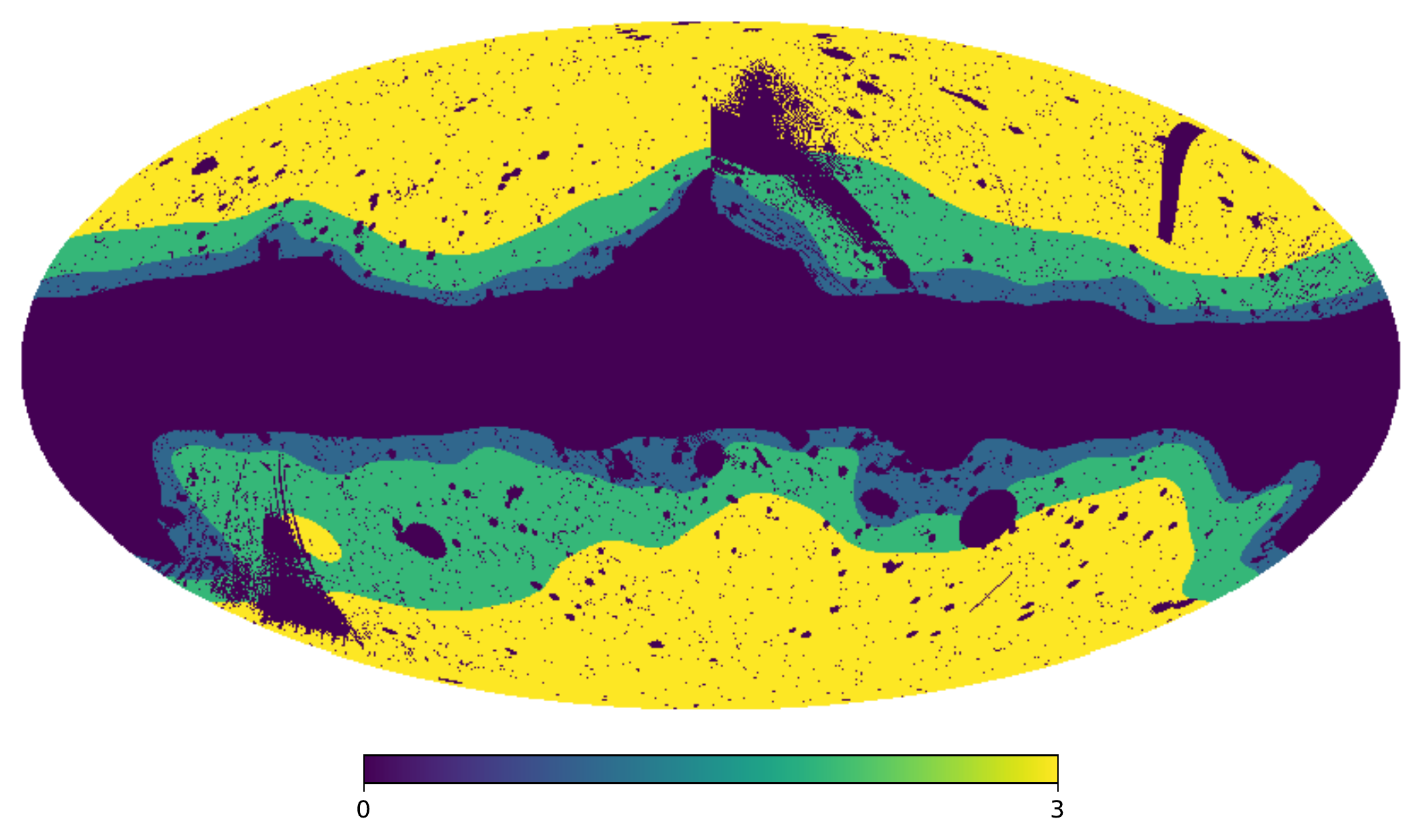}
\caption{This figure depicts the different masks used for the $\mu T$ and $\mu E$ cross correlation analysis on component maps derived from \Planck data. The yellow, green and blue masks correspond to $f_{\rm sky}=0.37,0.54$ and $0.62$ respectively.}
\label{fig:planck_fnl_mask}
\end{figure}
%

We begin by carrying out the component separation analysis on the full mission as well as the two half mission data sets. We  re-validate our NILC pipeline by comparing the power spectra estimated from the component separated maps generated during our analysis with the corresponding power spectra derived from SMICA component separation maps. We find good consistency and these comparisons are reported in \app{app:smica_spec_compare}. 
In all the analyses presented below we work with a union of the common intensity and polarization masks, HFI mask as well as a point source mask made available by the \Planck collaboration. Since we work with half mission maps we also duly account for the missing pixel mask. We also test the sensitivity of our measurements to using different sky fractions in the analysis by using HFI masks that retain differing portions of the sky. The different masks used in the analyses are depicted in \fig{fig:planck_fnl_mask}. 
In the following sections we discuss the $\mu T$ and $\mu E$ measurements.

\vspace{-3mm}
\subsection{$\mu T$ analysis}
\label{sec:planck_muT}
\begin{figure*}
\hspace*{-0.18cm} 
\subfigure[\label{fig:muT_smica_compare}]{\includegraphics[width=\columnwidth]{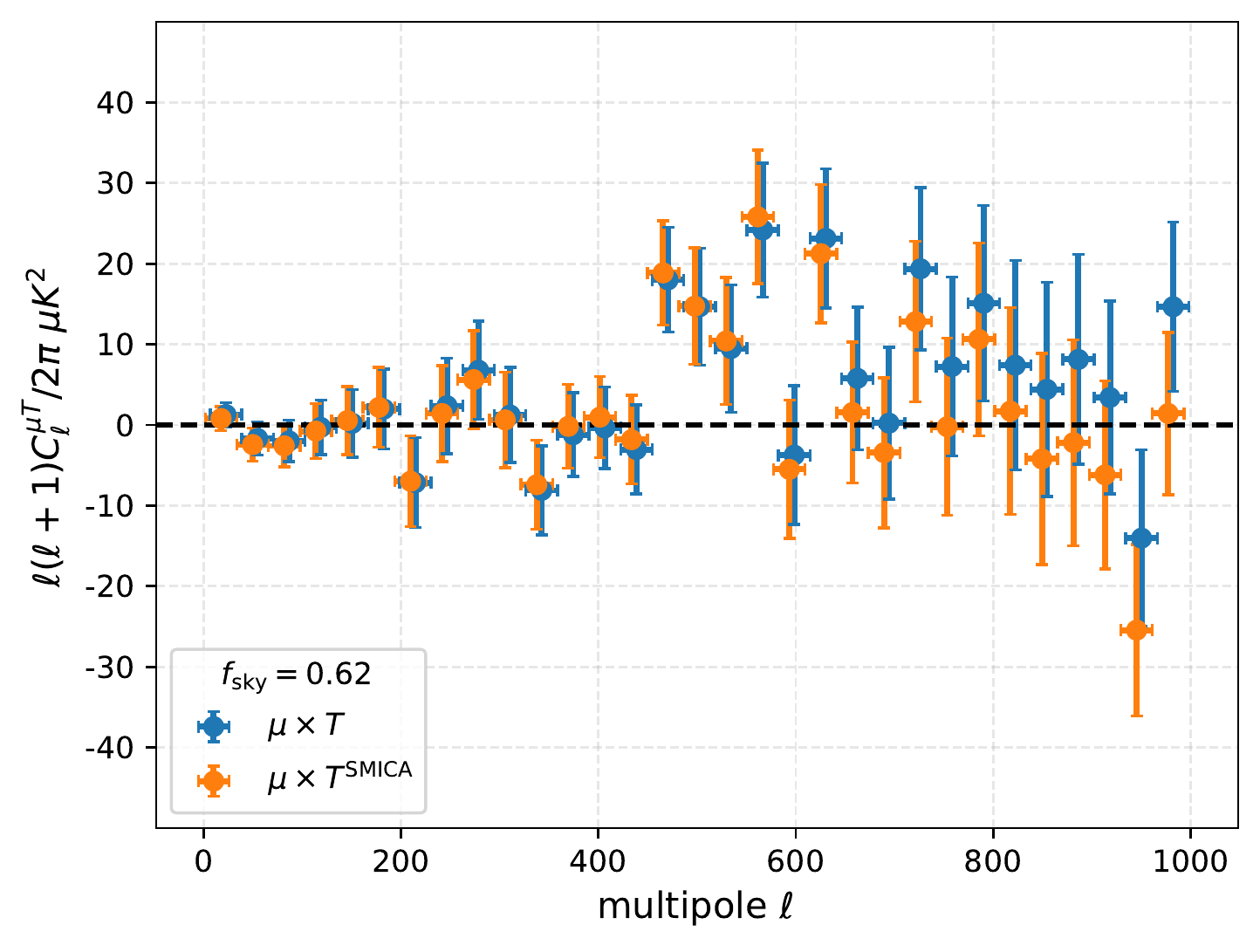}}
\subfigure[\label{fig:muT_fsky_vary}]{\includegraphics[width=\columnwidth]{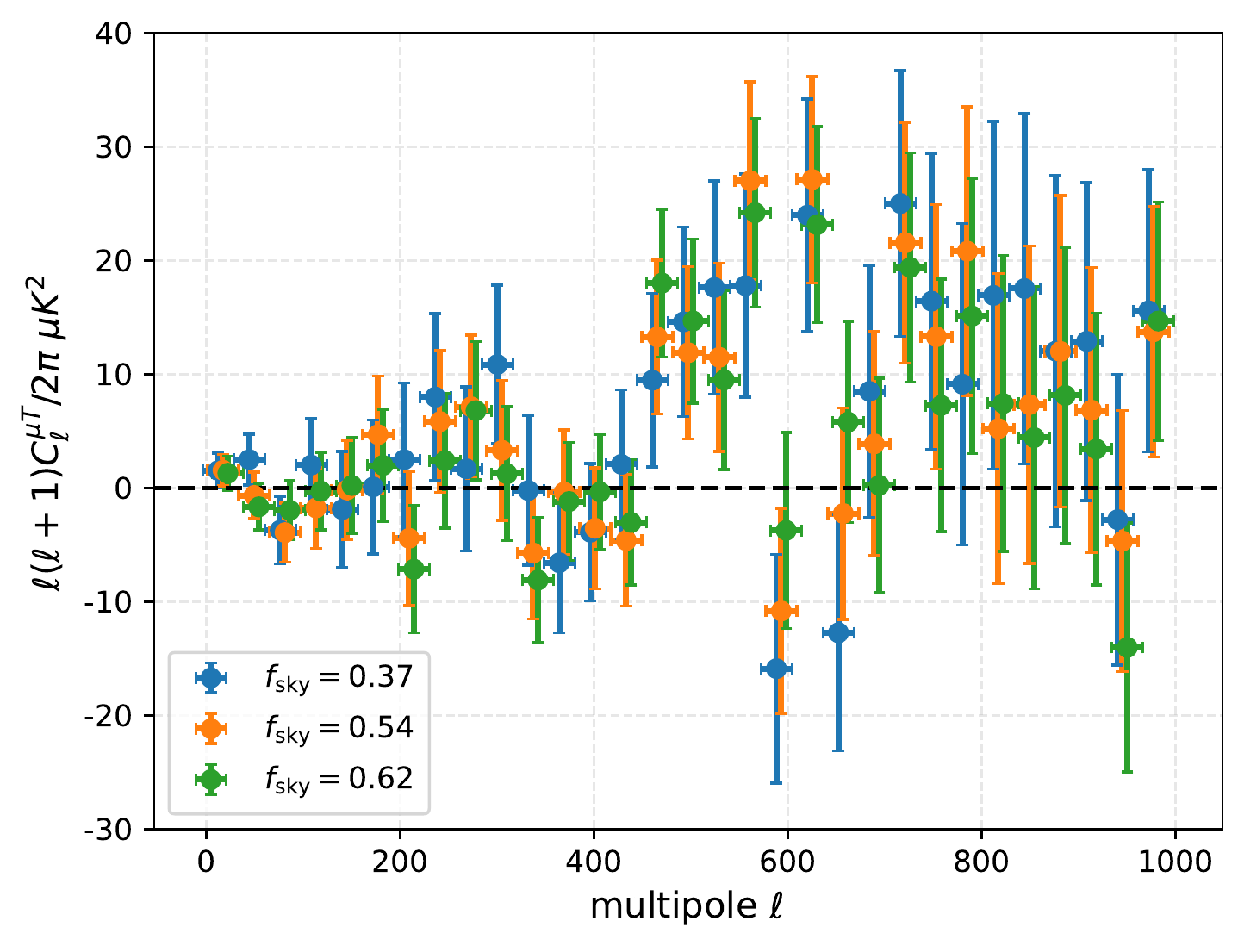}}
\caption{The figure on the left compares the $C_{\ell}^{\mu T}$ measurements derived using the temperature maps derived in this work with that derived from using SMICA temperature maps. The figure on the right shows the $C_{\ell}^{\mu T}$ derived from retaining different fraction of the sky.}
\label{fig:planck_muT_spectra}
\end{figure*}

\begin{figure}
\hspace*{-0.18cm} 
\includegraphics[width=\columnwidth]{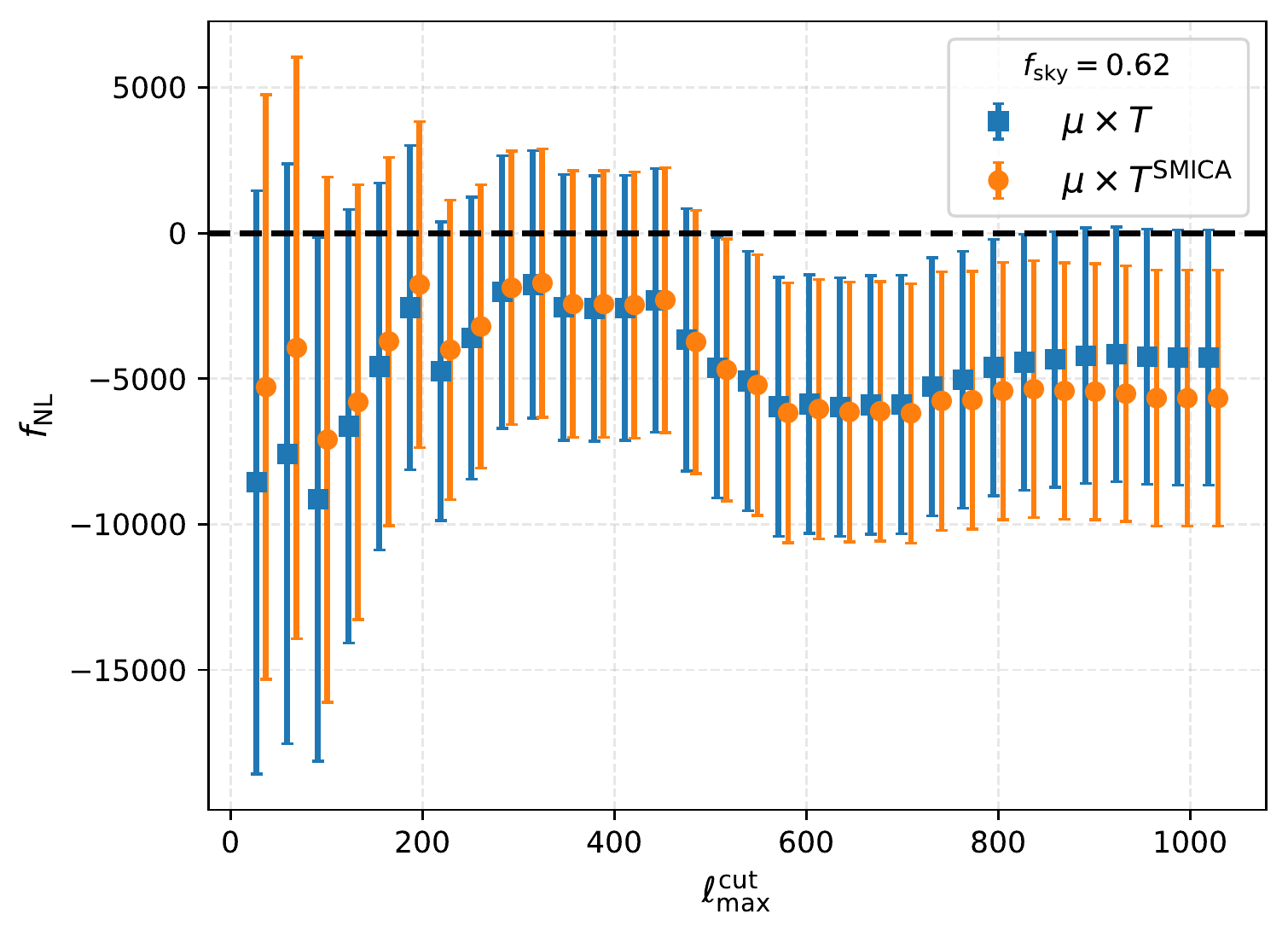}
\caption{This figure depicts the evolution of \fnl~ error as a function of maximum multipole used in the likelihood analysis.}
\label{fig:muT_fnl_lmax}
\end{figure}

We reiterate that since $\mu$ and $T$ maps have the same origin of noise we carry out our analysis by cross correlating complementary half mission component maps so as to avoid noise bias in the measurements of the $\mu T$ spectrum.
Here, we focus on discussing results derived from CMB deprojected $\mu$ maps. Motivated by the results seen in the analysis on realistic \Planck simulations, we first carry out the analysis without performing SZ deprojection. The spectral measurements are carried out using the maps derived from our component separation pipeline. Additionally, we also derive the spectral measurements using the corresponding SMICA component separation map\citep{Planck2018_diffuse_comp_sep} which serves as an additional consistency check. The $C_{\ell}^{\mu T}$ measurements derived from the different combination of component maps and for varying fraction of the sky are depicted in \fig{fig:planck_muT_spectra}. We begin by noting that $C_{\ell}^{\mu T}$ measurement derived using the temperature anisotropy maps estimated using the NILC implementation in this work and from using the SMICA temperature maps are highly consistent with each other as seen in \fig{fig:muT_smica_compare}. We note that the measurement are also very consistent on using varying fractions of the sky in the analysis as seen in \fig{fig:muT_fsky_vary}. 
Next we pass these measurements through our likelihood code to derive constraints on the \fnl parameter. A summary of how the constraint on \fnl evolve as a function of the maximum multipole used in the likelihood analysis is depicted in \fig{fig:muT_fnl_lmax}. We find the \fnl measurements to be consistent with a null measurement. A more quantitative summary of the \fnl statistics derived from assuming $\ell_{\rm max}=1024$ in the likelihood analysis for different sky fractions is given in \tab{tab:muT_fnl_stat}. We note that the measurements derived using SMICA $T$ maps as well as those derived using $T$ maps recovered using our pipeline yield highly consistent measurements. While this serves as a check, it is not surprising since the consistency of the recovered CMB maps was already demonstrated in \app{app:smica_spec_compare} and the the same $\mu$ map enters both the analyses. 

However, we note that the estimated errors are about $\simeq 15 $\% worse than those estimated from the simulated analysis presented in \sec{sec:realistic_planck_sim}. This is likely because our simulations do not fully incorporate the foreground complexity and use too simplistic noise estimates for each channel (e.g., by omitting inhomogeneous noise from the scanning etc).
\begin{table}
\centering
\groupedRowColors{1}{2}{gray!10}{white!10}
\begin{tabular}{lllll}
\toprule
 $f_{\rm sky}$    &    Data     & $f_{\rm NL}$ & $\sigma_{f_{\rm NL}}$ &   SNR \\
\midrule
0.37 & $T$ &         -866 &                  5078 & -0.17 \\
     & SMICA $T$ &        -1249 &                  5096 & -0.25 \\
0.54 & $T$ &        -2513 &                  4459 & -0.56 \\
     & SMICA $T$ &        -3649 &                  4473 & -0.82 \\
0.62 & $T$ &        -4273 &                  4382 & -0.98 \\
     & SMICA $T$ &        -5670 &                  4395 & -1.29 \\
\bottomrule
\end{tabular}

\caption{This table summarizes the statistics of the measured \fnl parameters from $\mu T$ analyses that use different fractions of the sky. These likelihood analyses used multipoles $\ell \in [2,1024]$.}
\label{tab:muT_fnl_stat}
\end{table}

\vspace{-3mm}
\subsubsection{Comparison with previous works}
\label{sec:planck_muT_compare}
Previous work \citep{Khatri2015mu} has derived constraints on primordial non-Gaussianity via measurements of the $\mu T$ correlation function using \Planck data. However, when constructing the $\mu$ map they did not deproject the CMB. As we have demonstrated in this work and so have other works \citep{Remazeilles2018mu,Remazeilles2021mu}, this procedure is critical to making an unbiased measurement of the $\mu T$ power spectrum. Owing to this, \citet{Khatri2015mu} do indeed find the $C_{\ell}^{\mu T}$ measurements to not be consistent with zero (Fig.~6 of their paper). This potentially led the authors to quote overly conservative limits in their final conclusions, not truly exploiting the full potential of \Planck data.

\begin{figure}
\hspace*{-0.18cm} 
\includegraphics[width=\columnwidth]{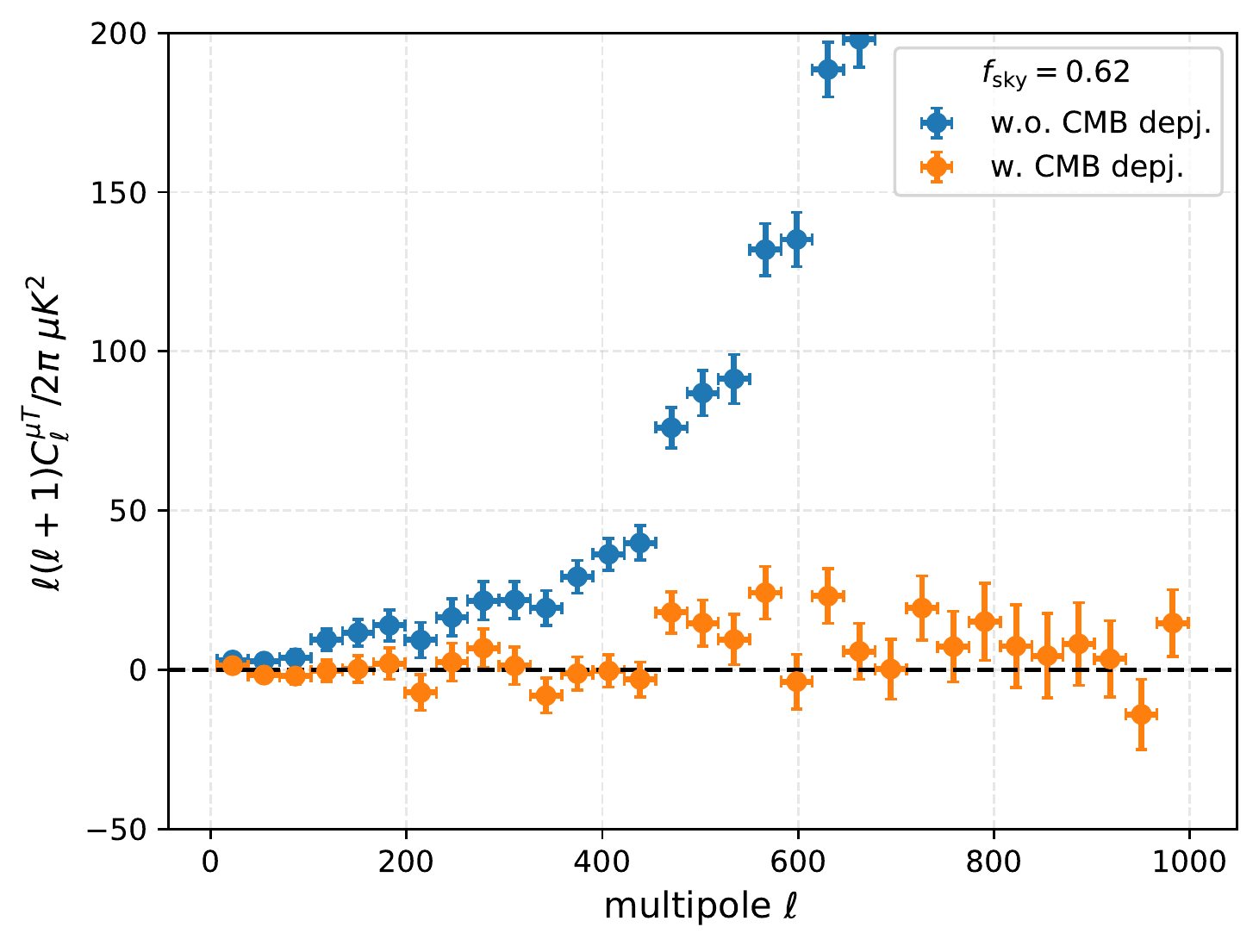}
\caption{This figure highlights the importance of deprojecting CMB from the $\mu$-map by comparing the relevant $C_{\ell}^{\mu T}$ spectra. The first is estimated using the conventional $\mu$ map and the other uses the CMB deprojected $\mu$ map, both derived from \Planck data.}
\label{fig:biased_muT_compare}
\end{figure}

To re-emphasize the importance of this deprojection we compare the $C_{\ell}^{\mu T}$ spectra derived using the conventional $\mu$ maps to that estimated using the CMB deprojected $\mu$ maps from \Planck data and these are depicted in \fig{fig:biased_muT_compare}.
Specifically the blue points in \fig{fig:biased_muT_compare} represent a qualitative reproduction of Fig.~6 of \citet{Khatri2015mu}. The authors, however, interpret this measurement to be consistent with a null measurement focussing only on the lowest ($\ell < 32$) multipoles. Furthermore, in addition to ignoring the two highest frequency \Planck channels (i.e 545 GHz and 857 Ghz) while reconstructing the $\mu$ map, \citet{Khatri2015mu} also drop the low-frequency channels in their analysis, thereby severely limiting the ability of their setup to distinguish between the $\mu$ and CMB spectra as one may expect from observing \fig{fig:spectra}. This is also potentially reflected in the enhanced errors reported in their work. 

Bearing in mind these caveats, we note that they report a upper limit on the \fnl $\lesssim 10^5$. Assuming this to be the 95\% upper limit, in comparison we find a null detection of the \fnl parameter and set a $2\sigma$ upper limit of \fnl $\lesssim 8800$, an improvement of over an order of magnitude. In addition to this \Planck data is expected to improve the constraints on \fnl further via measurements of $\mu E$ correlations, which we discuss next.
\subsection{$\mu E$ analysis}
\label{sec:planck_muE}
\begin{figure}
\hspace*{-0.18cm} 
\includegraphics[width=\columnwidth]{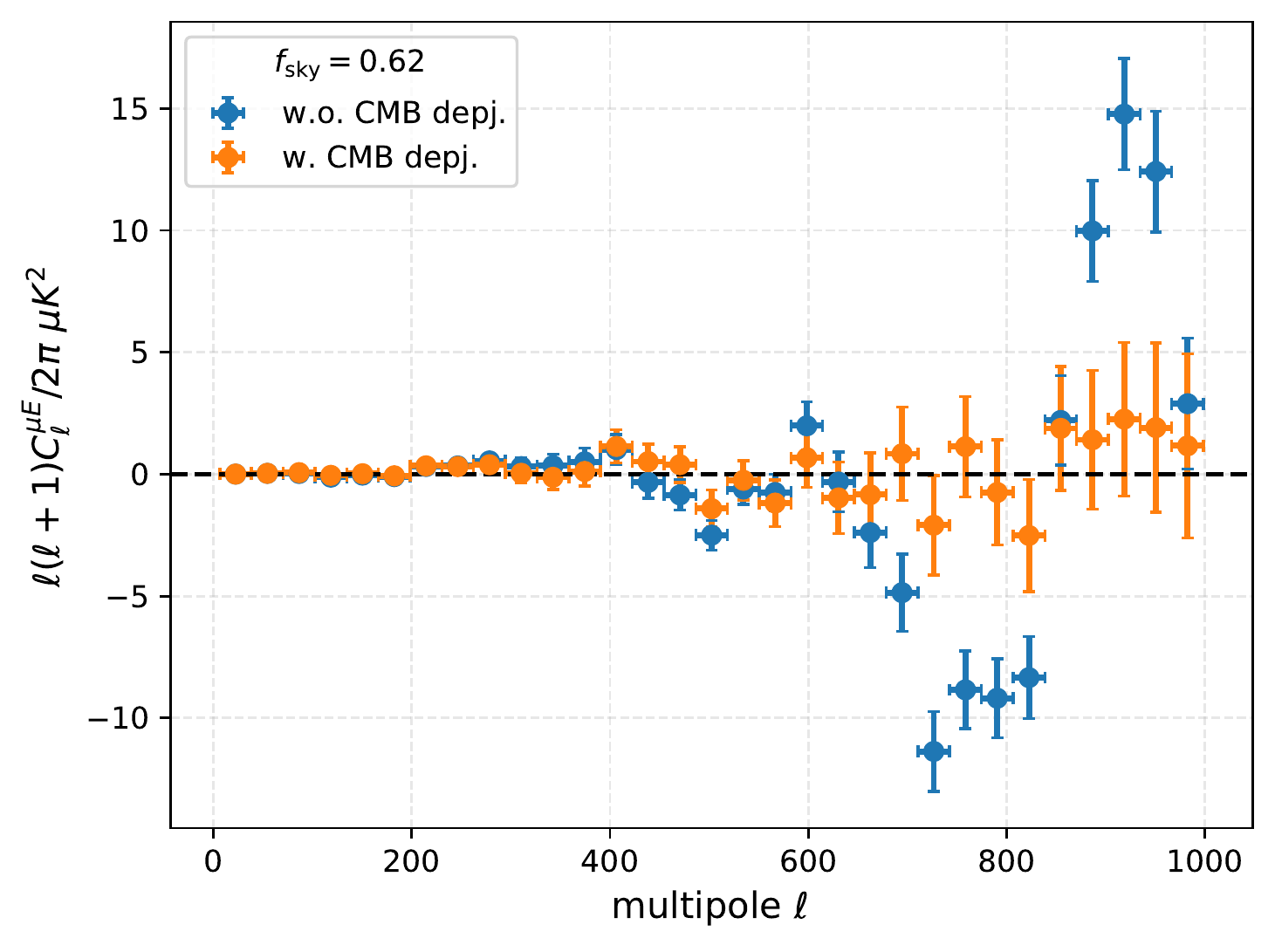}
\caption{This figure depicts a comparison of the two $C_{\ell}^{\mu E}$ spectra measurement, the first estimated using the conventional $\mu$ map and the other uses CMB deprojected $\mu$ map, both derived from \Planck data.}
\label{fig:biased_mue_compare}
\end{figure}

\begin{figure*}
\hspace*{-0.18cm} 
\subfigure[\label{fig:muE_smica_compare}]{\includegraphics[width=\columnwidth]{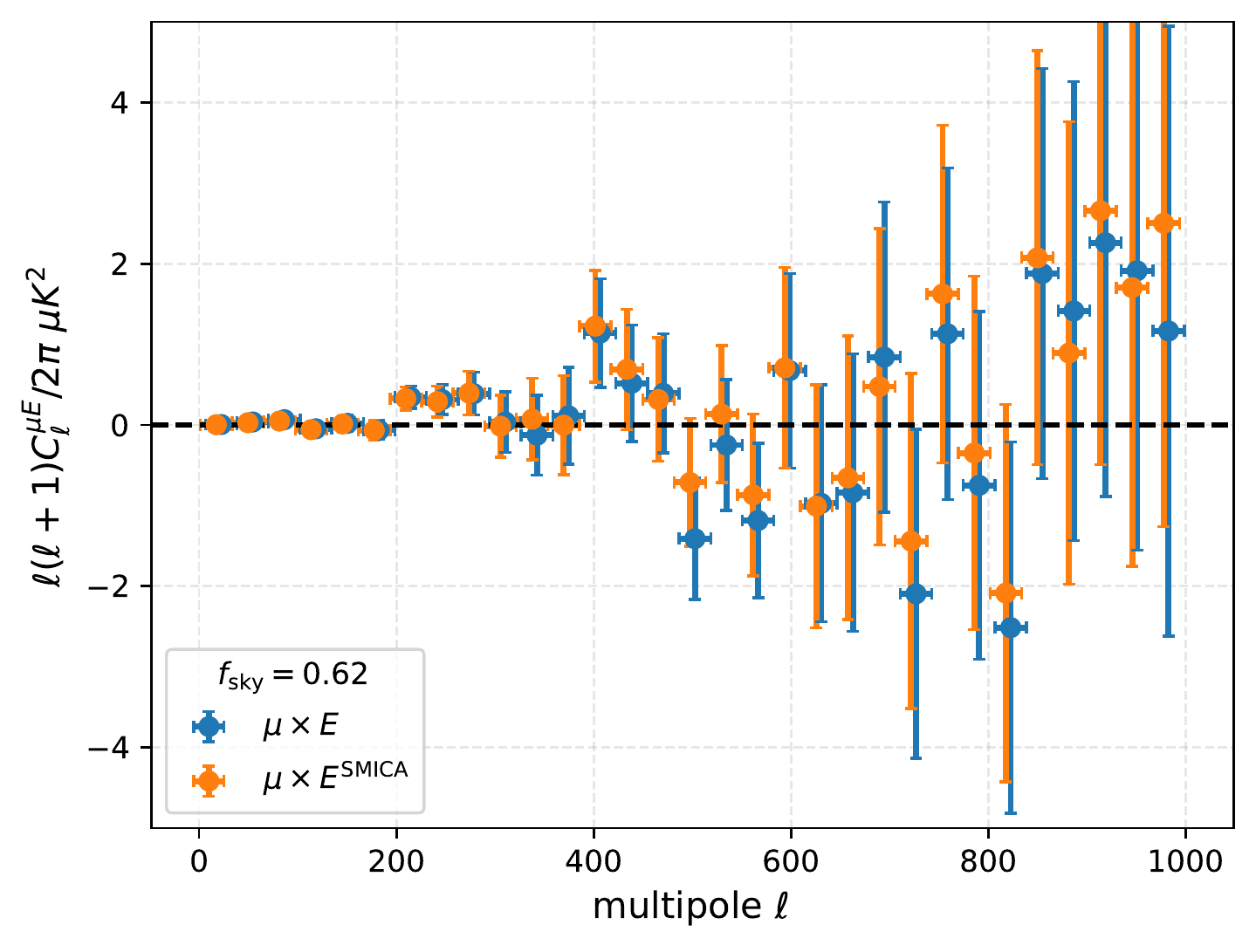}}
\subfigure[\label{fig:muE_fsky_vary}]{\includegraphics[width=\columnwidth]{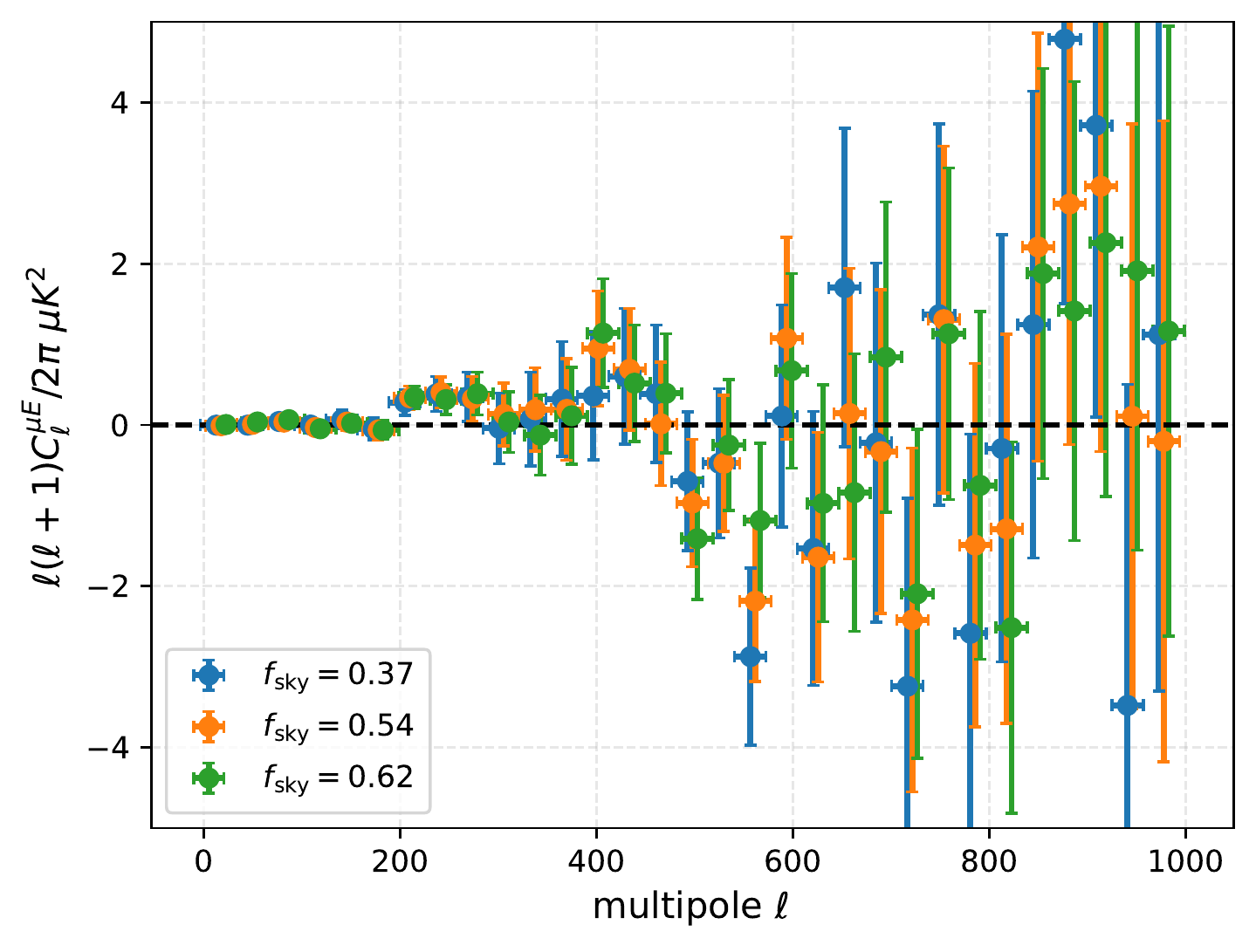}}
\caption{The figure on the left compares the $C_{\ell}^{\mu E}$ measurements derived using the $E$ mode maps derived in this work with that derived from using SMICA temperature maps. The figure on the right shows the $C_{\ell}^{\mu E}$ derived from retaining different fraction of the sky.}
\label{fig:planck_muE_spectra}
\end{figure*}

The measurements of the $\mu E$ spectra are derived from full mission component maps. We begin by showcasing the biased spectra that arise when not performing the critical CMB deprojection step. This is depicted in \fig{fig:biased_mue_compare}. 
Even for this analysis we therefore work with the CMB deprojected $\mu$ maps. Like in the previous analysis we measure $C_{\ell}^{\mu E}$ spectra from a combination of  component maps derived in this work and SMICA E-mode maps \citep{Planck2018_diffuse_comp_sep}. We find excellent consistency between $C_{\ell}^{\mu E}$  measurements derived using E-mode maps derived from our component separation pipeline and that of SMICA as seen in \fig{fig:muE_smica_compare}. We also carry out the measurements by retaining different fractions of the sky and again find very good consistency between the different measurements as seen in \fig{fig:muE_fsky_vary}.

\begin{figure}
\hspace*{-0.18cm} 
\includegraphics[width=\columnwidth]{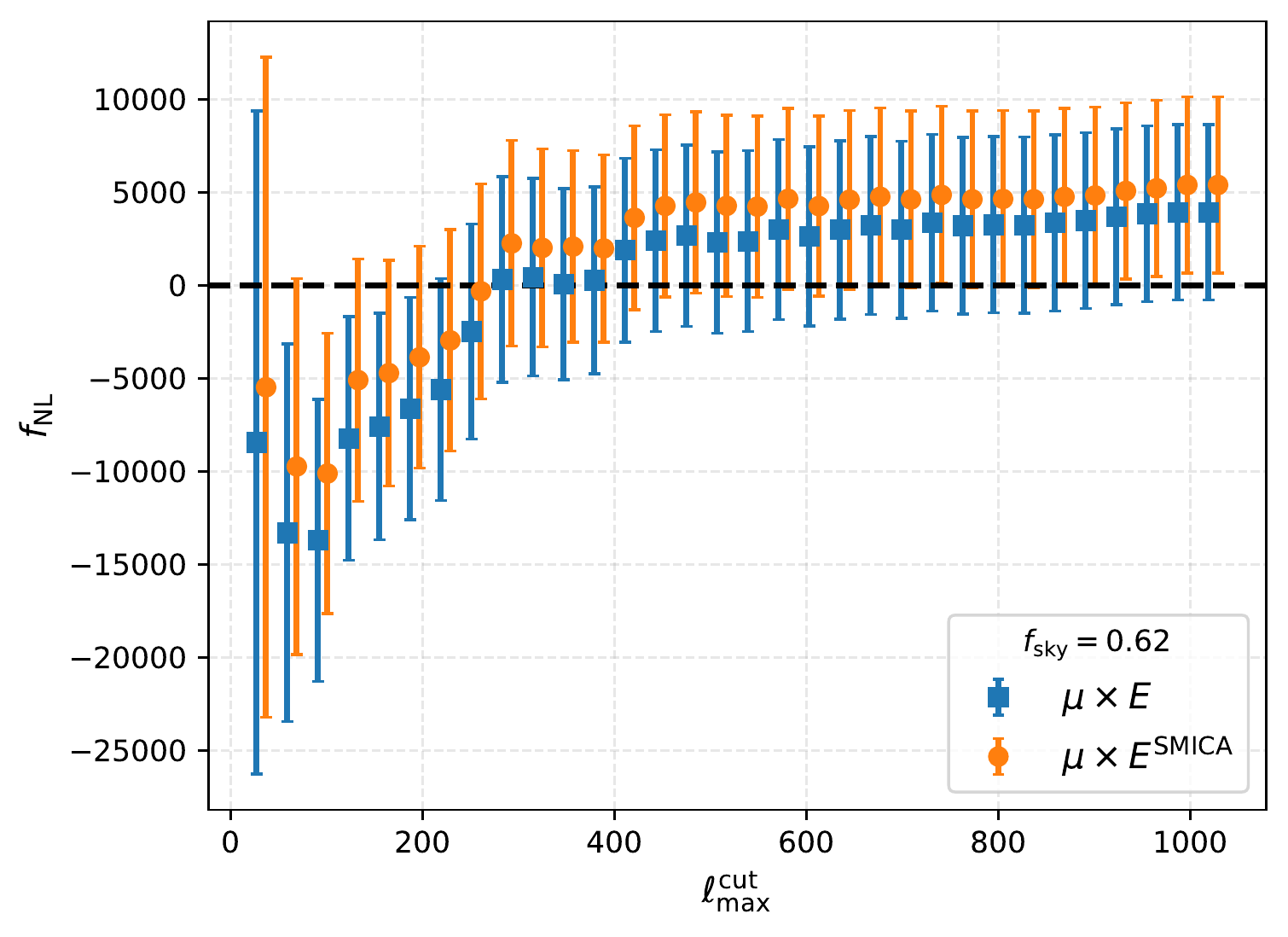}
\caption{This figure depicts the evolution of \fnl measurement as a function of maximum multipole used in the analysis.}
\label{fig:muE_fnl_lmax}
\end{figure}

We then pass these measured spectra through the likelihood pipeline to derive constraints on the \fnl parameter from CMB distortion anisotropies. We find little variation in the mean of the \fnl measurements on varying the maximum multipole used in the likelihood analysis as seen in \fig{fig:muE_fnl_lmax}. We again find excellent consistency with the measurements derived from SMICA $E$ mode maps. Finally we provide in \tab{tab:muE_fnl_stat} a quantitative summary of the \fnl statistics derived from the likelihood analysis assuming $\ell_{\rm max}=1024$  and on using different fractions of the sky in the analysis.
\begin{table}
\centering
\groupedRowColors{1}{2}{gray!10}{white!10}
\begin{tabular}{lllll}
\toprule
$f_{\rm sky}$     &   Data      & $f_{\rm NL}$ & $\sigma_{f_{\rm NL}}$ &   SNR \\
\midrule
0.37 & $E$ &         6995 &                  5347 &  1.31 \\
     & SMICA $E$ &         6092 &                  5441 &  1.12 \\
0.54 & $E$ &         9404 &                  4812 &  1.95 \\
     & SMICA $E$ &        10383 &                  4842 &  2.14 \\
0.62 & $E$ &         3937 &                  4719 &  0.83 \\
     & SMICA $E$ &         5409 &                  4733 &  1.14 \\
\bottomrule
\end{tabular}

\caption{This table summarizes the statistics of the measured \fnl parameters from $\mu T$ analyses that use different fractions of the sky. These likelihood analyses used multipoles $\ell \in [2,1024]$.}
\label{tab:muE_fnl_stat}
\end{table}
%

\subsection{Joint $\mu T$ \& $\mu E$ constraints on \fnl}
\label{sec:planck_muT_muE}
\begin{figure}
\hspace*{-0.18cm} 
\includegraphics[width=\columnwidth]{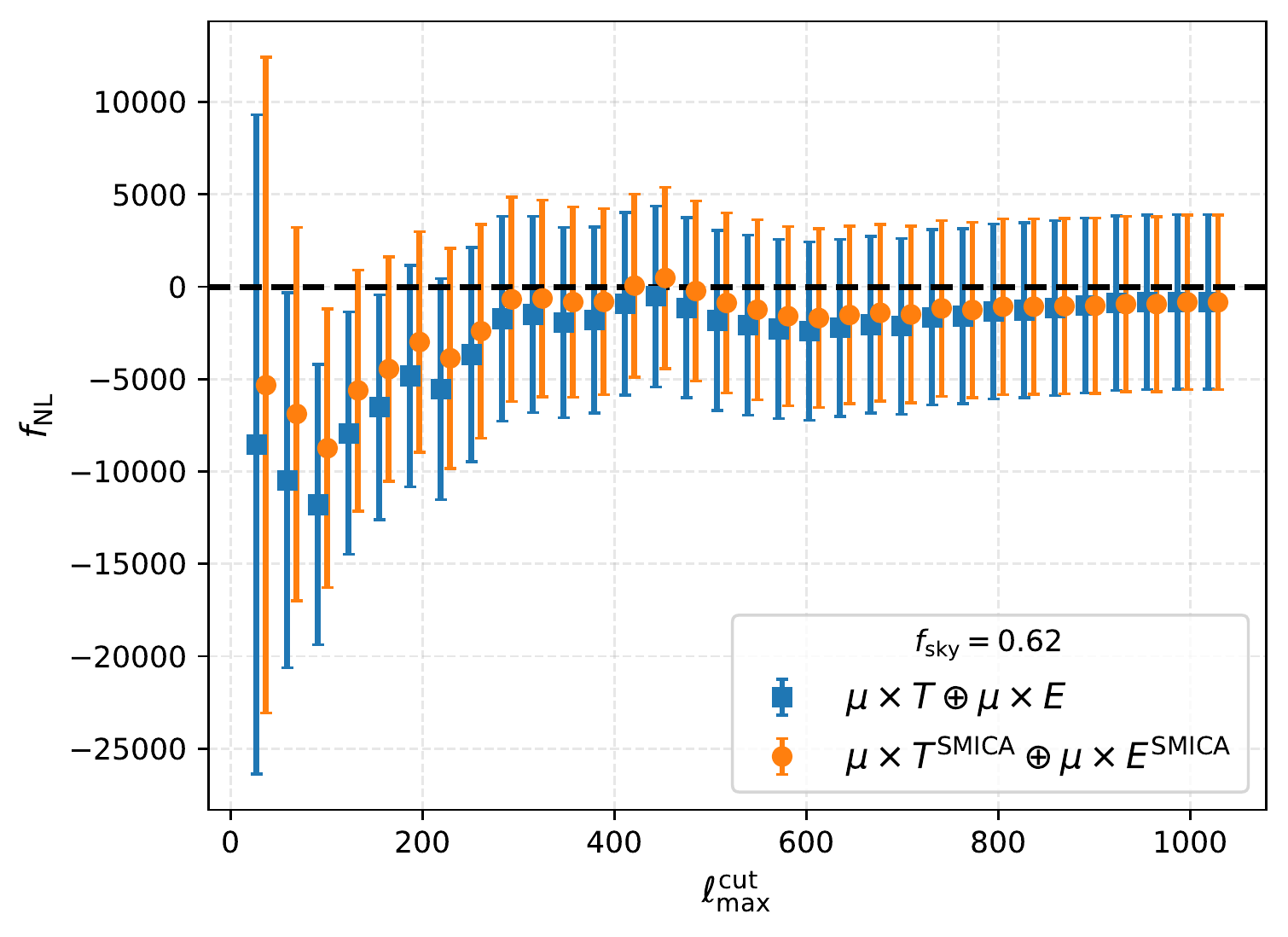}
\caption{This figure depicts the evolution of the inferred \fnl and its error from combining the $\mu T$ and $\mu E$ measurements, as a function of the maximum multipole used in the analysis.}
\label{fig:muT_muE_fnl_lmax}
\end{figure}

We find that the \fnl measurements inferred from $\mu T$ as well as the $\mu E$ measurements to be consistent with zero with in errors.
We now combine these two measurements to provide the best constraints on the \fnl parameter duly accounting for the covariance between these measurements as detailed in \sec{sec:fnl_lkl} and \app{app:FM_HM_comparison}. First we derive the \fnl constraints as a function of maximum multipole used in the likelihood analysis and this is depicted in \fig{fig:muT_muE_fnl_lmax}.  Finally to fully exploit the data, we use the maximum sky and multipole coverage to provide the best \fnl constraints from \Planck to date which are summarized in \tab{tab:muT_muE_fnl_stat}.

\begin{table}
\centering
\groupedRowColors{1}{2}{gray!10}{white!10}
\begin{tabular}{lllll}
\toprule
Data & $f_{\rm NL}$ & $\sigma_{f_{\rm NL}}$ &   SNR \\
\midrule
$T$         &        -4273 &                  4382 & -0.98 \\
SMICA $T$   &        -5670 &                  4395 & -1.29 \\
$E$         &         3937 &                  4719 &  0.83 \\
SMICA $E$   &         5409 &                  4733 &  1.14 \\
$T$+$E$       &         -812 &                  3398 & -0.24 \\
SMICA $T$+$E$ &         -840 &                  3410 & -0.25 \\
\bottomrule
\end{tabular}

\caption{This table summarizes the best constraints on the \fnl parameter derived from \Planck data. These constraints correspond to $f_{\rm sky}=0.62$ and $\ell_{\rm max}=1024$, fully exploiting the constraining power of the data.}
\label{tab:muT_muE_fnl_stat}
\end{table}
%

%

\subsection{Why does tSZ deprojection bias the $\mu T$ measurement?}
\label{sec:sz_cib_role}
We demonstrated in \sec{sec:ideal_planck_sim} where we analysed idealised simulations that it was important to deproject CMB as well as the tSZ spectra in order to yield unbiased measurements of the $\mu T$ spectra. However in the subsequent section where we discussed analyses on realistic \Planck simulations we observed that SZ deprojection actually led to biased measurements of the $\mu T$ spectrum. This motivated carrying out the analysis on \Planck data  without performing a SZ deprojection. Here we explore the reason for this intriguing observation. We essentially find that the reason we do not see a negative bias sourced by tSZ contamination is because there is a positive bias sourced by the cosmic infrared background anisotropies which coincidentally compensates for the negative bias induced by tSZ leakages in the component maps. We validate this explanation using two different analyses. 

\begin{figure*}
\hspace*{-0.18cm} 
\subfigure[\label{fig:planck_comp_corr_cib}]{\includegraphics[width=\columnwidth]{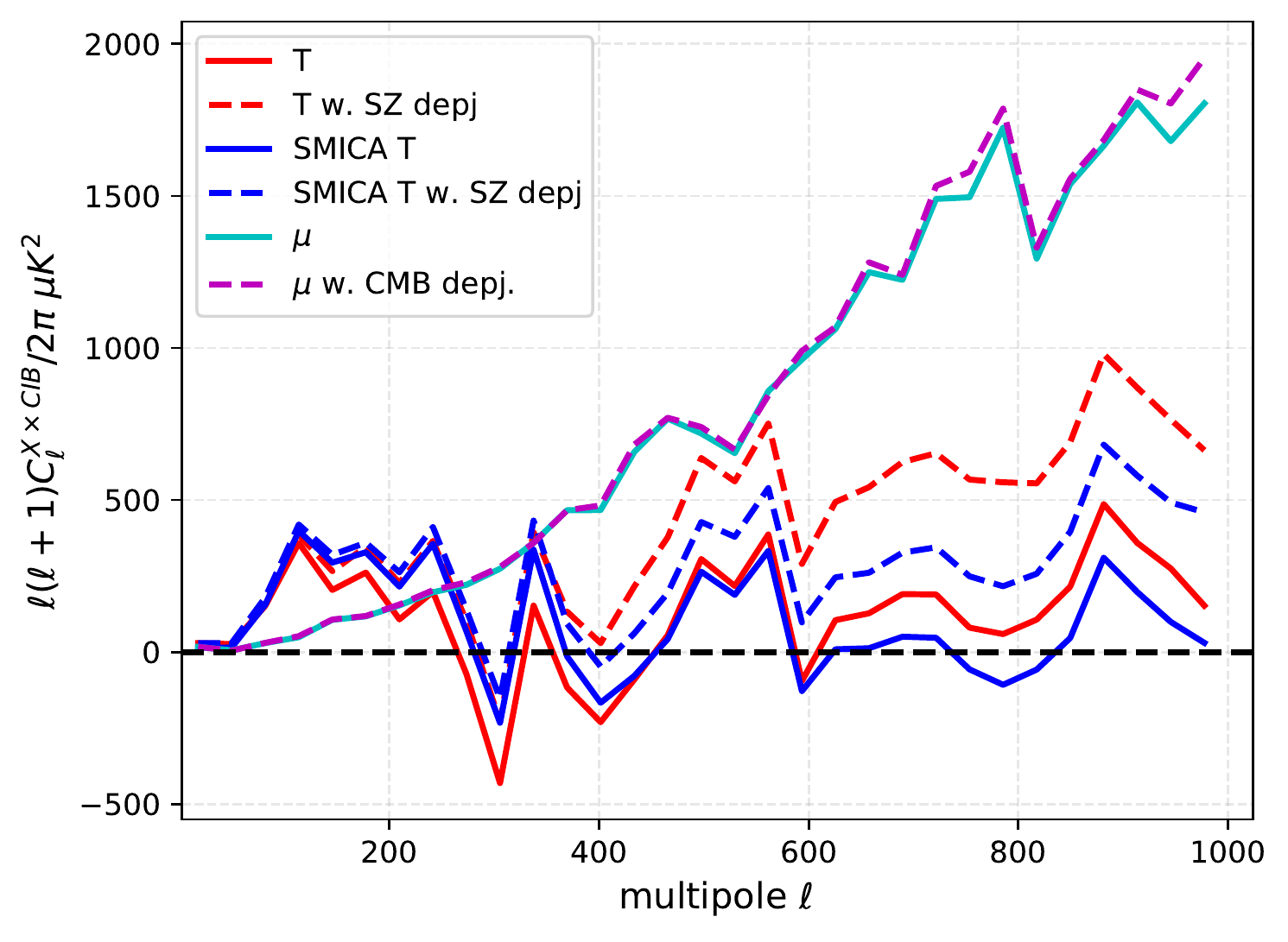}}
\subfigure[\label{fig:planck_comp_corr_tsz}]{\includegraphics[width=\columnwidth]{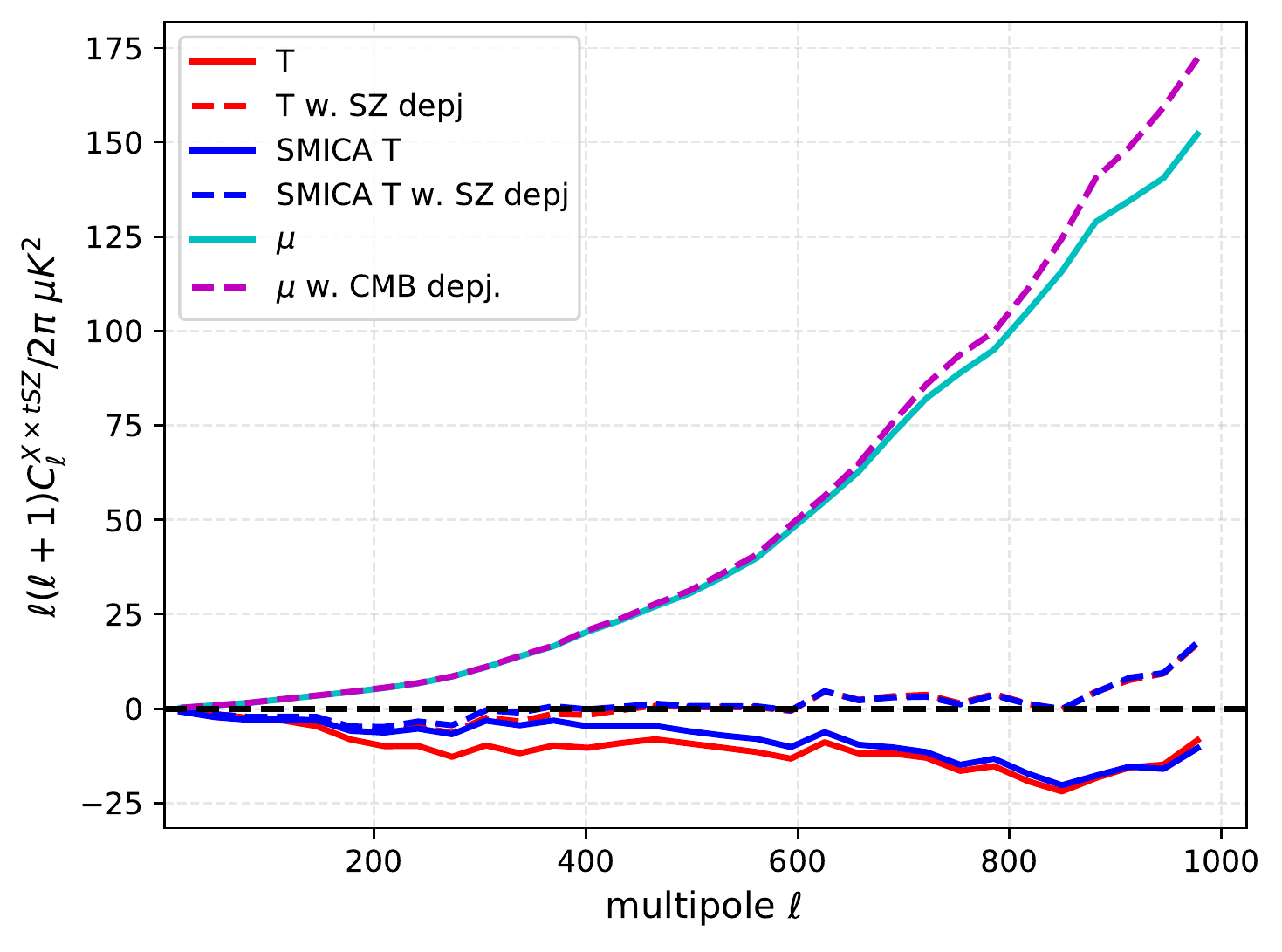}}
\subfigure[\label{fig:sim_comp_corr_cib}]{\includegraphics[width=\columnwidth]{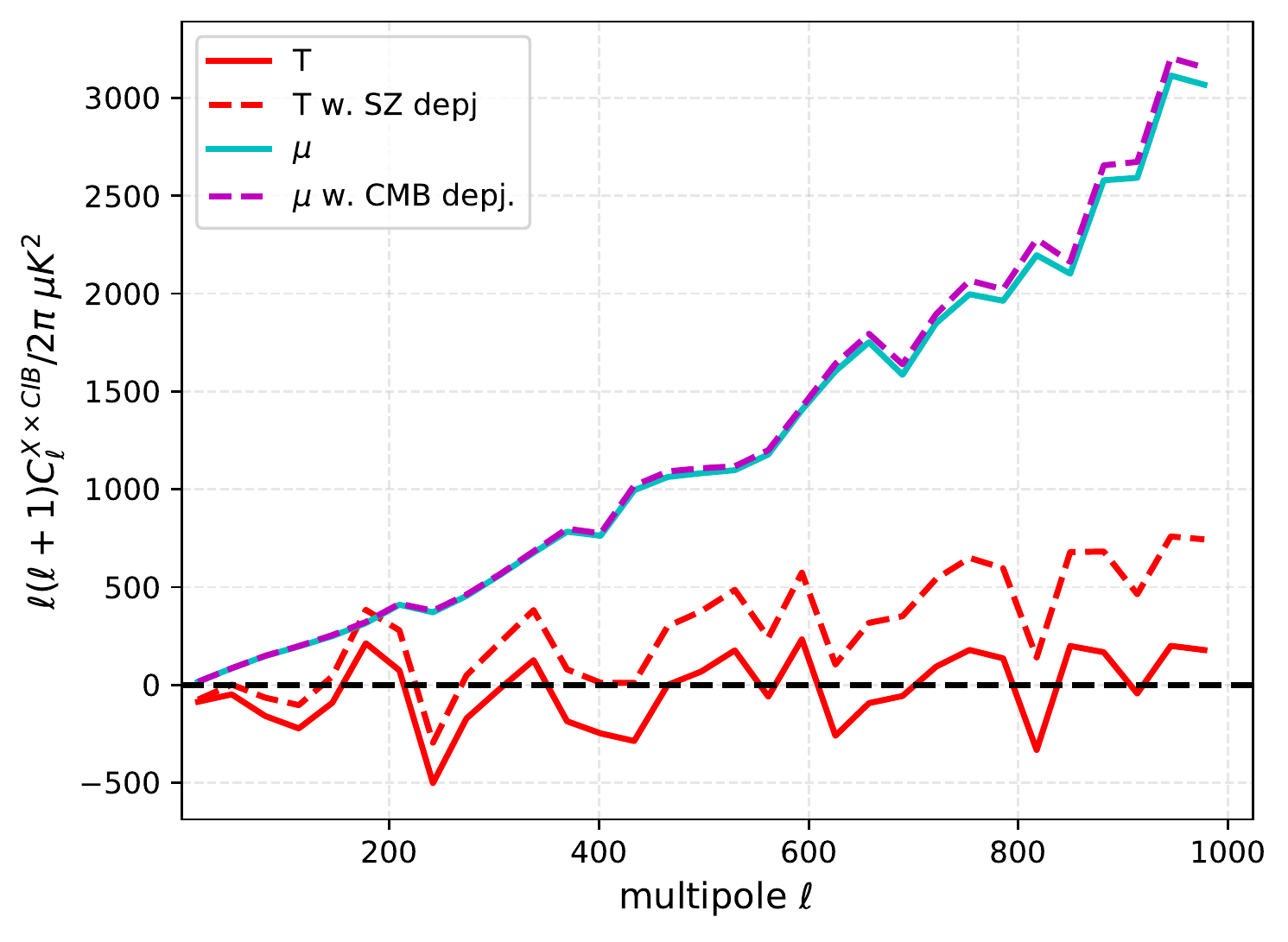}}
\subfigure[\label{fig:sim_comp_corr_tsz}]{\includegraphics[width=\columnwidth]{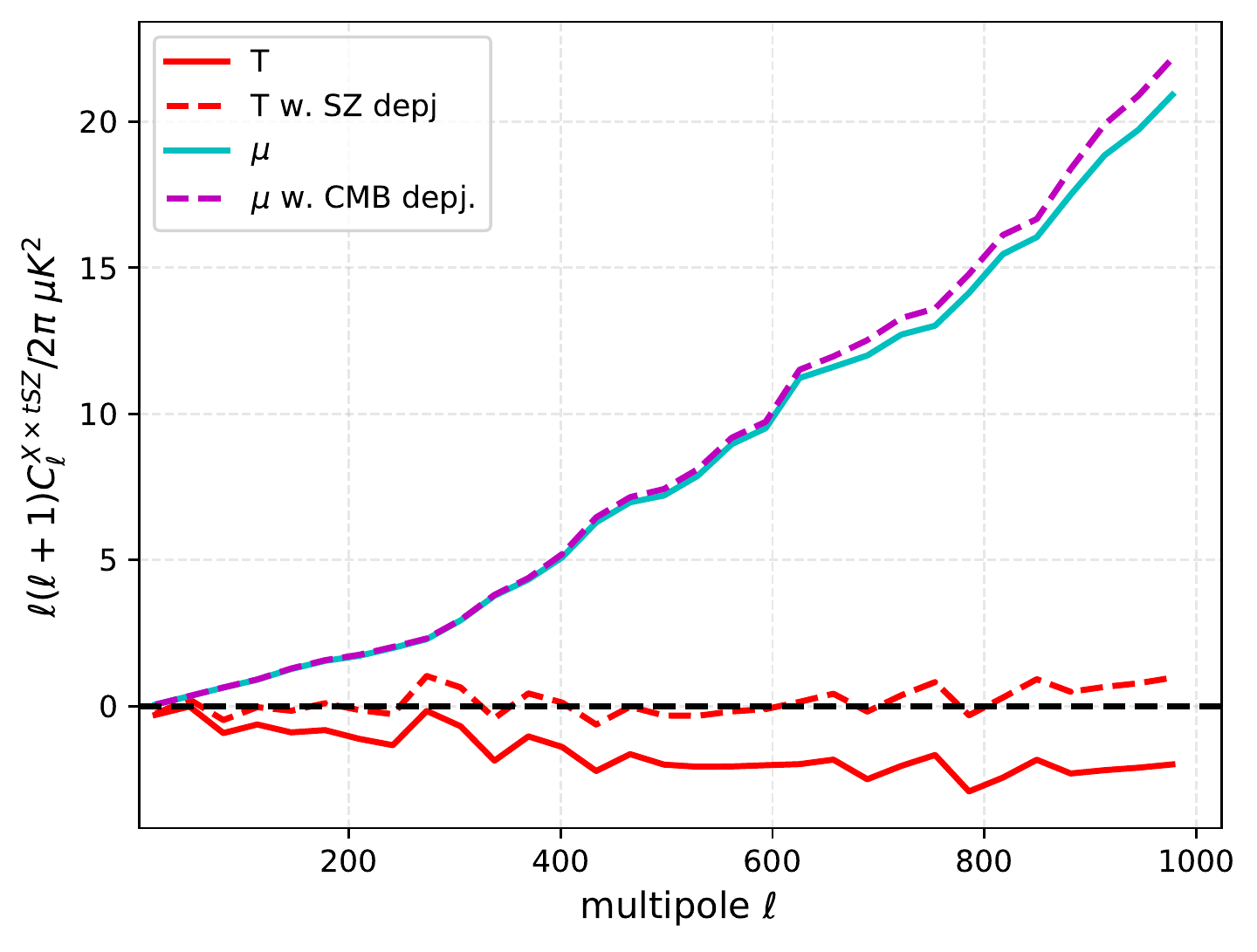}}
\caption{The top left and right panels depict the cross power spectrum measurements between recovered $T$ and $\mu$ maps with CIB and tSZ respectively. The bottom panels depict the same except these measurements are derived from simulations.}
\label{fig:cib_sz_corr}
\end{figure*}

First we compute the cross power spectrum between the recovered component maps, $\mu$ \& $T$ and the GNILC CIB \citep{Planck2016_GNILC} at 545 GHz\footnote{Using CIB at 353 GHz and 857 GHz yields the similar trends, though the amplitudes are understandably different.}.  We also estimate these cross power spectra using the SZ deprojected $T$ and $\mu$ maps. Since these spectra are measured from component maps extracted from full mission \Planck data, these suffer from noise bias which has not been corrected for in the spectra shown in the top panels of \fig{fig:cib_sz_corr}.  To complement and help interpret these potentially biased measurements we also carry out an analogous exercise on simulations where we correlated the extracted $T$ and $\mu$ maps with the true injected tSZ and CIB skies. Since these simulated spectral measurements do not suffer from noise bias, they are more reliable than those measured from data.  While there are differences in the spectra measured from data and simulations, owing to noise biases and details of simulations, we note that they essentially follow the same trends, which is all we need to construct the arguments we present next.

\begin{figure*}
\hspace*{-0.18cm} 
\subfigure[\label{fig:planck_mut_szdep}]{\includegraphics[width=\columnwidth]{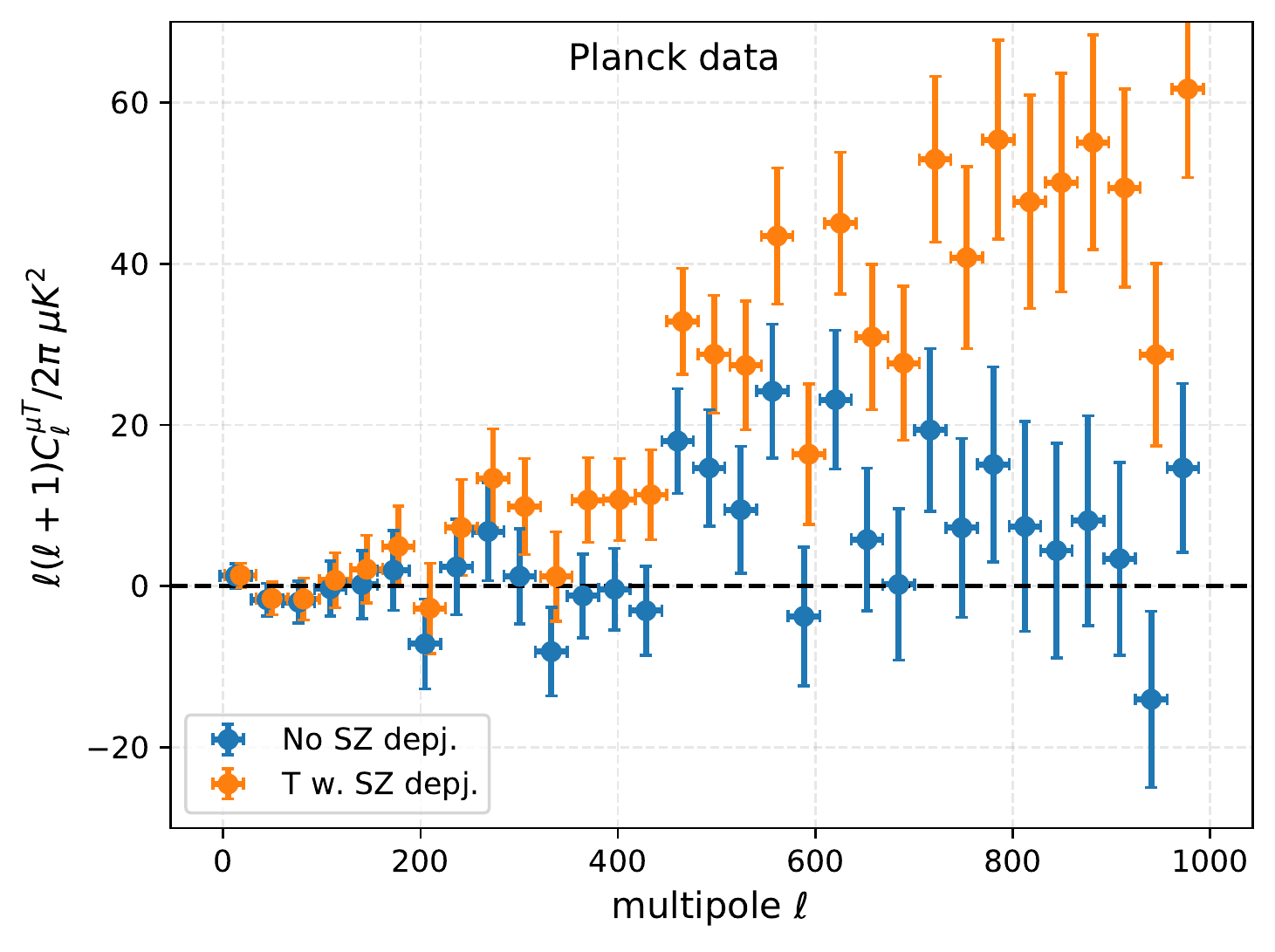}}
\subfigure[\label{fig:sim_mut_szdep}]{\includegraphics[width=\columnwidth]{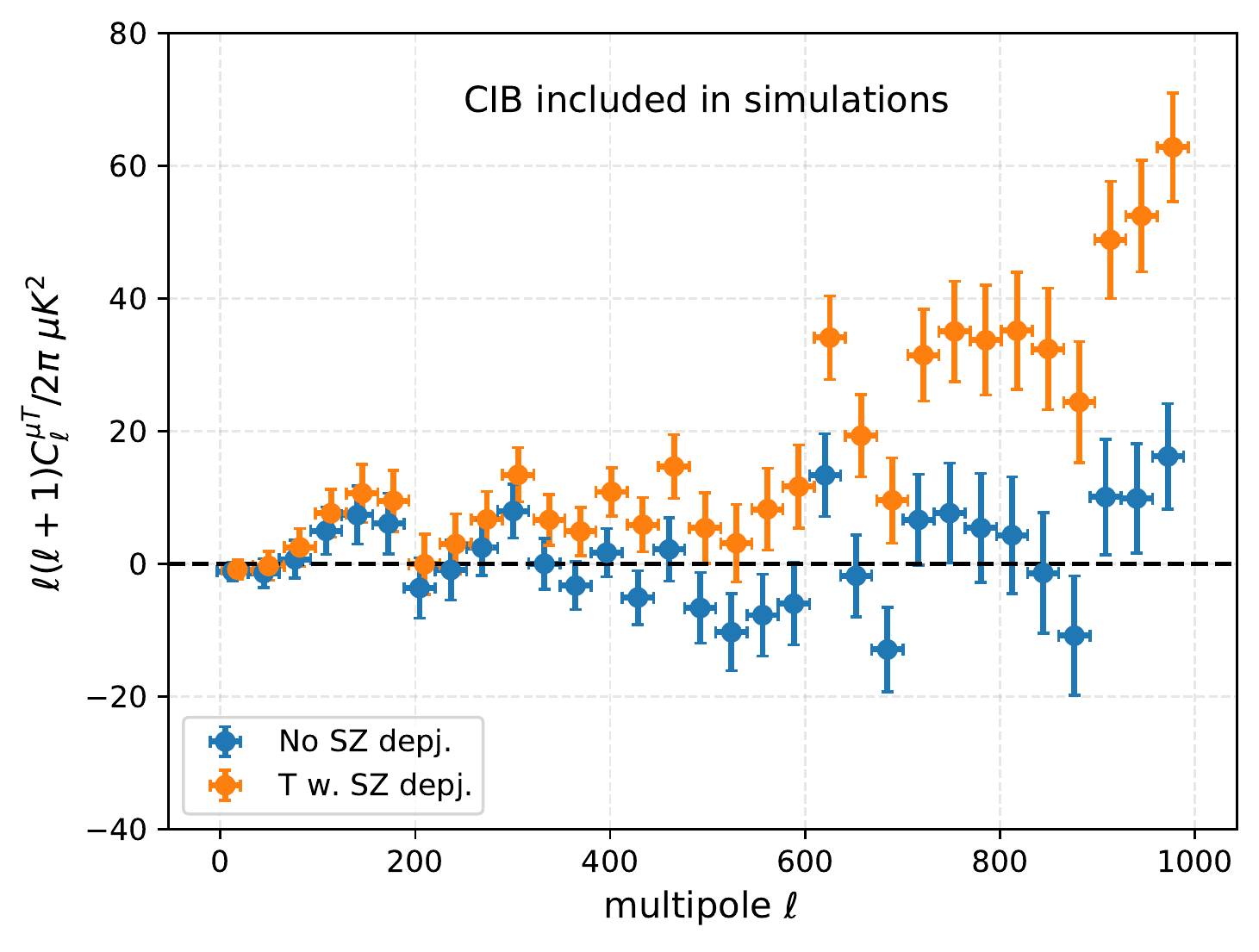}}
\caption{Same as \fig{fig:ideal_ilc_mt_me}, except the analysis is done on realistic sky simulations that include CIB.}
\label{fig:mut_szdep}
\end{figure*}

We begin by noting that while the $T$ map is negatively correlated with the tSZ map, the $\mu$ map is in fact positively correlated. This is the primary reason why the $\mu T$ measurement was expected to be biased negative. However we find both the CMB deprojected $\mu$ as well as the $T$ maps to be positively correlated with CIB as seen in the top panel of \fig{fig:cib_sz_corr}. We argue that this positive correlation with CIB is what cancels the negative bias expected from correlations with tSZ and results in an unbiased measurement of the $\mu T$ spectrum, even when not carrying out the SZ deprojection as seen in \fig{fig:mut_szdep}. Furthermore on performing SZ deprojection, while the $T$ correlation with tSZ is expectedly removed as seen in  \fig{fig:planck_comp_corr_tsz} and \fig{fig:sim_comp_corr_tsz}, this procedure enhances the $T$-CIB correlation as shown in the left panels of \fig{fig:cib_sz_corr}. It is this excess correlation in the SZ deprojected $T$ and CIB maps that results in the excess positive correlation seen in resultant $\mu T$ spectrum as shown in \fig{fig:mut_szdep}. We find these exact trend even on using the SMICA component maps in our analysis. We also find compatible trends both in data as well as in simulated analysis which validates this explanation. Furthermore on performing SZ deprojection when recovering the $\mu$ maps (instead of performing the deprojection when recovering $T$), we actually find the $\mu$ CIB correlation to be enhanced but it also changes sign to become  negatively correlated. This results in a $\mu T$ measurement which is biased negative and this is confirmed by our analysis, though not shown here for brevity. 

\begin{figure*}
\hspace*{-0.18cm} 
\includegraphics[width=\columnwidth]{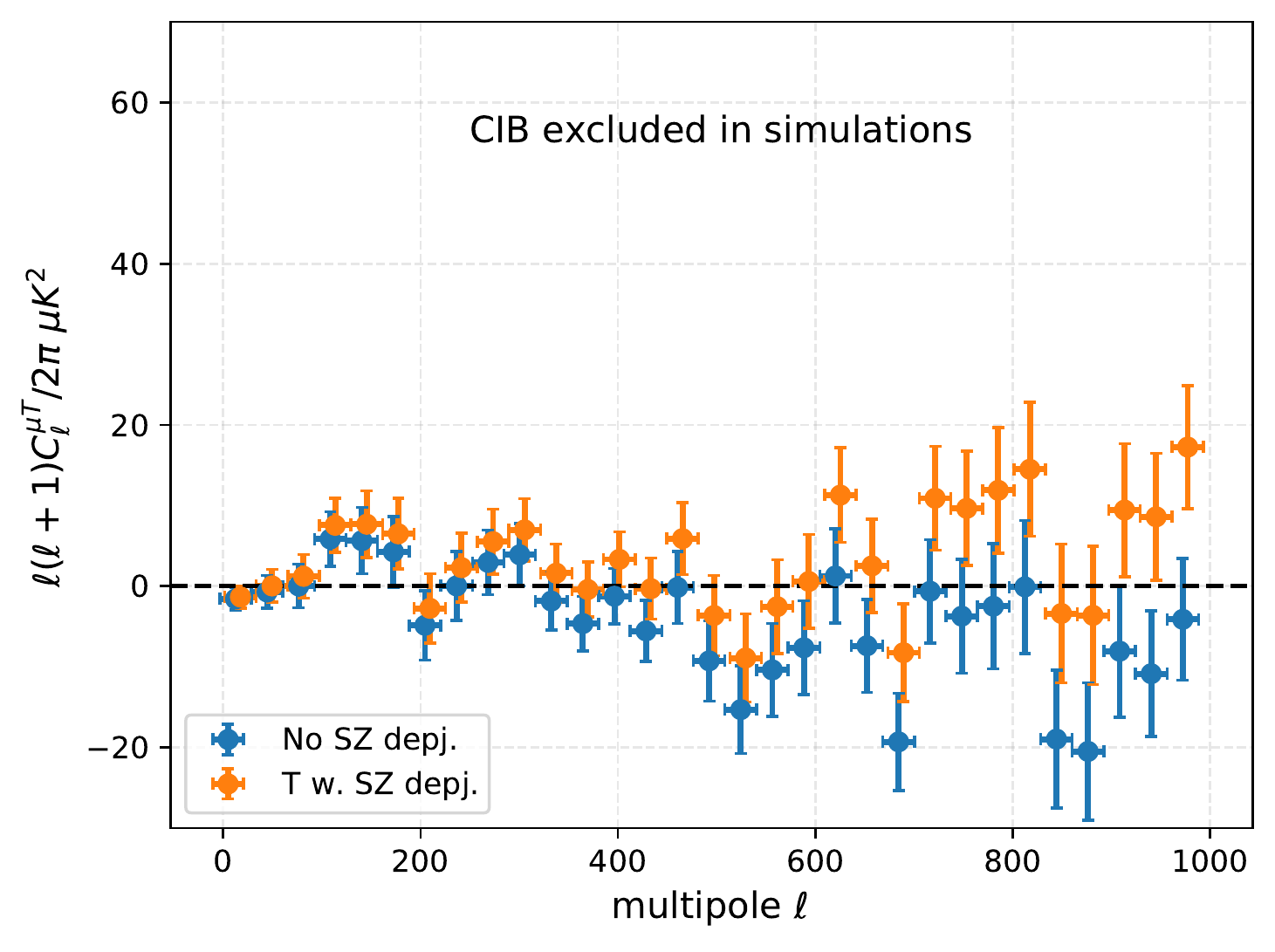}
\caption{Same as \fig{fig:ideal_ilc_mt_me}, except the analysis is done on realistic sky simulations that exclude CIB.}
\label{fig:muT_no_cib}
\end{figure*}

While these correlation certainly indicate that CIB could contribute a positive bias it is not proof that it does. To solidify this claim we perform another exercise where in we generate simulations that do not include CIB. Now we expect the negative SZ bias to re-appear as we have completely removed the source of the positive bias. The negative SZ bias can now be dealt with by performing a SZ deprojection when reconstructing the temperature anisotropy map as we had demonstrated on idealised simulations in \sec{sec:ideal_planck_sim}. This is exactly what we find and the results of this exercise are summarized in \fig{fig:muT_no_cib}.

While we have answered why there is no SZ bias in the $\mu T$ measurements in data, it is still very intriguing that the biases induced by CIB and SZ should near perfectly cancel each other. A detailed exploration of this coincidence we leave to future work.

\vspace{-3mm}
\section{Future perspective}
\label{sec:NextGenerationForecast}
Given the accuracy of the Fisher forecast based on power-spectrum-based cILC, we venture into forecasting the constraining power of future surveys (see Table~\ref{tab:LiteBIRD_Forecasts} for a summary of all results).
We consider the baseline \LiteBIRD configuration \citep{Hazumi2019} to compare with the results of a more accurate analysis presented in \citep{Remazeilles2021mu}, finding excellent agreement ($\Delta \sigma_{\fNL} \! = 7\%$).
The errors from the analysis on actual \Planck data is found to be larger than about $\approx 40\%$ than those forecasted using Fisher and therefore we expect the forecasts presented in this section be too optimistic by a similar amount.

\begin{table}
\centering
\begin{tabular}{llll}
\toprule
configuration & \multicolumn{3}{c}{$\sigma_{f_{\rm NL}}$} \\
    & $T$ & $E$ & $T\!+\!E$ \\
\midrule
LiteBIRD                                    & 1194  & 992  &	878 \\
LiteBIRD + \Planck 30 GHz                   & 1166  & 971  &	860 \\
LiteBIRD + \Planck 545 GHz 875 GHz          & 1049  & 859  &	764 \\
LiteBIRD + all \Planck channels             & 1015  & 835  &	742 \\
\bottomrule
\end{tabular}
\caption{Forecasted \fnl errors  from $\mu T$ and $\mu E$ measurements for \LiteBIRD and combinations of \LiteBIRD and \Planck channels assuming $\fsky = 0.65$.}
\label{tab:LiteBIRD_Forecasts}
\end{table}

\begin{figure*}
\includegraphics[width=\columnwidth]{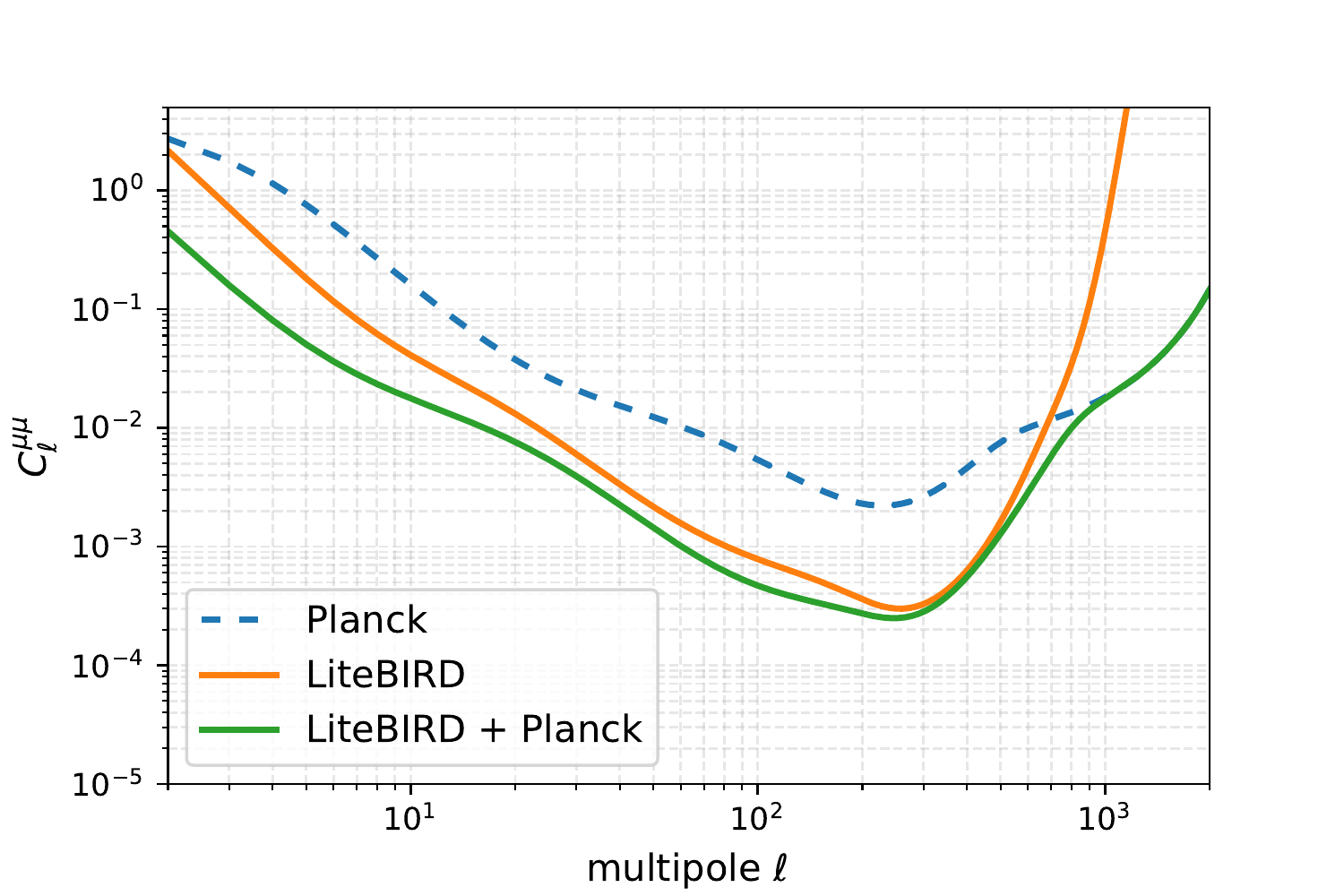}
\caption{$\mu$-distortion power spectrum for different instrumental configurations. The spectrum is dominated by measurement noise and foreground residuals.}
\label{fig:mumu_LB_Planck}
\end{figure*}

While $T$ measurements from \Planck data are already cosmic variance limited, \LiteBIRD will deliver cosmic variance measurements of the E-mode of CMB polarization.
\LiteBIRD is expected to improve the \fnl constraints by a factor of $\sim 3$ owing to its enhanced sensitivity \citep{Remazeilles2021mu}. We report that complementing future \LiteBIRD measurements with existing \Planck data can enhance the \fnl constraining power by $\sim 18 \%$.
These improvements in the \fnl constraining power are  driven by reduction in noise power in the reconstructed $\mu$ maps, as shown in \fig{fig:mumu_LB_Planck}. Further exploration reveal that since \Planck low frequency channels measure the sky at  almost 5 times poorer sensitivity as compared to \LiteBIRD and it does not extend the low frequency lever arm by much, it only contributes to improving the \fnl constraints by a mere $\sim 2 \%$. On the other hand, we find most of the gains are driven by the inclusion of the two highest frequency \Planck channels, which serve as dust and CIB monitors. This overall reduction in $\mu$ power is reflected in the $\sim 18 \%$ lower errors on \fnl, both from $\mu T$ as well as $\mu E$ measurements as summarized in \tab{tab:LiteBIRD_Forecasts}.
%


\vspace{-3mm}
\section{Discussion and conclusions}
\label{sec:conclusions}
While measurements of CMB anisotropies constrain the properties of the Universe at $k \simeq 0.005 \,{\rm Mpc}^{-1}$, the $\mu$ distortions are sourced by dissipation damping at large wavenumber corresponding to $k \simeq 10^3 {\rm Mpc}^{-1}$. \Planck being a differential experiment does not see the sky monopole and hence does not allow us to directly constrain the global $\mu$ signal. However, primordial non-Gaussianity can induce correlations between small- and large-$k$ modes, and source $\mu$ distortion anisotropies. These correlations between large and small wavelength modes can be probed by measuring cross correlating the $\mu$ anisotropies with CMB temperature as well as the E-mode maps. In this work we have delivered the first robust constraints from \Planck data on these correlations allowing us to constrain the highly squeezed state bispectrum characterized by \fnl.

We developed and validated our component separation pipeline by rigorously testing its performance on simulations and ensuring that we obtain an unbiased detection of the injected signal in a variety of cases.
After these tests, we finally apply this analysis pipeline to \Planck data to derive the first $\mu$ map using {\it all} the \Planck channels as well as re-extracting the standard CMB $T$ and $E$ mode maps. We validate the recovered $T$ and $E$ maps by comparing them against \Planck SMICA maps. Using the measurement of the $\mu T$ spectrum we derive an \fnl measurement of $\fNL = -4273 \pm 4383$ and similarly using the $\mu E$ spectrum we find $\fNL = 3937 \pm 4719 $. We find the $\mu T$ and $\mu E$ measurements to yield fully compatible measurements on the \fnl parameter with no notable trend when we vary the multipoles used in the analysis. We find the best constraint of $\fNL = -840 \pm 3398$, by combining the $\mu T$ and $\mu E$ measurements that used 62\% of the sky and multipoles up to $\ell_{\rm max}=1024$. Given that the measurements are highly consistent with zero we set a 95\% $(2 \sigma)$ upper bound on on $|\fNL| \leq 6800$, improving the previously-quoted conservative estimates by a factor of $\simeq 15$ (see \sec{sec:planck_muT_compare} for discussion).

We highlight that in reality the amplitude of the $\mu T$ and $\mu E$ spectra is set by the product $\langle \mu \rangle \fNL$ \citep{Chluba:2016aln}. When deriving the constraints on \fnl we have assumed $\langle \mu \rangle = 2.3 \times 10^{-8}$, which corresponds to the amplitude expected from dissipation of acoustic modes in our fiducial cosmology \citep{Chluba2016}. This invariable implies that without an independent measurement of $\langle \mu \rangle$ one is indeed unable to interpret the obtained constraint in a model-independent way. In addition, models with significant $\fNL$ could also enhance the level of $\langle \mu \rangle$ further highlighting the necessity to perform absolute spectroscopy in tandem with future CMB imaging.

When we carry out this analysis we were met with a few surprises. We found that it was crucial to use the \Planck RIMO beams when performing the component separation as opposed to using an effective Gaussian beam. Not doing so leads to a $T$ to $\mu$ leakage at the level of 0.5\% which introduces a huge spurious signal in the $\mu T$ as well as the $\mu E$ measurements. This is indicative of this analysis exploiting the full capability of \Planck as it was important to account for percent level affects to obtain proper measurements. We also demonstrated that the $\mu T$ measurements benefit from a subtle and coincidental quasi-exact cancellation of biases sourced by the presence of residual tSZ and CIB in the component maps. The subtleties of this analysis put a spotlight on the level of detail one will have to cope with when actual spectral distortion measurements will be attempted in the future.

Fianlly we have shown that the Fisher ILC frame work is able to estimate the foreground contribution to the total noise budget rather accurately, matching the actual component separation estimates on errors at the level of few {10\%}. This is a very useful and powerful forecasting tool that can provide accurate forecasts in a fast and efficient manner. We have used this to asses that \Planck, owing primarily to its high angular resolution and high frequency coverage, will be able to help improve \LiteBIRD ability to constrain \fnl by $\simeq 18\%$, indicating that \Planck's measurement will continue to be very relevant in the future.


\section*{Data Availability} Reconstructed $\mu$, $T$ \& $E$ maps are available upon reasonable request.

\vspace{3mm}
{\noindent \small {\it Acknowledgments}:
The authors thank David Alonso and  Will Coulton for stimulating discussion of the results relating to SZ and CIB biases. The authors also thank useful correspondence with Eiichiro Komatsu and Reijo Keskitalo.
This work was supported by the ERC Consolidator Grant {\it CMBSPEC} (No.~725456) as part of the European Union's Horizon 2020 research and innovation program.
JC was also supported by the Royal Society as a Royal Society URF at the University of Manchester.
ARa acknowledges support by the project "Combining Cosmic Microwave Background and Large Scale Structure data: an Integrated Approach for Addressing Fundamental Questions in Cosmology", funded by the MIUR Progetti di Ricerca di Rilevante Interesse Nazionale (PRIN) Bando 2017 - grant 2017YJYZAH.}

\bibliographystyle{mnras}
\bibliography{Lit,Lit1,Lit1_ref,Bibliografia} 

\begin{appendix}

\vspace{-3mm}
\section{Comparison of $T$ and $E$ mode power spectra with the true injected signal}
\label{app:sim_spec_compare}
\begin{figure*}
\hspace*{-0.18cm} 
\subfigure[\label{fig:sim_tt}]{\includegraphics[width=\columnwidth]{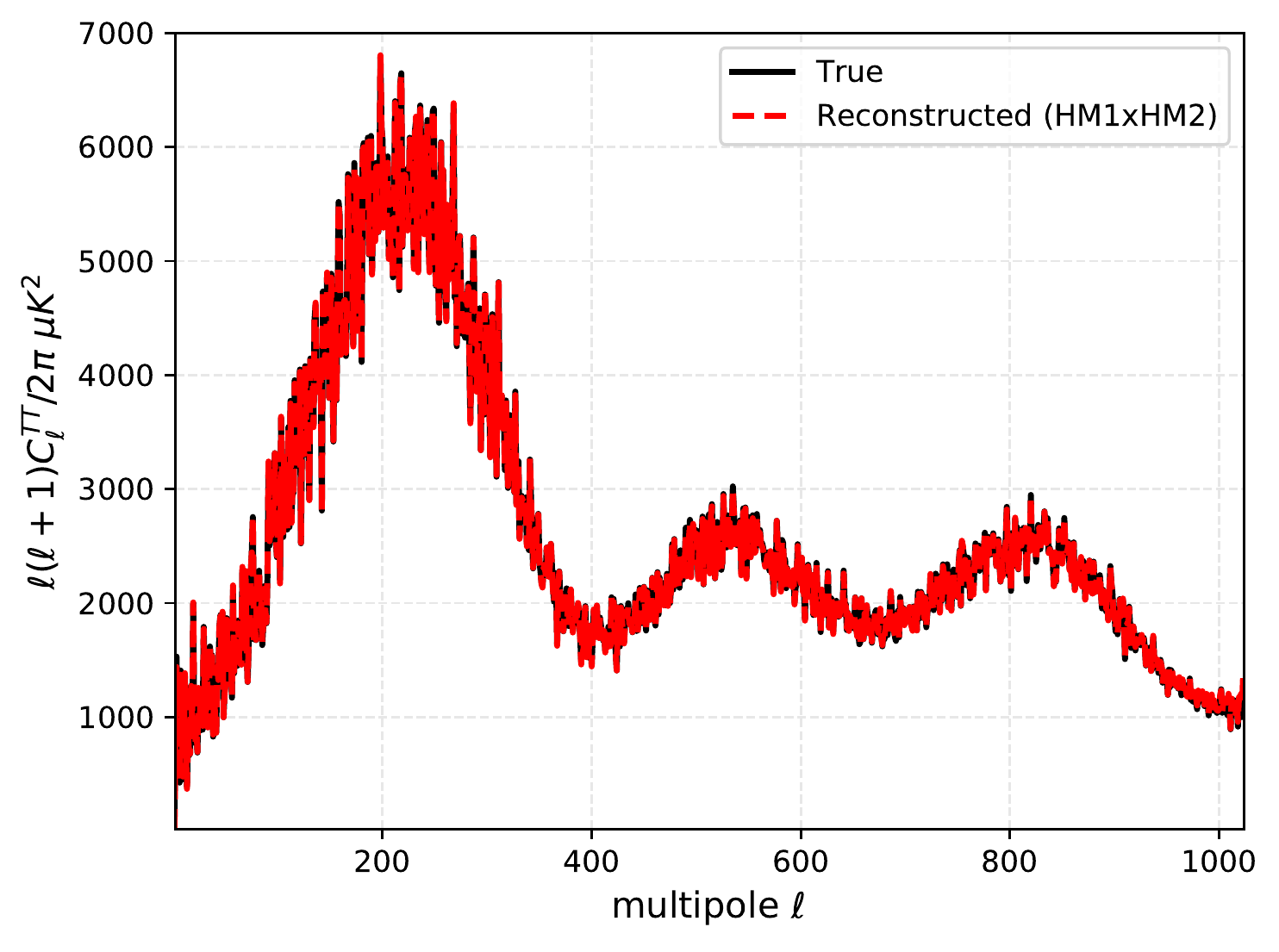}}
\subfigure[\label{fig:sim_tt_bias}]{\includegraphics[width=\columnwidth]{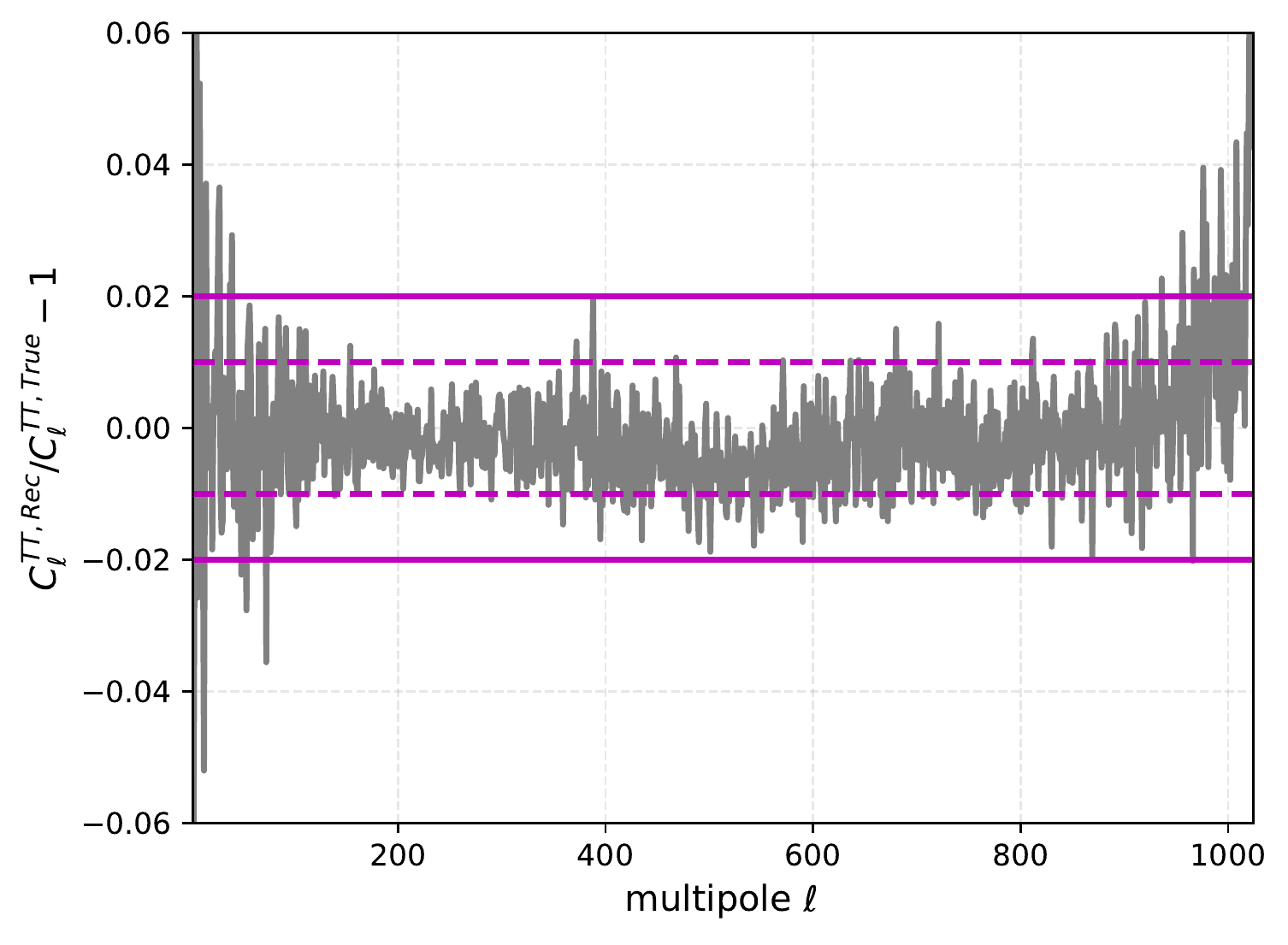}}
\subfigure[\label{fig:sim_ee}]{\includegraphics[width=\columnwidth]{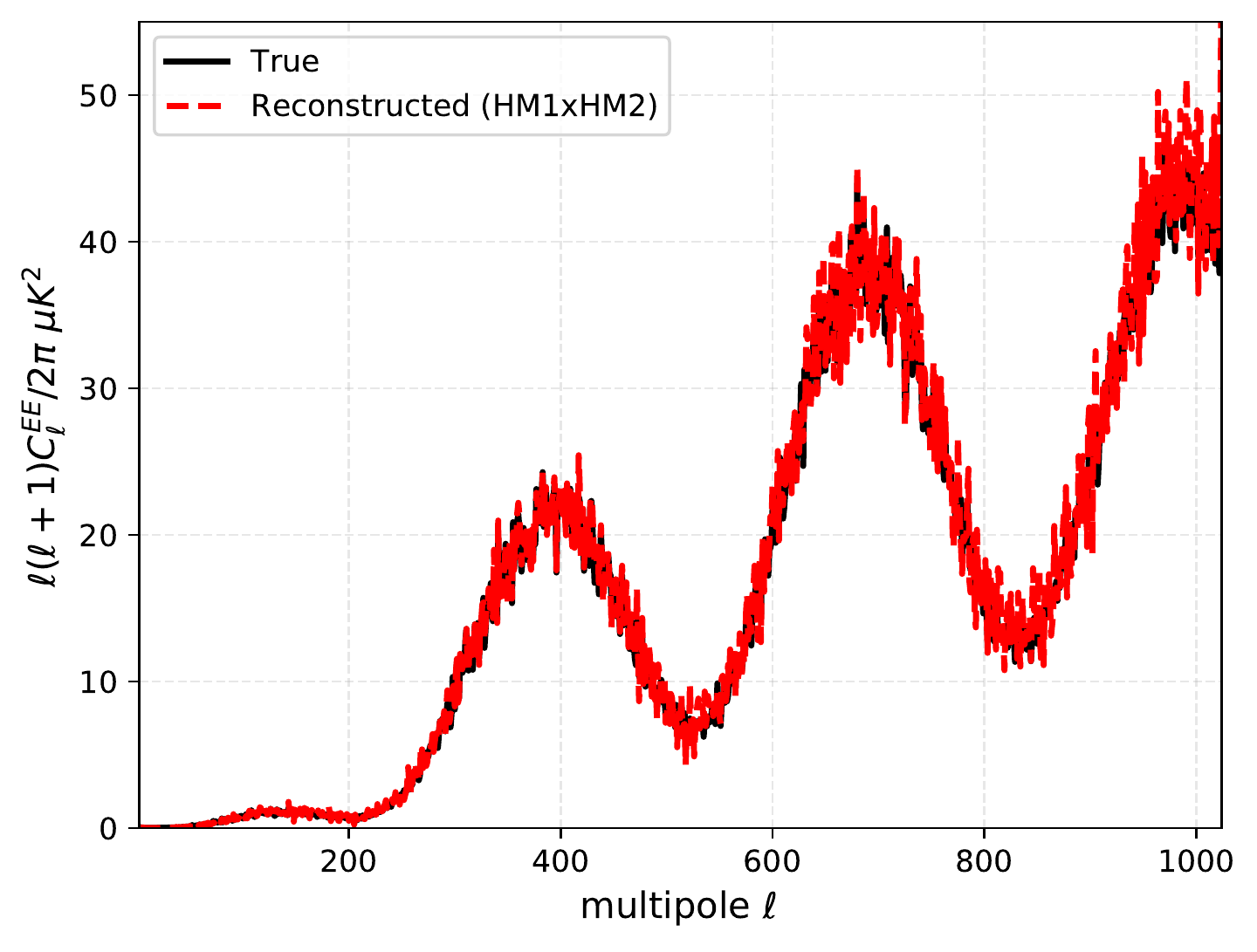}}
\subfigure[\label{fig:sim_ee_bias}]{\includegraphics[width=\columnwidth]{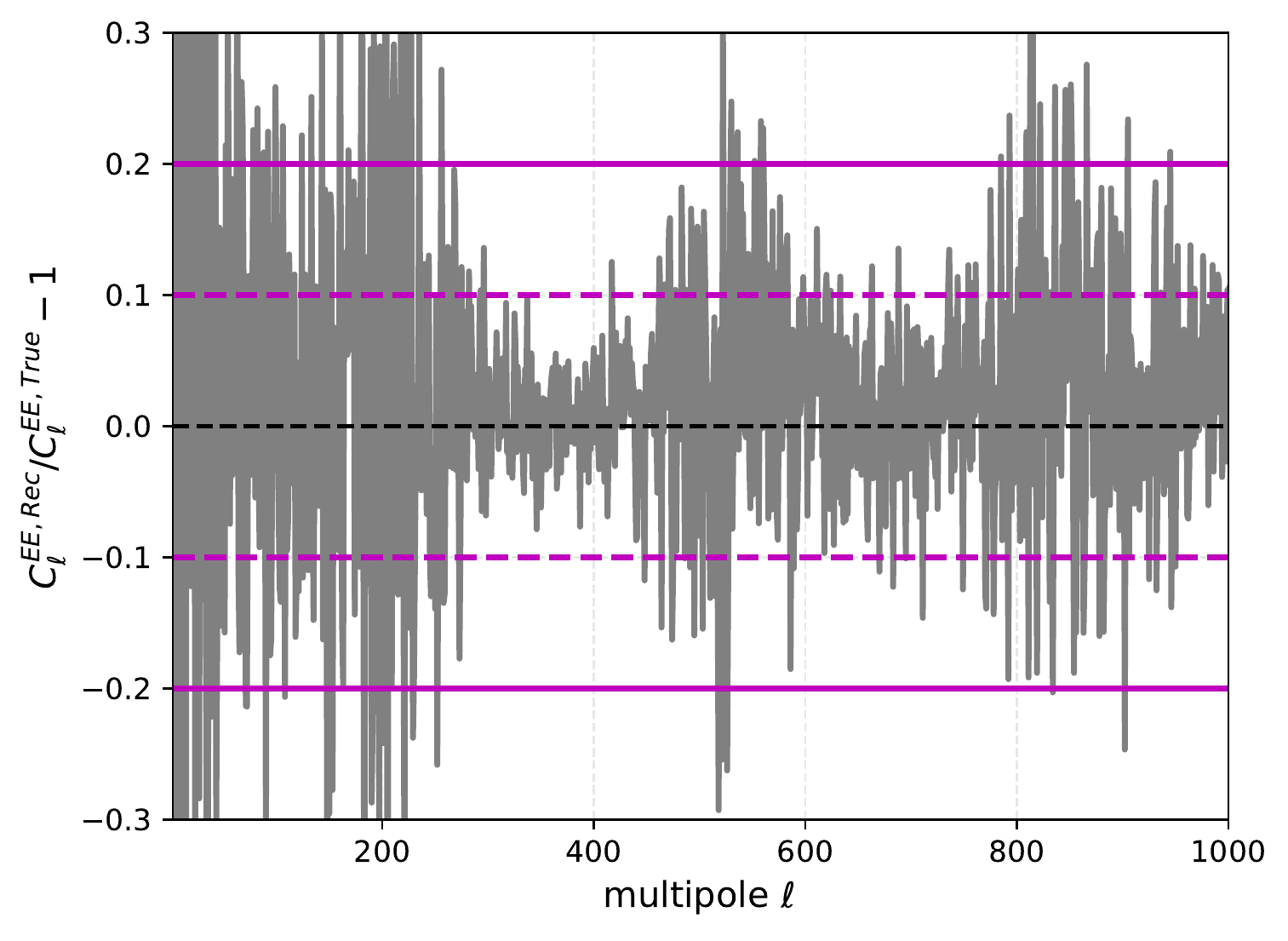}}
\subfigure[\label{fig:sim_te}]{\includegraphics[width=\columnwidth]{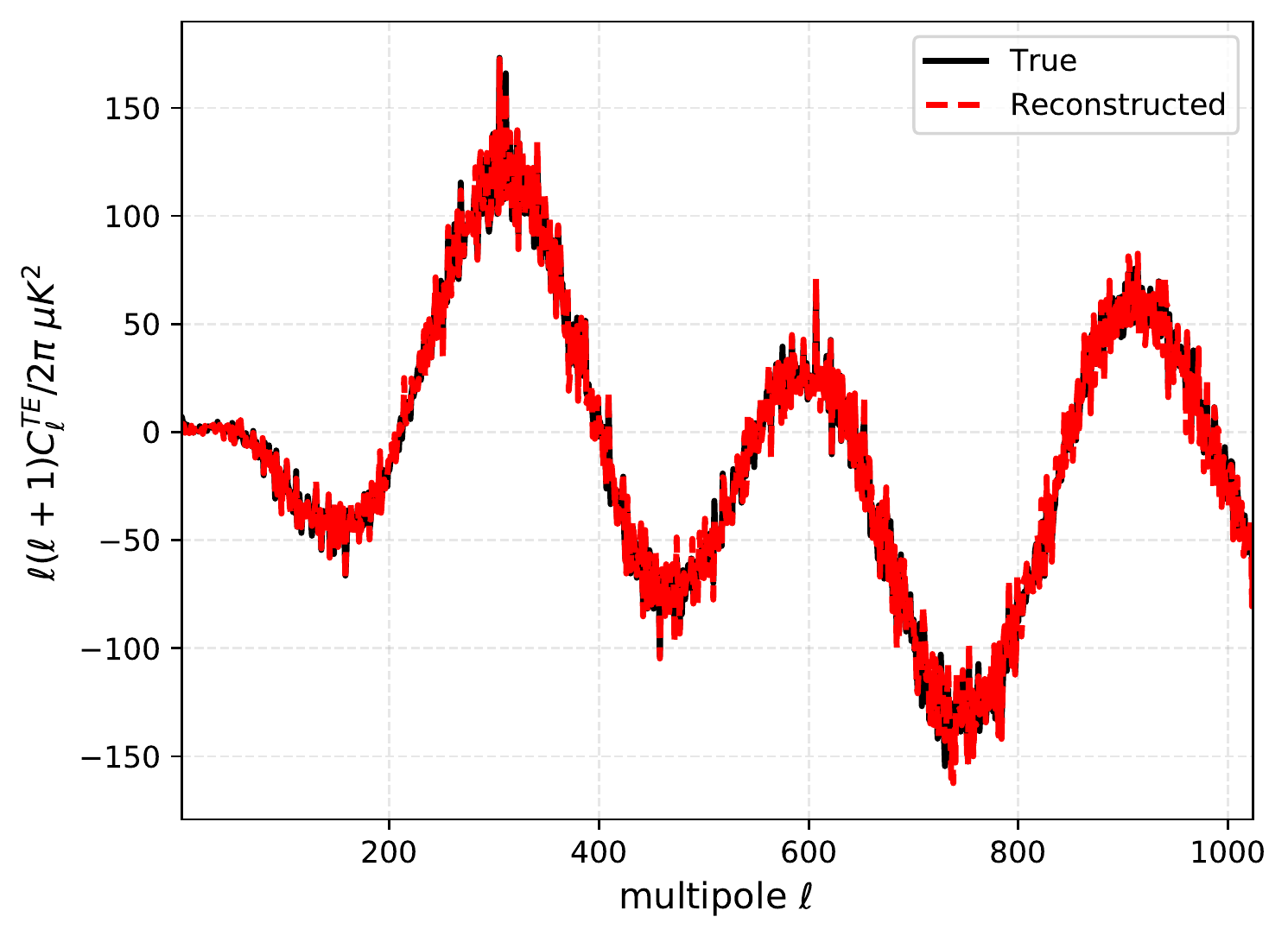}}
\subfigure[\label{fig:sim_te_bias}]{\includegraphics[width=\columnwidth]{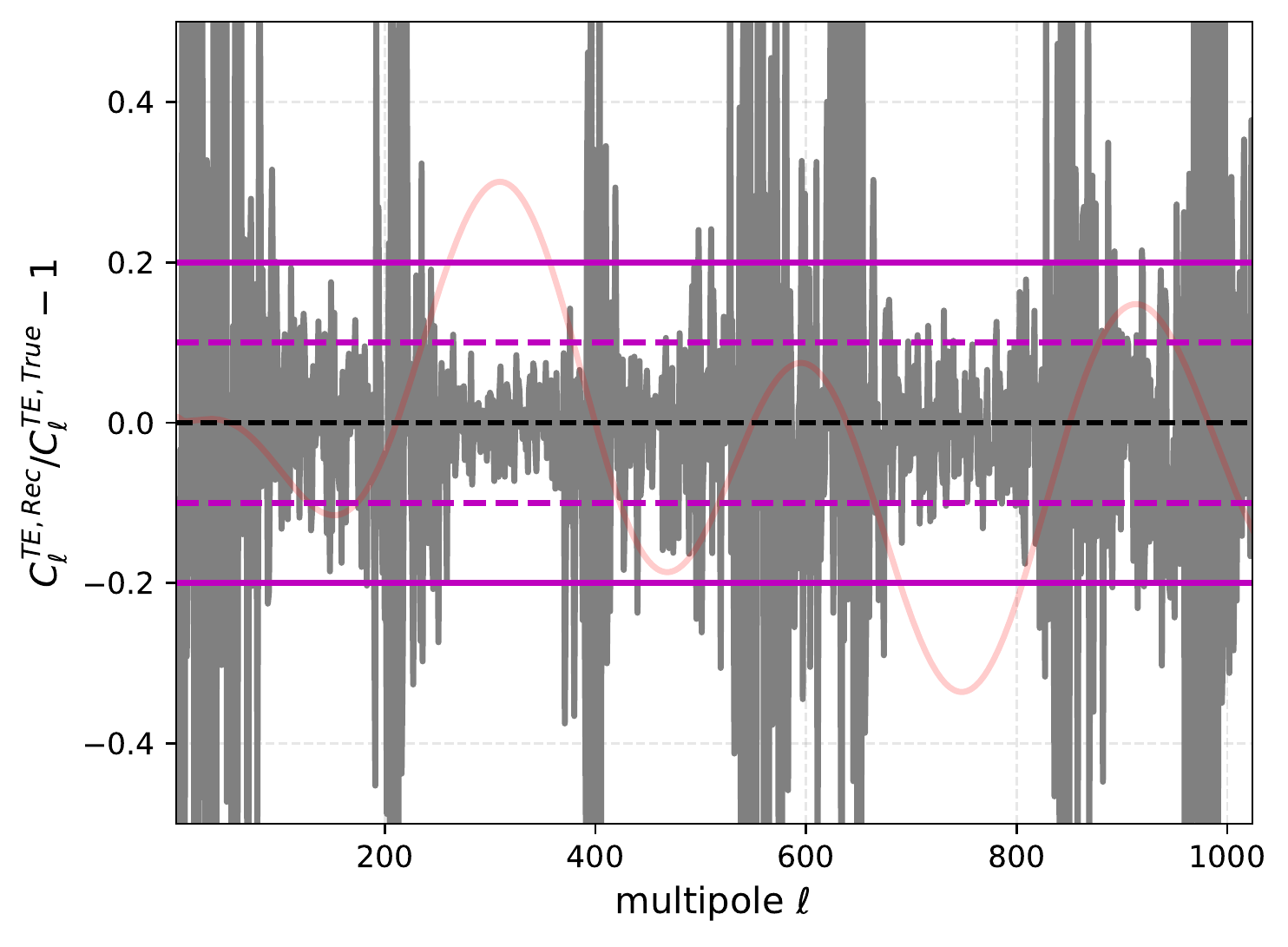}}
\caption{The figures showcases the recovery of the standard CMB spectra and their comparison to the true spectra estimated from the true injected maps. The panels shows the comparison of the recovered spectra to the true spectra, while the figures on the right show the relative error between the two.}
\label{fig:sim_fid_cmb_spec_compare}
\end{figure*}

The main signals are the $C_{\ell}^{\mu T}$ \& $C_{\ell}^{\mu E}$ spectra were extensively discussed in the main text. When carrying out these measurements, one of the key steps is the recovery of the CMB component maps, specifically $T$ and E. Here we showcase the quality of recovered component separation maps by comparing the power spectra with those of the true injected CMB maps as shown in \fig{fig:sim_fid_cmb_spec_compare}. The $TT$ and $EE$ spectra are estimated by cross correlating the component maps derived from analysis on the two half mission simulations and therefore do not need any noise bias corrections. We find very good agreement between the recovered and true spectra further validating our component separation pipeline. Since E-mode maps are noise dominated at most multipoles, the relative errors in the $C_{\ell}^{EE}$ and $C_{\ell}^{TE}$ are significantly larger than those observed for the recovered $C_{\ell}^{TT}$  spectrum. 
%

\section{Comparison of $T$ and $E$ mode power spectra to those derived from SMICA}
\label{app:smica_spec_compare}
\begin{figure*}
\hspace*{-0.18cm} 
\subfigure[\label{fig:smica_tt}]{\includegraphics[width=\columnwidth]{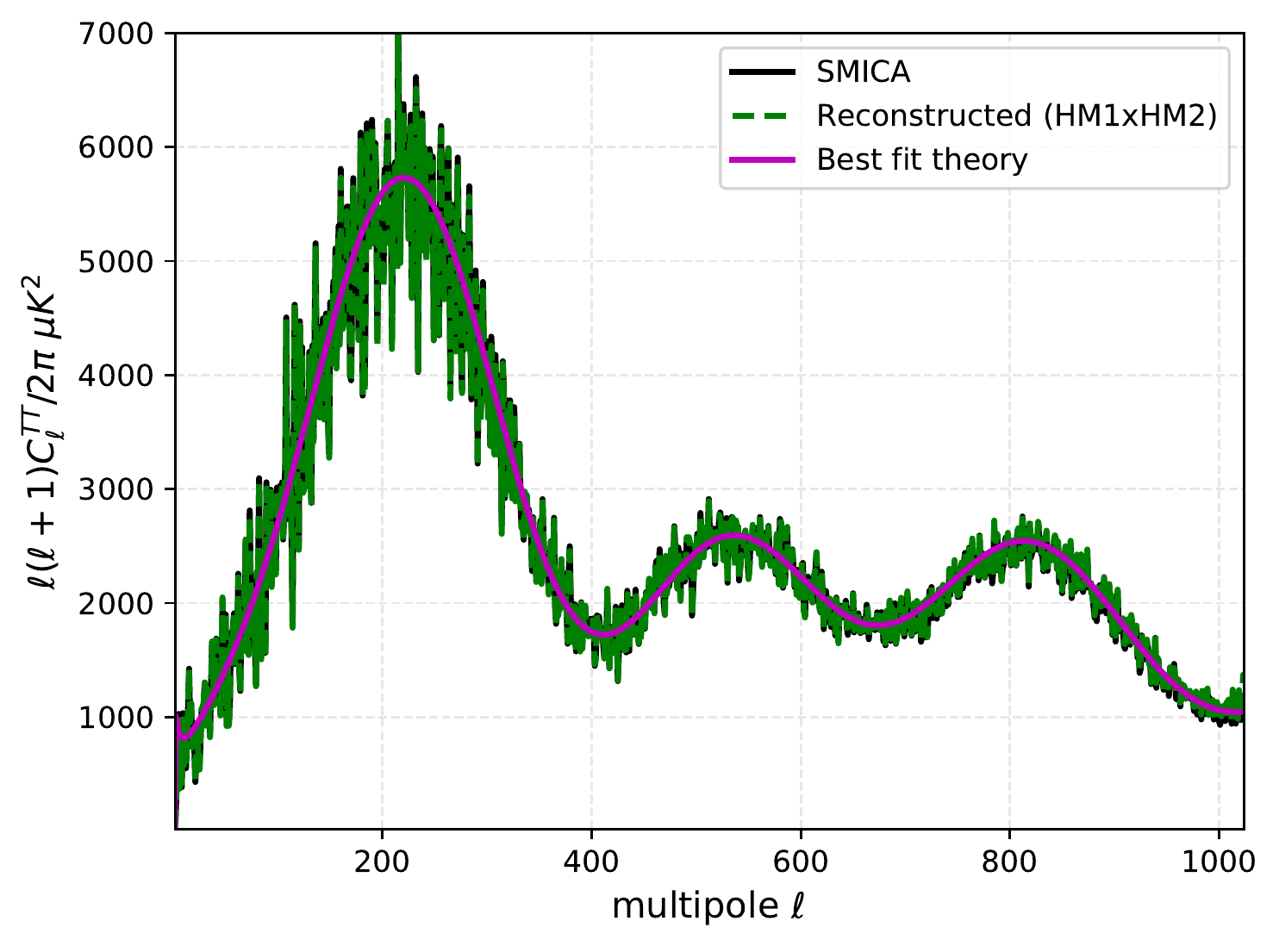}}
\subfigure[\label{fig:smica_tt_bias}]{\includegraphics[width=\columnwidth]{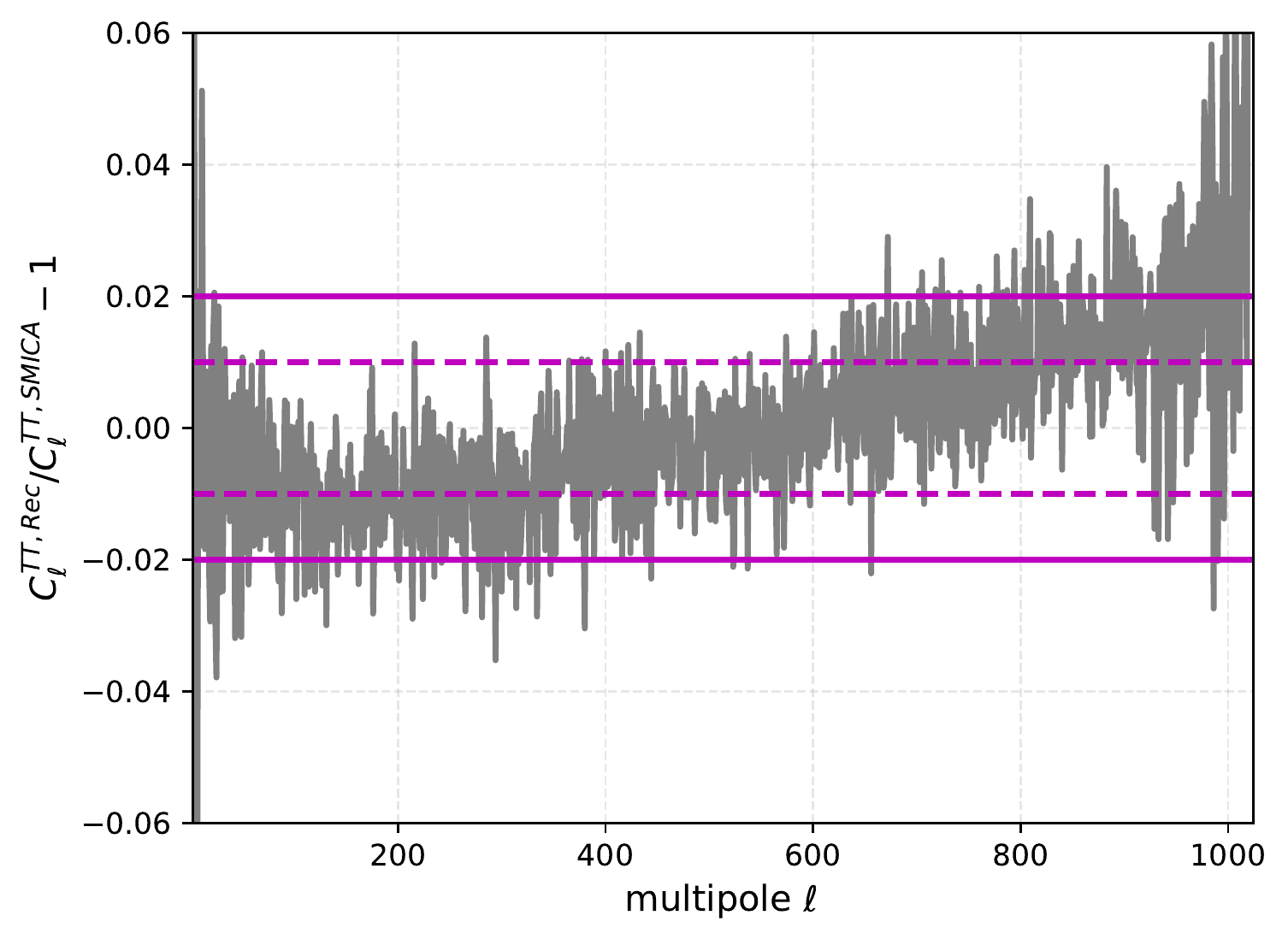}}
\subfigure[\label{fig:smica_ee}]{\includegraphics[width=\columnwidth]{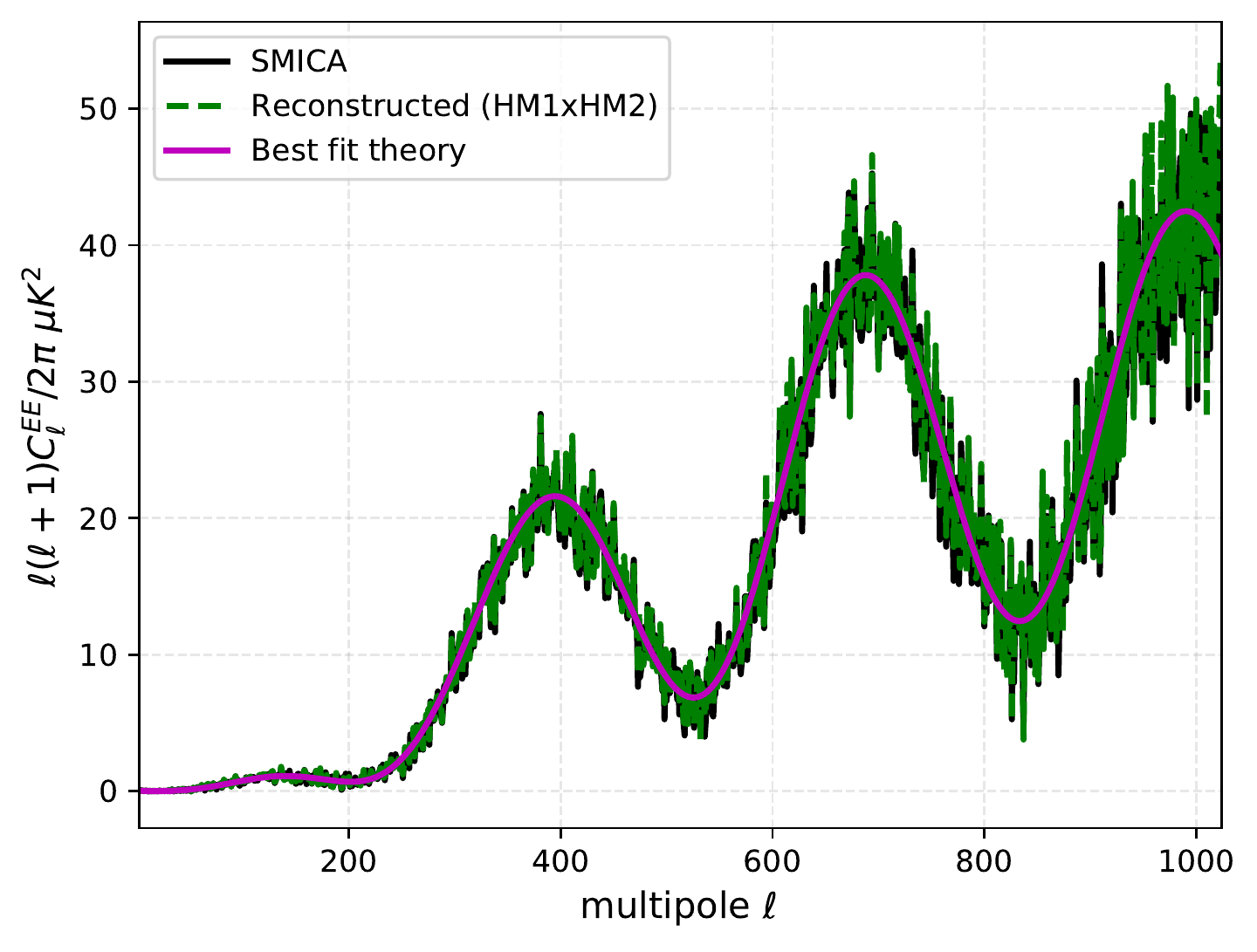}}
\subfigure[\label{fig:smica_ee_bias}]{\includegraphics[width=\columnwidth]{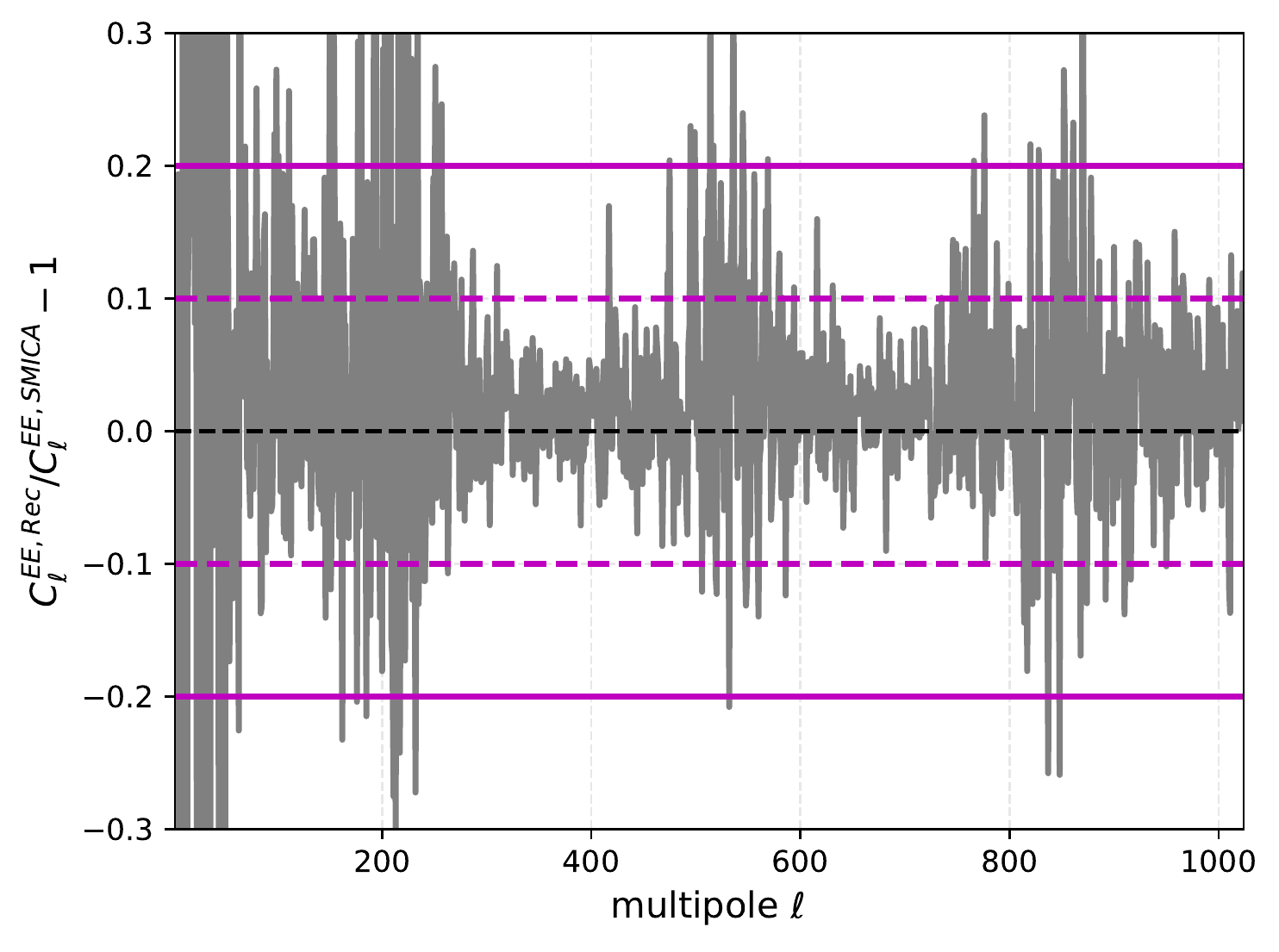}}
\subfigure[\label{fig:smica_te}]{\includegraphics[width=\columnwidth]{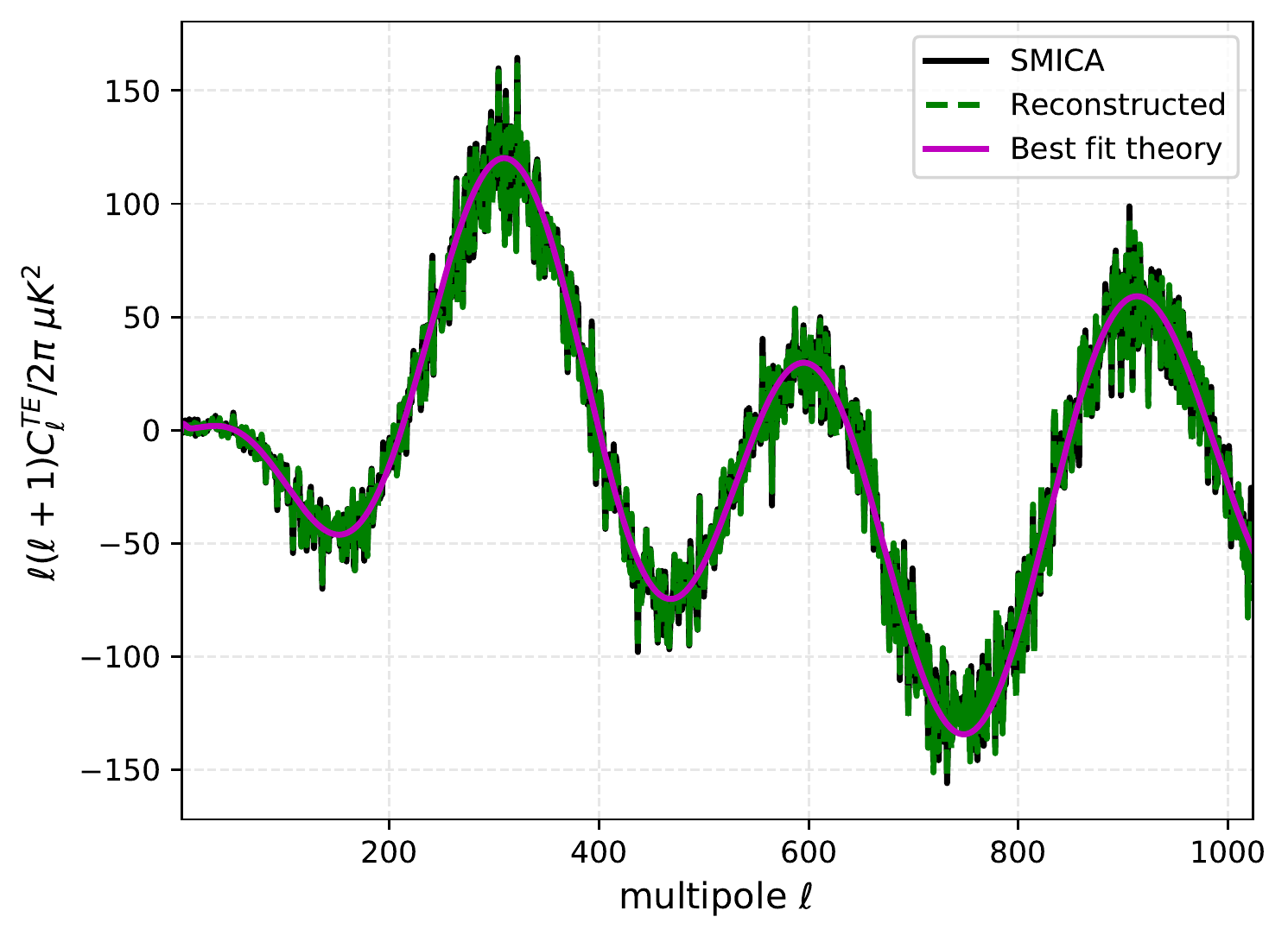}}
\subfigure[\label{fig:smica_te_bias}]{\includegraphics[width=\columnwidth]{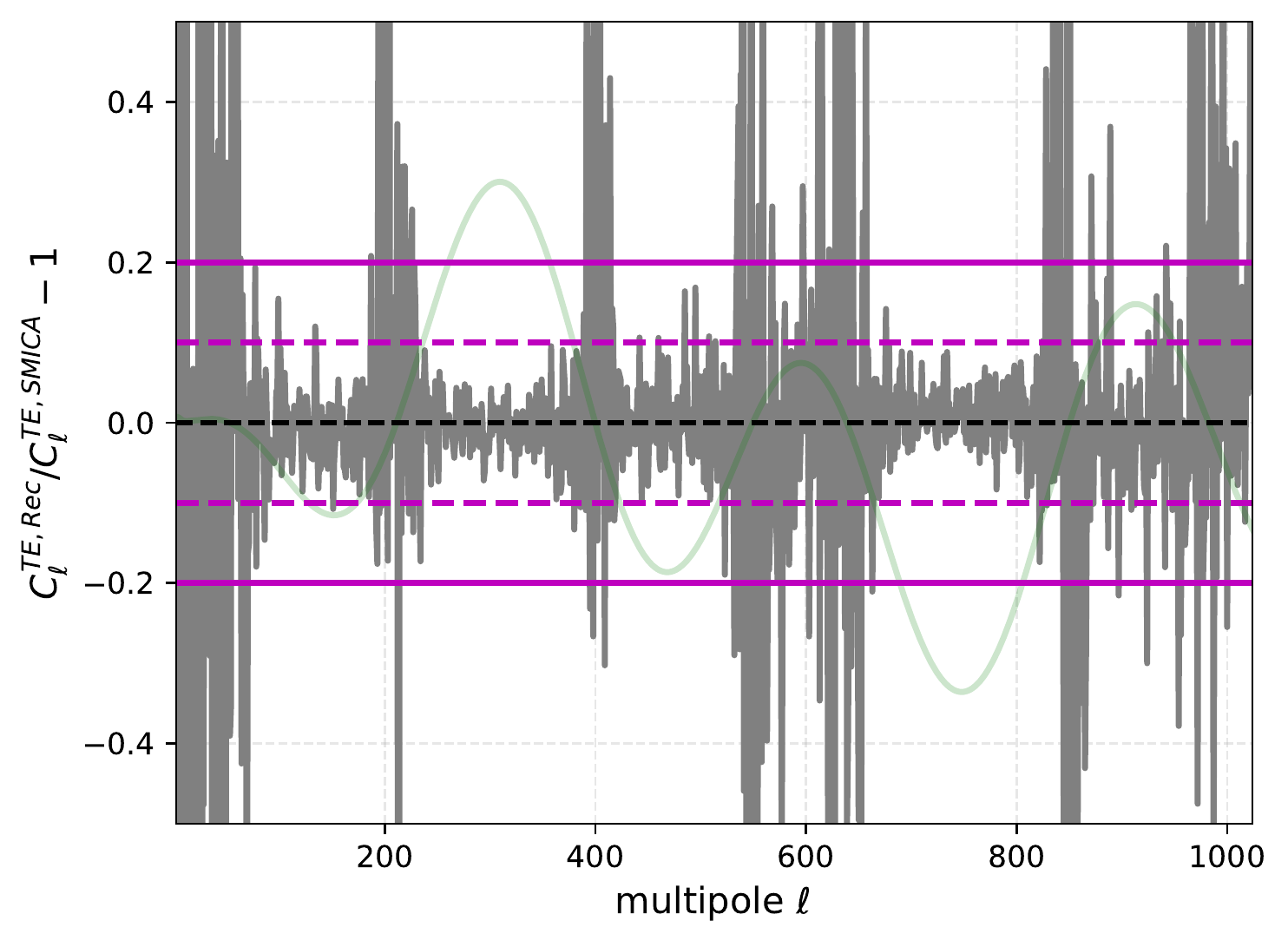}}
\caption{The figures showcases the recovered the CMB spectra and their comparison to the spectra derived from the corresponding SMICA component maps. The panels on left depict a comparison of the two recovered spectra, while the figures on the right show the relative different between the two.}
\label{fig:smica_cmb_spec_compare}
\end{figure*}

Here we compare the power spectra corresponding to the component separation maps derived from \Planck data to the equivalent maps delivered by the \Planck collaboration using the SMICA pipeline. Since we showcase the spectra derived from cross correlating component maps derived from half mission data, these spectra represent an unbiased measurement of the true sky CMB maps. A direct comparison of the power spectra derived from the NILC algorithm employed in this work and analogous quantities derived from SMICA maps as well as the relative difference between the two estimated spectra are show in \fig{fig:smica_cmb_spec_compare}. These spectral estimates from our component separated maps are very consistent to those derived from SMICA component maps. Notably the scatter seen in the fractional difference between the two estimates is larger than that seen in the corresponding figures in \sec{app:sim_spec_compare}. This is because here we compare two noisy estimates of the corresponding power spectra, while in \fig{fig:sim_fid_cmb_spec_compare} we compare the spectra estimated from the component separated maps to the those corresponding to the true injected CMB skies.
%
\section{Instrumental effects}
\label{app:instr_model}
It is important to accurately correct for the different instrument properties in the analysis. The \textit{Planck} collaboration makes available the Reduced Instrument Model (RIMO) which includes a beam profiles in multipole space as well as the respective band pass corrections. 

\begin{figure*}
\hspace*{-0.18cm} 
\subfigure[\label{fig:rimo_vs_fid}]{\includegraphics[width=\columnwidth]{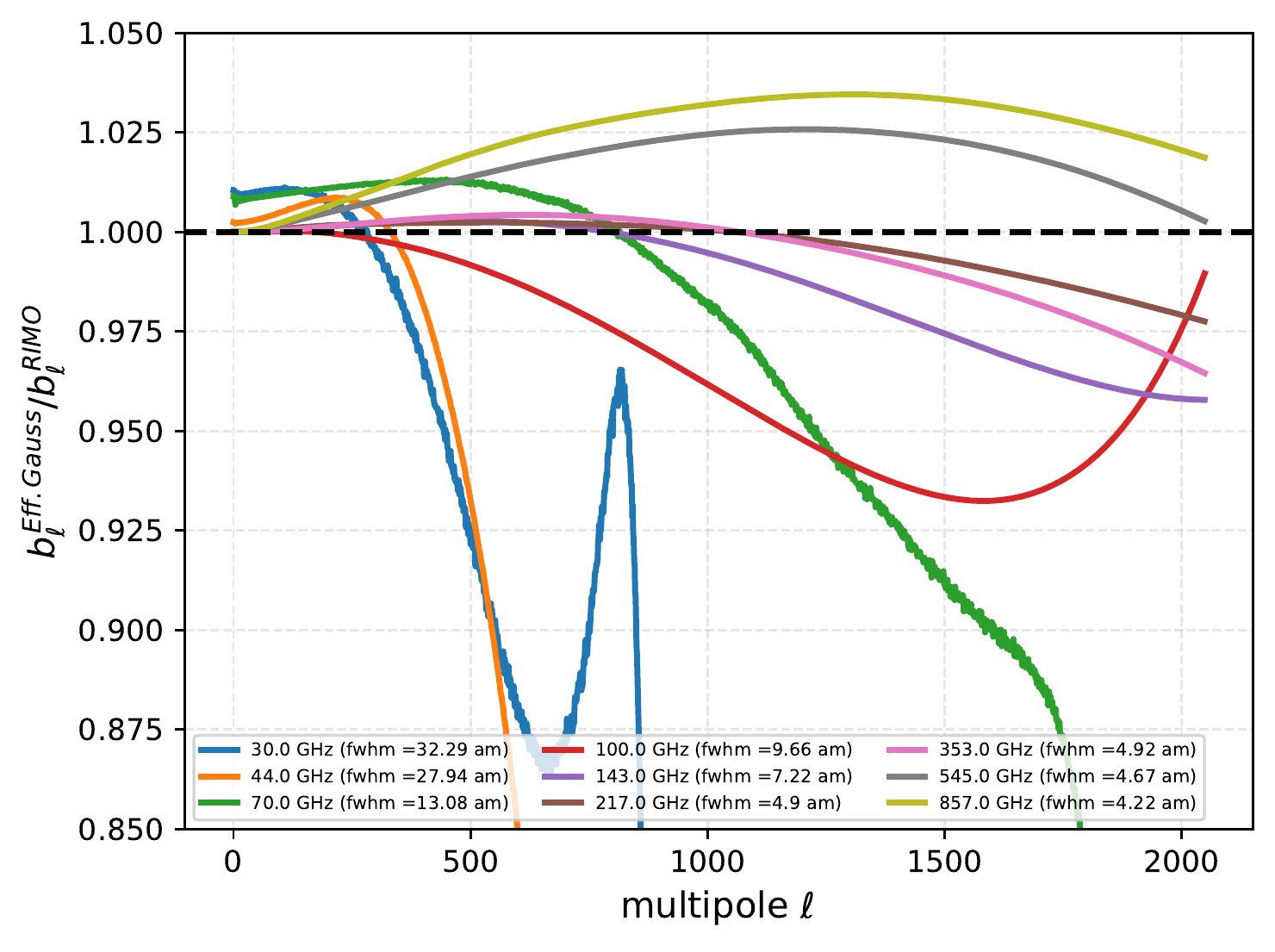}}
\subfigure[\label{fig:pwc}]{\includegraphics[width=\columnwidth]{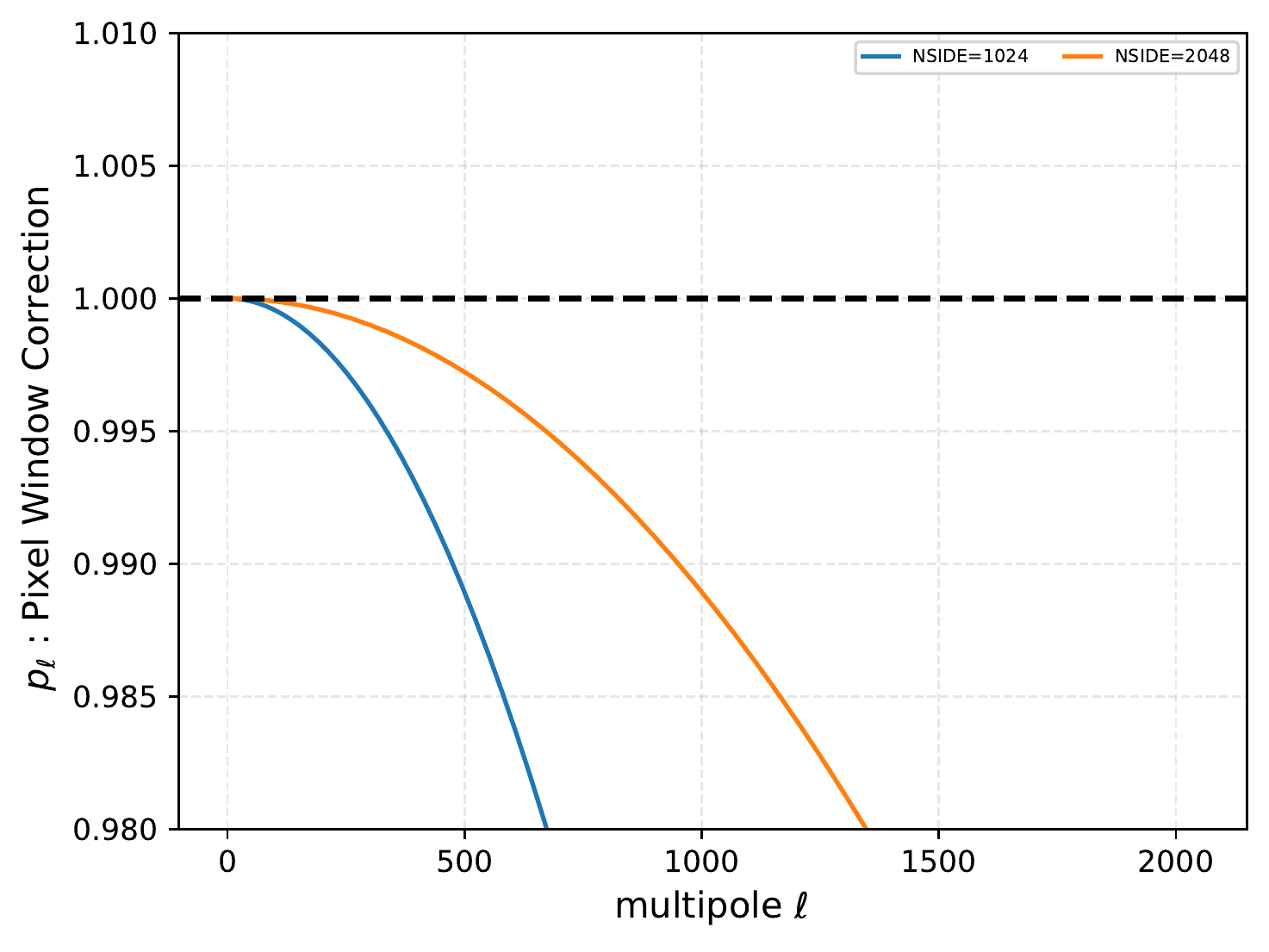}}
\caption{\textit{Left: }The figure depicts the ratio of the RIMO beam and the effective Gaussian beam characterized by the \textrm{FWHM} for the respective \textit{Planck} channels.  \textit{Right: } This figure depicts the pixel window correction for the two resolutions at which the different \textit{Planck}  maps are made available.}
\label{fig:instr_corr}
\end{figure*}

In this work we account for the band pass correction as well as the instrumental beam which changes as a function of frequency.  We incorporate the band pass corrections by forward modeling their effects on the spectra used in the component separation pipeline. We find this correction to be less critical to the analysis. 
We find using the correct instrumental beams to perform the deconvolution of the \Planck maps to be particularly important to the analysis presented here. The instrumental beam is corrected for by de-convolving the observed maps with the respective RIMO beam functions. The RIMO instrument beam profiles have important differences from effective Gaussian beams modeled characterized by their \textrm{FWHM} as shown in \fig{fig:rimo_vs_fid}.  Therefore using the effective Gaussian beams instead of the RIMO beams can be thought of as inducing a multipole dependent mis-calibration. 

\begin{figure*}
\hspace*{-0.18cm} 
\subfigure[\label{fig:muT_wo_rimo}]{\includegraphics[width=\columnwidth]{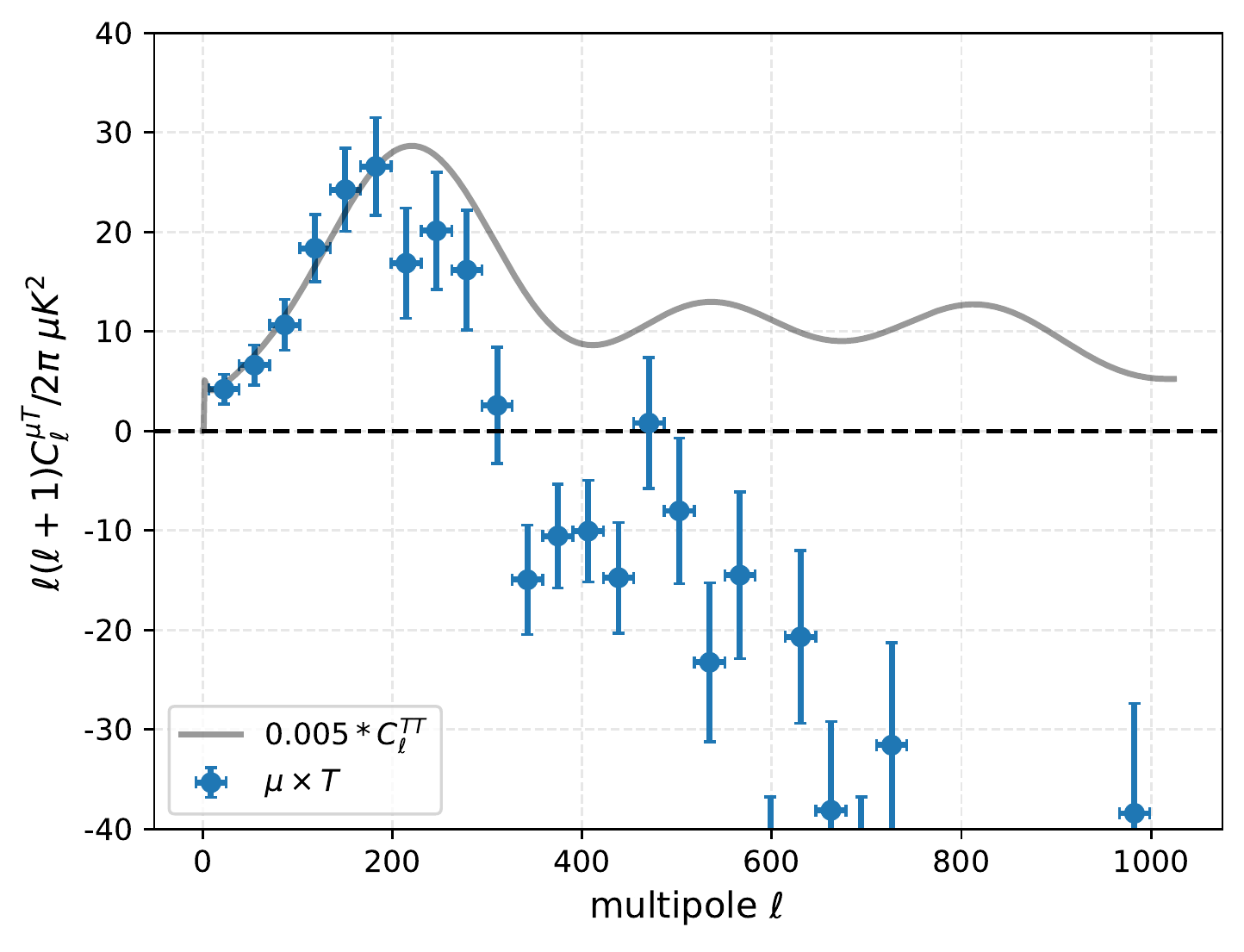}}
\subfigure[\label{fig:muE_wo_rimo}]{\includegraphics[width=\columnwidth]{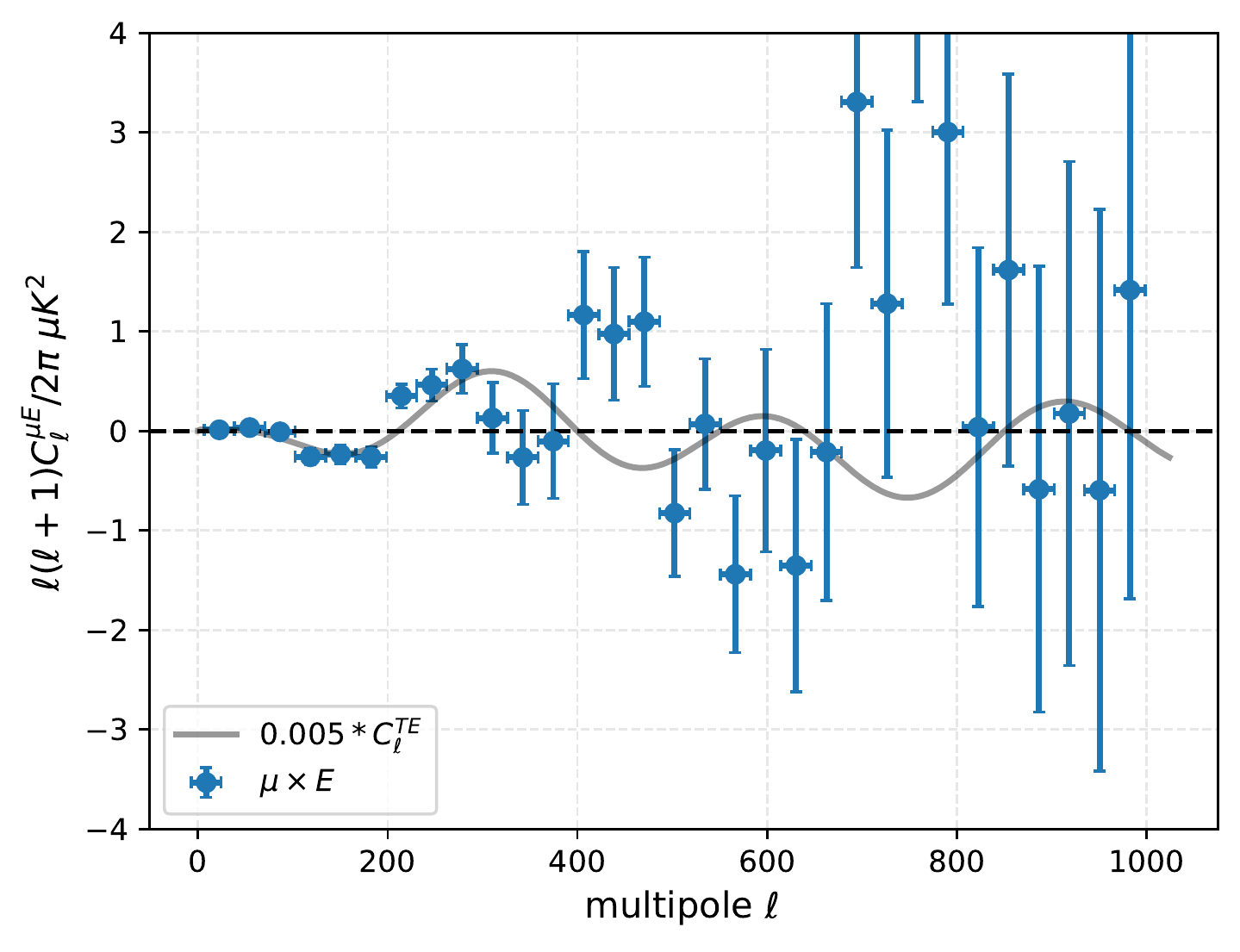}}
\caption{These figures depict the $\mu T$ and $\mu E$ measurements on using effective Gaussian beams to deconvolve \Planck maps as opposed to using the RIMO beams. The fiducial CMB spectra rescaled by a factor of 0.005 plotted for reference. Using approximate Gaussian beams as opposed to RIMO beams leads to 0.5\% leakage of $T$ to $\mu$ which explains the measurements at low multipoles $\ell \lesssim 250$}
\label{fig:RIMO_beam_importance}
\end{figure*}

The dominant effect of using the improper beams can be thought of as sourcing a $T$ to $\mu$  leakage which causes the $\mu T$ and $\mu E$ correlations to look like the $TT$ and $TE$ correlations, though with much smaller amplitudes, as seen in \fig{fig:RIMO_beam_importance}. 
We also find a similar effect when not properly correcting for the pixel window corrections, depicted in \fig{fig:pwc}. 
Note that though the most stringent constraints on \fnl are derived from the lowest multipoles $\ell \lesssim 400$ where the corrections due to the beam and the pixel window correction are only at the $\simeq 1\%$ level, these are enough to induce a non-vanishing signals.

%

\section{Covariance estimation}
\label{app:FM_HM_comparison}
As discussed, it is important to cross correlate different half mission data to remove the noise bias when estimating the $\mu T$ cross power spectrum. On the other hand, when estimating the $\mu E$ spectrum, we can work with full mission data, since $T$ and $E$ noise are independent and hence the measured spectrum has no noise bias. To infer \fnl measurements and the statistics on it we require the covariance of the measured spectra. We outline the key steps involved in covariance calculation below.

Our spectral measurements are of the form,
\begin{equation}
    \hat{C}_\ell^{\mu X}
    =
    \sum_m \frac{a^{\mu}_{\ell m} a^{X}_{\ell m}}{(2\ell + 1)}
\end{equation}
for $X\in[T,E]$. The variance for $\hat{C}_\ell^{\mu X}$ is then given by,
\begin{align}
\text{Var}\left(\hat{C}_\ell^{\mu X}\right) = & \,
    \left<\hat{C}_\ell^{\mu X} \hat{C}_\ell^{\mu X} \right>
    - \left<\hat{C}_\ell^{\mu X} \right>^2
\\
    = & \,
    \left<
    \sum_m \frac{a^{\mu}_{\ell m} a^{X}_{\ell m}}{(2\ell + 1)}
    \sum_{m'} \frac{a^{\mu}_{\ell m'} a^{X}_{\ell m'}}{(2\ell + 1)}
    \right> -
    \left<
    \sum_m \frac{a^{\mu}_{\ell m} a^{X}_{\ell m}}{(2\ell + 1)}
    \right>^2 \nonumber
\end{align}
Using the Wick theorem to expand the four point functions in products of two point functions, the above expression can be reduced to a simple form given in the main text. Here we focus on providing the specifics for variance estimate for the the half mission cross analysis that is used for $\mu T$ spectra measurement as well as its covariance with the full mission $\mu E$ measurements.

The estimator for the bias free $\mu T$ measurement is given by the following expression,
\begin{align}
\hCl^{\mu T} = \frac{1}{2} \left[ \hCl^{\mu_1 T_2} +  \hCl^{\mu_2 T_1} \right] \,,
\end{align}
where $[\mu_1,T_1]$ and $[\mu_2, T_2]$ denote the component maps estimated from the two half mission \Planck data respectively. Using the same prescription as described above, it can be shown that the variance of this spectrum is given by,
\begin{equation}
\label{eq:cross_half_mission_estimator_variance}
\begin{split}
	& \,
	4 \left(2\ell+1\right) \text{Var}\left( \hCl^{\mu T} \right)
	= 
	\left(\hCl^{\mu_1 T_2}\right)^2 + \left(\hCl^{\mu_2 T_1}\right)^2 + 2\, \hCl^{\mu_1 T_1}\hCl^{\mu_2 T_2}
\\
	& \, +
	\hCl^{\mu_1 \mu_1} \hCl^{T_2 T_2} + \hCl^{\mu_2 \mu_2} \hCl^{T_1 T_1} + 2 \hCl^{\mu_1 \mu_2} \hCl^{T_1 T_2} \,
\end{split}
\end{equation}
The best \fnl estimate is derived by combining the $\mu T$ and $\mu E$ spectral measurements. These spectra are not independent of each other and hence it is important to include the covariance of these spectra in the final likelihood. It can be shown that the covariance between these two spectral measurements is given by the following expression,
\begin{equation}
\begin{split}
    & \,
    2 \left(2\ell + 1 \right)\text{Cov}\left(\hCl^{\mu T}, \hCl^{\mu E} \right)
    =
\\
    & \,
    \hCl^{\mu_1 E} \hCl^{\mu T_2}
    +\hCl^{\mu_2 E} \hCl^{\mu T_1}
    +\hCl^{T_1 E} \hCl^{\mu \mu_2}
    +\hCl^{T_2 E} \hCl^{\mu \mu_1} .
\end{split}
\end{equation}
where $[\mu, T]$ refer to component maps derived from full mission data.

\end{appendix}

\bsp	
\label{lastpage}
\end{document}